\newcommand{\Ab}{\mathbf{A}}
\newcommand{\Bb}{\mathbf{B}}
\newcommand{\Cb}{\mathbf{C}}
\newcommand{\Db}{\mathbf{D}}
\newcommand{\Eb}{\mathbf{E}}
\newcommand{\Fb}{\mathbf{F}}
\newcommand{\Gb}{\mathbf{G}}
\newcommand{\Kb}{\mathbf{K}}
\newcommand{\Mb}{\mathbf{M}}
\newcommand{\Ib}{\mathbf{I}}
\newcommand{\Lb}{\mathbf{L}}
\newcommand{\Jb}{\mathbf{J}}
\newcommand{\Rb}{\mathbf{R}}
\newcommand{\Sbb}{\mathbf{S}}
\newcommand{\Tb}{\mathbf{T}}

\newcommand{\Vb}{\mathbf{V}}
\newcommand{\Wb}{\mathbf{W}}

\newcommand{\ab}{\mathbf{a}}
\newcommand{\bb}{\mathbf{b}}

\newcommand{\eb}{\mathbf{e}}

\newcommand{\ub}{\mathbf{u}}
\newcommand{\vb}{\mathbf{v}}
\newcommand{\xb}{\mathbf{x}} 
\newcommand{\mb}{\mathbf{m}} 

\newcommand{\bzero}{\mathbf{0}}

\newcommand{\Tbb}{\bar{\mathbf{T}}}

\newcommand{\Abm}{\mathsf{A}}

\newcommand{\Pbm}{\mathsf{P}}
\newcommand{\Qbm}{\mathsf{Q}}

\newcommand{\Gbm}{\mathsf{G}}

\newcommand{\Tbm}{\mathsf{T}}
\newcommand{\Hbm}{\mathsf{H}}

\newcommand{\Ael}{\mathbb{A}}
\newcommand{\Bel}{\mathbb{B}}
\newcommand{\Cel}{\mathbb{C}}
\newcommand{\Del}{\mathbb{D}}

\newcommand{\Jel}{\mathbb{J}}

\newcommand{\caA}{\mathcal A}
\newcommand{\caB}{\mathcal B}
\newcommand{\caF}{\mathcal F}
\newcommand{\caE}{\mathcal E}
\newcommand{\caG}{\mathcal G}
\newcommand{\caL}{\mathcal L}
\newcommand{\caM}{\mathcal M}
\newcommand{\caN}{\mathcal N}
\newcommand{\caV}{\mathcal V}
\newcommand{\caU}{\mathcal U}

\newcommand{\caZ}{\mathcal Z}

\newcommand{\caH}{\mathfrak H}

\newcommand{\fraC}{\mathfrak C}

\newcommand{\Lin}{\mathop{\rm{Lin}}}
\newcommand{\LLin}{\mathfrak{Lin}}
\newcommand{\Sym}{\mathop{\rm{Sym}}}
\newcommand{\Skw}{\mathop{\rm{Skw}}}
\newcommand{\dif}{\mathrm{d}}
\newcommand{\dive}{\mathop{\rm{div}}}
\newcommand{\grad}{\mathop{\rm{grad}}}
\newcommand{\skw}{\mathop{\rm{skw}}}
\newcommand{\sym}{\mathop{\rm{sym}}}
\newcommand{\spn}{\mathop{\rm{span}}}
\newtheorem{rmk}{Remark}

\documentclass[preprint, 12pt]{elsarticle}
\usepackage{epsfig}
\usepackage{epic}
\usepackage{amssymb}
\usepackage{amsmath}
\usepackage{amscd}

\usepackage{tikz}
\usepackage{tikz-cd}
\usetikzlibrary{matrix, calc, arrows}

\usepackage{pgfplots}
\usepackage{pgf}
\pgfplotsset{compat=1.14}

\begin{document}
\begin{frontmatter}
\title{Wave propagation in micromorphic anisotropic continua with an application to tetragonal crystals}
\author{Fabrizio Dav\'{\i}\footnote{e-mail: davi@univpm.it}}
\address{DICEA and ICRYS, Universit\'a Politecnica delle Marche,\\
via Brecce Bianche, 60131 Ancona, Italy}

\begin{abstract} 
We study the coupled macroscopic and lattice wave propagation in anisotropic crystals seen as continua with affine microstructure (or micromorphic). In the general case we obtain qualitative informations on the frequencies and the dispersion realtions. These results are then specialized to crystals of the tetragonal point group for various propagation directions: exact representation for the acoustic and optic frequencies and for the coupled vibrations modes are obtained for propagation directions along the tetragonal $c$-axis.
\end{abstract}

\end{frontmatter}

\section{Introduction}
Let me say from the beginning that this paper was motivated by second toughts: indeed my research deals mostly with the photoelastic properties of scintillating crystals, that is crystals which convert ionizing radiations into photons within the visible range. Massive scintillating crystals were used to detect particle collisions in the CMS calorimeter at CERN, Geneve \cite{LEC06} and shall be used in the FAIR accelerator at GSI, Darmstadt \cite{ER13}  and can also be used into security and medical imaging devices. Amongst many other problems concerning quality control and efficency, one of the major issues related with the prolongated use of these crystals is the radiation damage which displaces atoms and reduces crystal efficiency and the radiation/photons ratio (\emph{vid. e.g.\/} \cite{DO05}).

In order to recover the radiation damage many techniques were used: between them one of the most promising is the laser-induced ultrasound lattice vibration, which can be a really efficient way to recover the damage. It becomes mandatory, therefore, to study the problem of coupled bulk/lattice waves propagation in crystals in order to evaluate the frequency range of bulk and lattice vibrations, the amount of energy which is lost in the coupling between lattice and bulk and how much of the incoming energy makes the lattice to vibrate and around which modes. The problem is remarkably complex since most scintillators exhibit strong anisotropy, like the monoclinic Cerium doped  {L}u$_{x}${Y}$_{2-x}${S}i{O}$_5$:Ce (LYSO) \cite{ME15}, the hexagonal LaBr$_3$ \cite{LAB} and the tetragonal PbWO$_4$ (PWO) \cite{AKL02}.

Such problem can of course be studied with a classical lattice dynamics approach \cite{BH54}, \cite{DO93}: however, since scintillators generally are massive crystal whose size is in the range of decimeters, the continuum mechanics approach looks more suited to describe the interactions between the crystal lattice vibrations and the macroscopic vibrations of the crystal specimen. Thus it seem natural to model the crystal as a \emph{continuum with affine structure} \cite{CA00} or \emph{micromorphic continuum} \cite{ER99}, since it appears as a reasonable compromise between the microscopical aspects related to lattice vibrations and the macroscopic vibrations. 

Micromorphic continua have attracted a never-fading attention since the pionieering work of Mindlin \cite{MI64}, the treatises \cite{CA00} and \cite{ER99} and the revamped attention in the recent years, motivated by the study of metamaterials, by the means of both the classical approach as in \cite{BBEN11} or the relaxed one first proposed into \cite{NE13}. The  majority of these results however concern isotropic materials, with some limited exceptions. Here for classical micromorphic continua, we extend to the general case of anisotropic materials the previously known results of wave propagation into isotropic material, and then specialize them to crystals of the tetragonal group.

The paper is organized as follows: in \S.2 we write the balance law as proposed into \cite{CA00} and then, by using the results of \cite{CPW82} we show that they are fully equivalent to those given in \cite{MI64}; upon the assumption of linearized kinematics and by using linear constitutive relations as in \cite{MI64}, we arrive at the propagation condition for the macroscopic progressive waves coupled with microdistortions  lattice waves. Such propagation condition, which depends on the \emph{wavenumber} $\xi$ and on two dimensionless parameters which  relates the various length-scales of the problem,  is  completely described by a 12$\times$12 Hermitian matrix whose blocks represents various kind of generalized acoustic tensors and whose eigenvector represents macroscopic displacements coupled with lattice microdistortions. In the general case of triclinic crystals we show that there ever exist three acoustic and nine optic waves and we also give an insight into the structure of dispersion relations: moreover by a suitable scaling in terms of the dimensionless parameters we show that the problem admits two physically meaningful limit cases, namely the \emph{Long wavelength approximation} which represents the propagation phenomena in a body in which we are ''zooming-out'' away from the crystal lattice, and the \emph{Microvibration} case, where on the converse we are ''zooming-in'' into the crystal lattice. These two limit cases first introduced into \cite{MI64}, besides representing two physical picture of the phenomena, give an insight into the general propagation problem with the cut-off optical frequencies given by the microvibration frequencies.

In the following \S.3 these general results are specialized to crystals of the Tetragonal point group and the reason for such a choice is two-fold: first of all we are interested to damage recovery in the tetragonal PWO crystals which are currently used in the FAIR accelerator  \cite{NKDM11}; second, the reduction of independent  constitutive parameter for tetragonal micromorphic bodies allow for some explicit solutions of the propagation condition and to a at least qualitative representation of the dispersion relations. We study separately the low-symmetric tetragonal classes $4$, $\bar{4}$, $4/m$ and the high-symmetric classes $4mm$, $422$, $\bar{4}2m$ and $4/mm$, in the case of waves propagating along the tetragonal $c$-axis. For the other relevant case, that of waves propagating in directions orthogonal to the $c$-axis we give  an insight to the solution structure since along such directions the crystal behaves in a fully coupled manner as in triclinic crystal: hence it is not possible to give exact closed-form solutions.

It there exists a major criticism to such an approach: as it was correctly pointed out into \cite{BANE17}, such a formal treatment has some limitations since it depends indeed on a large number of parameters whose experimental identification can be  both difficult and elusive. With respect to such a correct criticism we can say that in any case we have a  general framework to design correct experiments aimed to parameter identification and moreover,  as it is show into two recent papers \cite{MS19}, \cite{MS20}, by homogenization techniques we can estimate the micromorphic model constitutive parameters by the means of classical lattice dynamics.

In the  final \S.4, the results for the full propagation condition and for the two limiting cases are given in tabular form;  as far as we know this is the most complete analysis of wave propagation in micromorphic tetragonal crystals up to now and it could be the starting point for both experiment design and homogenization techniques based on lattice dynamics.

\subsection{Notation}

Let $\caV$ be the three-dimensional vector space whose elements we denote $\vb\in\caV$ and let $\Lin$ be the space of the second order tensors $\Ab\in\Lin$, $\Ab:\caV\rightarrow\caV$. We denote $\Ab^{T}$ the transpose of $\Ab$ such that $\Ab\ub\cdot\vb=\Ab^{T}\vb\cdot\ub$, $\forall\ub\,,\vb\in\caV$; we shall also denote $\Sym$ and $\Skw$ the subspaces of $\Lin$ of the symmetric $(\Ab=\Ab^{T})$ and skew-symmetric $(\Ab=-\Ab^{T})$ tensors respectively.

Let $\LLin$ be the space of third-order tensors $\Pbm:\Lin\rightarrow\caV$ and for all $\Pbm\in\LLin$ we denote the transpose $\Pbm^{T}:\caV\rightarrow\Lin$ as:
\begin{equation}
\Pbm[\Ab]\cdot\vb=\Pbm^{T}\vb\cdot\Ab\,,\quad\forall\vb\in\caV\,,\forall\Ab\in\Lin\,.
\end{equation}

We shall also make use of fourth-order tensors $\Cel:\Lin\rightarrow\Lin$, fifth-order tensors $\caF:\LLin\rightarrow\Lin$ and sixth-order tensors $\caH:\LLin\rightarrow\LLin$ whose transpose are defined by the means of
\begin{eqnarray}
\Cel[\Ab]\cdot\Bb&=&\Cel^{T}[\Bb]\cdot\Ab\,,\quad\forall\Ab\,,\Bb\in\Lin\,,\nonumber\\
\caF[\Ab]\cdot\Pbm&=&\caF^{T}[\Pbm]\cdot\Ab\,,\quad\forall\Ab\in\Lin\,,\forall\Pbm\in\LLin\,,\\
\caH[\Pbm]\cdot\Qbm&=&\caH^{T}[\Qbm]\cdot\Pbm\,,\quad\forall\Pbm\,,\Qbm\in\LLin\,.\nonumber
\end{eqnarray}

For $\{\eb_{k}\}\,,k=1,2,3$ an orthonormal basis in $\caV$, we define the components of the aforementioned elements by:
\begin{eqnarray}
v_{k}&=&\vb\cdot\eb_{k}\,,\nonumber\\
A_{kj}&=&\Ab\eb_{j}\cdot\eb_{k}=\Ab\cdot\eb_{k}\otimes\eb_{j}\,,\nonumber\\
\Pbm_{ihk}&=&\Pbm[\eb_{h}\otimes\eb_{k}]\cdot\eb_{i}=\Pbm\cdot\eb_{i}\otimes\eb_{h}\otimes\eb_{k}\,,\nonumber\\
\Cel_{ijhk}&=&\Cel[\eb_{h}\otimes\eb_{k}]\cdot\eb_{i}\otimes\eb_{j}\,,\quad i,j,h,k,m,p=1,2,3\,,\\
\caF_{ijhkm}&=&\caF[\eb_{h}\otimes\eb_{k}\otimes\eb_{m}]\cdot\eb_{i}\otimes\eb_{j}\,,\nonumber\\
\caH_{ijkhmp}&=&\caH[\eb_{h}\otimes\eb_{m}\otimes\eb_{p}]\cdot[\eb_{i}\otimes\eb_{j}\otimes\eb_{k}]\,.\nonumber
\end{eqnarray}
We shall also made use of the orthonormal base $\{\Wb_{k}\}\,,k=1,\ldots,9$ in $\Lin$:
\begin{eqnarray}\label{baseninedim}
&\Wb_{1}=\eb_{1}\otimes\eb_{1}\quad\Wb_{2}=\eb_{2}\otimes\eb_{2}\quad\Wb_{3}=\eb_{3}\otimes\eb_{3}\,,\nonumber\\
&\Wb_{4}=\eb_{2}\otimes\eb_{3}\quad\Wb_{5}=\eb_{3}\otimes\eb_{1}\quad\Wb_{6}=\eb_{1}\otimes\eb_{2}\,,\\
&\Wb_{7}=\eb_{3}\otimes\eb_{2}\quad\Wb_{8}=\eb_{1}\otimes\eb_{3}\quad\Wb_{9}=\eb_{2}\otimes\eb_{1}\,,\nonumber
\end{eqnarray}
whereas the orthogonal bases $\{\hat{\Wb}_{k}\}\,,k=1,\ldots,6$ in $\Sym$ and $\{\bar{\Wb}_{k}\}$, $k=4,5,6$ in $\Skw$ are respectively defined as:
\begin{eqnarray}\label{basesym}
&\hat{\Wb}_{k}=\Wb_{k}\,,\quad k=1,2,3\,,\\
&\hat{\Wb}_{4}=\frac{1}{2}(\Wb_{4}+\Wb_{7})\,,\quad\hat{\Wb}_{5}=\frac{1}{2}(\Wb_{5}+\Wb_{8})\,,\quad\hat{\Wb}_{6}=\frac{1}{2}(\Wb_{6}+\Wb_{9})\,,\nonumber
\end{eqnarray}
and
\begin{equation}\label{baseskw}
\bar{\Wb}_{4}=\frac{1}{2}(\Wb_{4}-\Wb_{7})\,,\quad\bar{\Wb}_{5}=\frac{1}{2}(\Wb_{5}-\Wb_{8})\,,\quad\bar{\Wb}_{6}=\frac{1}{2}(\Wb_{6}-\Wb_{9})\,.
\end{equation}
In terms of the bases (\ref{baseninedim})-(\ref{baseskw}) we can also represent the fourth-order tensors components with the so-called Voigt two-index notation, namely \emph{e.g.\/} for (\ref{baseninedim}):
\begin{equation}
\Cel_{ij}=\Cel[\Wb_{j}]\cdot\Wb_{i}\,,\quad i,j=1\,,\ldots\,,9\,.
\end{equation}

Finally, in order to describe the infinitesimal lattice vibrations we shall make use of the following seven modes:
\begin{itemize}
\item Non uniform dilatation:
\begin{equation}
\Db_{1}=\alpha\Wb_{1}+\beta\Wb_{2}+\gamma\Wb_{3}\,;
\end{equation}
\item Dilatation along $\eb_{3}$ and uniform plane strain in the plane orthogonal to $\eb_{3}$:
\begin{equation}
\Db_{2}=\alpha(\Ib-\Wb_{3})+\gamma\Wb_{3}\,;
\end{equation}
\item Traceless plane  strain orthogonal to $\eb_{3}$:
\begin{equation}
\Db_{3}=\Wb_{1}-\Wb_{2}\,;
\end{equation}
\item Shear in the plane orthogonal to $\eb_{3}$:
\begin{equation}
\Sbb_{1}=\alpha\hat{\Wb}_{6}\,;
\end{equation}
\item Shear between $\eb_{3}$ and the direction $\eb_{\perp}=\alpha\eb_{1}+\beta\eb_{2}$:
\begin{equation}
\Sbb_{2}=-\alpha\hat{\Wb}_{4}+\beta\hat{\Wb}_{5}\,;
\end{equation}
\item Rigid rotation around the direction $\eb_{3}$:
\begin{equation}
\Rb_{1}=\omega_{3}\bar{\Wb}_{6}\,;
\end{equation}
\item Rigid rotation around the direction $\eb^{\perp}=\omega_{1}\eb_{1}+\omega_{2}\eb_{2}$:
\begin{equation}
\Rb_{2}=\omega_{2}\bar{\Wb}_{5}-\omega_{1}\bar{\Wb}_{4}\,.
\end{equation}
\end{itemize}

\section{Crystal as a micromorphic continuum}

\subsection{Balance laws}

Let $\caB$ a region of the Euclidean three-dimensional space we pointwise identify with the reference configuration of a crystal, and let $x$ be a point of $\caB$. We assume that at each point $x\in\caB$ is defined a crystal lattice $\{\ab_{1}\,,\ab_{2}\,,\ab_{3}\}$ with $\ab_{1}\times\ab_{2}\cdot\ab_{3}\geq 1$.

As in  \cite{MI64} we assume that at each point $x\in\caB$ and at each time $t\in[0\,,\tau)$ it is well-defined the \emph{motion} by the means of the two fields:
\begin{equation}\label{motion}
y=y(x\,,t)\,,\quad\Gb=\Gb(x\,,t)
\end{equation}
provided $y$ is locally injective and orientation-preserving and $\Gb$ is orientation-preserving, namely:
\begin{equation}
\det\Fb>0\,,\quad\Fb(x\,,t)=\nabla y(x\,,t)\,,\quad\det\Gb>0\,,
\end{equation}
in such a way that at at each point $y\in\caB_{t}\equiv y(\caB\,,t)$ the deformed crystal lattice  $\{\bar{\ab}_{1}\,,\bar{\ab}_{2}\,,\bar{\ab}_{3}\}$ is given by:
\begin{equation}
\bar{\ab}_{k}=\Gb\ab_{k}\,,\quad k=1,2,3\,.
\end{equation}

We identify $\caB$, endowed with the motion (\ref{motion}), with a \emph{continuum with affine structure}  \cite{CA00} or \emph{micromorphic} \cite{ER68}, whose underlying manifold is $\caM=\Lin^{+}$ and whose balance laws are given by:
\begin{itemize}
\item the balance of macroforces:
\begin{equation}\label{bala1}
\dive\Tb^{T}+\bb=\rho\dot{\vb}\,,
\end{equation}
where $\Tb$ is the (non-symmetric) Cauchy stress, $\bb$ is the volume force density, $\rho$ is the mass density and $\vb$ the material velocity;
\item the balance of microforces:
\begin{equation}\label{bala2}
\dive\Tbm-\Kb+\Bb=\rho\ddot{\Gb}\Jb\,,
\end{equation}
where the third-order tensor $\Tbm$ represents the microstress, $\Kb$ the interactive microforce, $\Bb$ the density of volume micro forces and $\Jb$ the microinertia tensor
\item the balance of couples:
\begin{equation}\label{bala3}
\skw(-\Tb+\Gb\Kb^{T}+(\grad\Gb)\Tbm^{T})=\bzero\,.
\end{equation}
\end{itemize}

A different set of balance laws was provided in \cite{MI64}: in order to recover these balance laws from (\ref{bala1})-(\ref{bala3}) first of all we notice that $\skw\Tb=-\skw\Tb^{T}$ and set:
\begin{equation}
\Tbb=\Tb^{T}+\Gb\Kb^{T}+(\grad\Gb)\Tbm^{T}\in\Sym\,;
\end{equation}
then, by following \cite{MI64}, we define the \emph{relative stress} as:
\begin{equation}\label{extrastress}
\Sbb=-(\Gb\Kb^{T}+(\grad\Gb)\Tbm^{T})\,,
\end{equation}
in such a way that
\begin{equation}\label{TYS}
\Tb^{T}=\Tbb+\Sbb\,.
\end{equation}
As second step we multiply the transpose of (\ref{bala2}) for $\Gb$ to obtain:
\begin{equation}\label{bala4}
\Gb(\dive\Tbm)^{T}-\Gb\Kb^{T}+\Gb\Bb^{T}=\rho\Gb(\ddot{\Gb}\Jb)^{T}\,,
\end{equation}
and since
\begin{equation}\label{12}
\Gb(\dive\Tbm)^{T}=\dive(\Gb\Tbm^{T})-(\grad\Gb)\Tbm^{T}\,,
\end{equation}
then with the aid of (\ref{TYS}) from (\ref{bala1}), (\ref{bala4}) and (\ref{12}) we recover equations (4.1) of \cite{MI64}:
\begin{eqnarray}\label{system1}
&\dive(\Tbb+\Sbb)+\bb=\rho\dot{\vb}\,,\nonumber\\
&\\
&\dive\Hbm+\Sbb+\bar{\Bb}=\rho\Gb\Jb\ddot{\Gb}^{T}\,,\nonumber
\end{eqnarray}
where 
\begin{equation}\label{newvariable}
\Hbm=\Gb\Tbm^{T}\,,\quad \bar{\Bb}=\Gb\Bb^{T}\,.
\end{equation}

\subsection{Linearized kinematics and constitutive relations}

As it is shown in \cite{CA00}, \cite{ER68}, \cite{ER99}, there are many appropriate kinematical measures for a constitutive theory of micromorphic continua; here we choose those proposed into \cite{ER99}, eqn. (1.5.11):
\begin{equation}\label{finitekinematics}
\Eb=\frac{1}{2}(\Fb^{T}\Fb-\Ib)\,,\quad\Mb=\Fb^{T}\Gb^{-1}-\Ib\,,\quad\Gbm=\Gb^{-1}\grad\Gb\,,
\end{equation}
where $\Eb$ is the Green-Lagrange deformation measure.

We assume that both the deformation gradient and the lattice deformation can be decomposed additively into:
\begin{equation}\label{additive}
\Fb=\Ib+\nabla\ub\,,\quad\Gb=\Ib+\Lb\,,
\end{equation}
where $\ub(x)=y(x)-x$ is the \emph{displacement vector} and $\Lb$ is the \emph{microdisplacement} or \emph{microdistortion} \cite{BANE17}. If we assume that 
\begin{equation}\label{normepsilon}
\varepsilon=\sup\{\|\nabla\ub\|\,,\|\Lb\|\}\,,
\end{equation}
then the kinematical measures (\ref{finitekinematics}) can be rewritten, to within higher-order terms into $\varepsilon$, as:
\begin{equation}\label{linearkinematics}
\Eb=\sym\nabla\ub\,,\quad\Mb=\nabla\ub^{T}-\Lb\,,\quad\Gbm=\nabla\Lb\,;
\end{equation}
the tensor $\Mb$ being called the \emph{relative strain} \cite{MI64} or \emph{relative distortion} \cite{BANE17}.

\begin{rmk}
By using (\ref{additive}) and (\ref{normepsilon}) into (\ref{extrastress}) and (\ref{newvariable}) we have:
\begin{equation}
\Hbm=\Tbm+O(\varepsilon)\,,\quad\Sbb=-\Kb^{T}+O(\varepsilon)\,,\quad\bar{\Bb}=\Bb^{T}+O(\varepsilon)\,,
\end{equation}
and:
\begin{equation}
\Gb\Jb\ddot{\Gb}^{T}=\Jb\ddot{\Lb}^{T}+O(\varepsilon^{2})\,.
\end{equation}
Accordingly, from the balance law (\ref{system1})$_{2}$ we obtain, to within higher-order terms, the balance of microforces (\ref{bala2}) with $\ddot{\,\Lb}$ in place of $\ddot{\,\Gb}$:
\begin{equation}
\dive\Tbm-\Kb+\Bb=\rho\ddot{\,\Lb}\Jb\,,
\end{equation}
whereas the macroforces balance (\ref{bala1}) can be rewritten as
\begin{equation}
\dive(\Tbb-\Kb^{T})+\bb=\rho\ddot{\ub}\,.
\end{equation}
In the sequel we shall use $\Hbm\,,\Sbb$ and $\Tbb$ as the mechanical descriptors associated to a linearized kinematics, since in this case they are equivalent to $\Tbm\,,\Kb$ and $\Tb$.
\end{rmk}

We assume a linear dependence of $\bar{\Tb}\,,\Sbb$ and $\Hbm$ on the linearized kinematical variables (\ref{finitekinematics}) and write (\emph{cf.\/} eqn. (5.3) of \cite{MI64}):
\begin{eqnarray}\label{constitutive}
\Tbb&=&\Cel[\Eb]+\Del[\Mb]+L_{c}\caF[\Gbm]\,,\nonumber\\
\Sbb&=&\Del^{T}[\Eb]+\Bel[\Mb]+L_{c}\caG[\Gbm]\,,\\
\Hbm&=&L_{c}\caF^{T}[\Eb]+L_{c}\caG^{T}[\Mb]+L_{c}^{2}\caH[\Gbm]\,,\nonumber
\end{eqnarray}
where:
\begin{itemize}
\item $\Cel:\Sym\rightarrow\Sym$, $\Cel=\Cel^{T}$ is the fourth-order elasticity tensor, whose components obey:
\begin{equation}
\Cel_{ijhk}=\Cel_{jihk}=\Cel_{ijkh}=\Cel_{hkij}\,,
\end{equation}
and there are at most 21 independent components.
\item The fourth-order tensor $\Bel:\Lin\rightarrow\Lin$, $\Bel=\Bel^{T}$,  whose independent components are at most 45:
\begin{equation}
\Bel_{ijhk}=\Bel_{hkij}\,.
\end{equation}
\item The fourth-order tensor $\Del:\Lin\rightarrow\Sym$ has 54 independent components:
\begin{equation}
\Del_{ijhk}=\Del_{jihk}\,,\quad(\Del^{T})_{ijhk}=\Del_{hkij}\,.
\end{equation}
\item The fifth-order tensors $\caF:\LLin\rightarrow\Sym$ and $\caG:\LLin\rightarrow\Lin$ have respectively 162 and 243 components which obey:
\begin{equation}
\caF_{ijhkm}=\caF_{jihkm}\,,\quad(\caF^{T})_{ijhkm}=\caF_{hkmij}\,,\quad(\caG^{T})_{ijhkm}=\caG_{hkmij}\,.
\end{equation}
\item The sixth-order tensor $\caH=\caH^{T}$ has at most 378 independent components $\caH_{ijkhmn}$ which obey:
\begin{equation}
\caH_{ijkhmn}=\caH_{hmnijk}\,.
\end{equation}
\item $L_{c}>0$ is the micromorphic \emph{correlation length} which makes all these tensorial quantities of the dimension of a stress (Force/Area)\,.
\end{itemize}
The correlation length $L_{c}$ is the first length scale we need to introduce into the model and is related to the non-local effects associated with the gradient of the microdistorsion tensor. For $L_{c}\rightarrow 0$ we are considering large samples of crystals \cite{BANE16} whereas the limit $L_{c}\rightarrow\infty$ acts as a zoom into the microstructure (\emph{cf.\/} \cite{BANE17}): we shall made these statements more rigorous in the next subsection.

As in \cite{MI64}, the requirement that the energy density is positive
\begin{equation}
2\caE(\Eb\,,\Mb\,,\Gbm)=\Tbb\cdot\Eb+\Sbb\cdot\Mb+\Hbm\cdot\Gbm>0\,,
\end{equation}
implies that $\Cel\,,\Bel$ and $\caH$ be positive definite:
\begin{eqnarray}\label{pdenergy}
&\Cel[\Ab]\cdot\Ab>0\,,\quad\Bel[\Ab]\cdot\Ab>0\,,\quad\forall\Ab\in\Lin/\{\bzero\}\,,\\
&\caH[\Abm]\cdot\Abm>0\,,\quad\forall\Abm\in\LLin/\{\bzero\}\,,\nonumber
\end{eqnarray}
as well as
\begin{eqnarray}\label{PDenergy}
&\det(\Bel-\Del^{T}\Cel^{-1}\Del)>0\,,\quad\det(\Cel-\Del\Bel^{-1}\Del^{T})>0\,,\\
&\det(\caH-\caF^{T}\Cel^{-1}\caF)>0\,,\quad\det(\caH-\caG^{T}\Bel^{-1}\caG)>0\,,\nonumber
\end{eqnarray}
and
\begin{equation}\label{pdenergy3}
\det
\renewcommand{\arraystretch}{1.5} 
\left[\begin{array}{@{}c|c@{}|c@{}}
\Cel&\Del &L_{c}\,\caF\\ \hline
\Del^{T} &\Bel &L_{c}\,\caG\\ \hline
L_{c}\,\caF^{T} &L_{c}\,\caG^{T}\, &L_{c}^{2}\,\caH
\end{array}\right]>0\,.
\end{equation}

In the most general case, that of crystals of the Triclinic group, these constitutive relations require the knowledge of 903 material constants, subject to the restrictions (\ref{pdenergy})-(\ref{pdenergy3}), whereas in the simplest case of Isotropic materials these constants reduce to 18 independent at most \cite{MI64}. In the next subsection we shall give a general and formal treatment of waves propagation in a crystal without any of the restrictions given by crystal symmetries.

\subsection{Wave propagation}

The balance laws (\ref{system1}) written in terms of the linearized kinematics (\ref{linearkinematics}) by the means of the constitutive relations (\ref{constitutive}) and zero volume macro- and microforces:
\begin{eqnarray}\label{systemwave}
&\dive(\Tbb+\Sbb)=\rho\ddot{\ub}\,,\nonumber\\
&\\
&\dive\Hbm+\Sbb=\rho\Jel[\ddot{\Lb}]\,,\nonumber
\end{eqnarray}
where 
\begin{equation}\label{Jeldef}
\Jel[\ddot\Lb]=\Jb\ddot{\Lb}^{T}\,,\quad\Jel_{ijhk}=\delta_{ih}J_{jk}\,,
\end{equation}
are  the starting point for the description of  the microscopic crystal lattice vibrations coupled with the macroscopic bulk vibrations. We seek for (\ref{systemwave}) progressive plane wave solutions of the form:

\begin{equation}\label{wavedispla}
\ub(x\,,t)=\ab e^{i\sigma}\,,\quad\Lb(x\,,t)=\Cb e^{i\sigma}\,,\quad\sigma=\xi\xb\cdot\mb-\omega t\,,
\end{equation}
where $\omega$ is the frequency, $\mb$ the direction of propagation, $\xi=\lambda^{-1}>0$ the wavenumber with $\lambda$ the wavelength and where $\ab\in\caV$ and $\Cb\in\Lin$ denote respectively the displacement and microdistortion amplitudes which in the general case are complex-valued.

We find at this point mandatory to introduce, besides the characteristic lenght scale $L_{c}$, two further length scales: the \emph{macroscopic length} $L_{m}>0$ associated to the displacement amplitude:
\begin{equation}
\ab=L_{m}\ab_{o}\,,
\end{equation}
with $\ab_{o}$ a dimensionless vector, and the \emph{lattice length} $L_{l}>0$ which allows to write
\begin{equation}\label{tensorHO2}
\Jel=L_{l}^{2}\Jel_{o}\,,
\end{equation}
with $\Jel_{o}$ a dimensionless fourth-order tensor \cite{BANE17}. If we define the two dimensionless parameters
\begin{equation}\label{dimensionless}
\zeta_{1}=\frac{L_{c}}{L_{m}}\,,\quad\zeta_{2}=\frac{L_{l}}{L_{m}}\,,
\end{equation}
then for $\zeta_{1}\rightarrow 0$ we are considering large samples of crystals and in the limit $\zeta_{1}\rightarrow\infty$ we are zooming into the microstructure. In the case $\zeta_{2}\rightarrow 0$ we are instead neglecting the contribution of microinertia with respect to the macroscopic inertia.

Since 
\begin{eqnarray}\label{wavederivative}
&\nabla\ub=i\xi L_{m}\ab_{o}\otimes\mb e^{i\sigma}\,,\quad\nabla\Lb=i\xi\Cb\otimes\mb e^{i\sigma}\,,\nonumber\\
&\\
&\ddot{\ub}=-\omega^{2} L_{m}\ab_{o} e^{i\sigma}\,,\quad\ddot{\Lb}=-\omega^{2}\Cb e^{i\sigma}\,,\nonumber
\end{eqnarray}
then we are led, by (\ref{linearkinematics}), (\ref{constitutive}),  (\ref{systemwave}), (\ref{dimensionless}) and (\ref{wavederivative}), to the propagation conditions written in terms of the two lengths $\lambda=\xi^{-1}\,,L_{m}$ and of the two dimensionless parameters $\zeta_{1}\,,\zeta_{2}$:
\begin{eqnarray}\label{propa1}
&\xi^{2}\Ab(\mb)\ab_{o}+(\xi^{2}\zeta_{1}\Pbm(\mb)+i\xi L_{m}^{-1}\Qbm(\mb))[\Cb]=\omega^{2}\ab_{o}\,,\nonumber\\
&\\
&(\xi^{2}\zeta_{1}\Pbm^{T}(\mb)-i\xi L_{m}^{-1}\Qbm^{T}(\mb))\ab+(\xi^{2}\zeta_{1}^{2}\Ael(\mb)+L_{m}^{-2}\bar{\Bel})[\Cb]=\omega^{2}\zeta_{2}^{2}\Jel_{o}[\Cb]\,,\nonumber
\end{eqnarray}
where 
\begin{eqnarray}\label{tensorHO1}
\Ab(\mb)\ab&=&\rho^{-1}(\Cel[\ab\otimes\mb]+\Del[\mb\otimes\ab]+\Del^{T}[\ab\otimes\mb]+\Bel[\mb\otimes\ab])\,,\nonumber\\
\Pbm(\mb)[\Cb]&=&\rho^{-1}(\caF+\caG)[\Cb\otimes\mb]\mb\,,\nonumber\\
\Qbm(\mb)[\Cb]&=&\rho^{-1}(\Del+\Bel)[\Cb]\mb\,,\\
\Ael(\mb)[\Cb]&=&\rho^{-1}\caH[\Cb\otimes\mb]\mb\,,\nonumber\\
\bar{\Bel}&=&\rho^{-1}\Bel\,,\nonumber
\end{eqnarray}
and whose representation in components are
\begin{eqnarray}\label{componentsHO}
A_{ij}&=&\rho^{-1}(\Cel_{iljk}m_{k}m_{l}+\Del_{ilhj}m_{h}m_{l}+\Del_{iljk}m_{k}m_{l}+\Bel_{ilhj}m_{h}m_{l})\,,\nonumber\\
\Pbm_{ihk}&=&\rho^{-1}(\caF_{ijhkp}+\caG_{ijhkp})m_{p}m_{j}\,,\nonumber\\
\Qbm_{ijh}&=&\rho^{-1}(\Del_{ijhk}+\Bel_{ijhk})m_{k}\,,\\
\Ael_{ijhk}&=&\rho^{-1}\caH_{ijlhkp}m_{l}m_{p}\,,\nonumber\\
\bar{\Bel}_{ijhk}&=&\rho^{-1}\Bel_{ijhk}\,.\nonumber
\end{eqnarray}
We notice that, since $\Cel=\Cel^{T}$, $\Bel=\Bel^{T}$ and $\caH=\caH^{T}$, from (\ref{componentsHO}) we have that:
\begin{equation}
\Ab\in\Sym\,,\quad\Ael=\Ael^{T}\,;
\end{equation}
we call $\Ab(\mb)$ the generalized acoustic tensor\footnote{Indeed in absence of the microstructure (\ref{tensorHO1})$_{1}$ reduces to the acoustic tensor for linearly elastic bodies, \emph{vid. e.g.\/} \cite{GU72}, \S.70.} and $\Ael(\mb)$ the microacoustic tensor.

We define the two $12\times 12$ hermitian block matrix $A(\xi)=A^{*}(\xi)$, which we call the \emph{Acoustic matrix},  the $12\times 12$ symmetric block matrix $J=J^{T}$ and the 12 dimension eigenvector $w$:
\begin{equation}\label{Agotica}
\renewcommand{\arraystretch}{1.5} 
A(\xi)\equiv
\left[\begin{array}{@{}c|c@{}}
\xi^{2}\Ab&\xi^{2}\zeta_{1}\Pbm+i\xi L_{m}^{-1}\Qbm \\ \hline
&\\
\xi^{2}\zeta_{1}\Pbm^{T}-i\xi L_{m}^{-1}\Qbm^{T} & \xi^{2}\zeta_{1}^{2}\Ael+L_{m}^{-2}\bar{\Bel}\\
&\\
\end{array}\right],
\renewcommand{\arraystretch}{1.5} 
J\equiv
\left[\begin{array}{@{}c|c@{}}
\Ib&\bzero \\ \hline
\bzero & \zeta_{2}^{2}\Jel_{o}
\end{array}\right]\,,
\renewcommand{\arraystretch}{1.5} 
w\equiv
\left[
\begin{array}{@{}c@{}}
\ab_{o}\\ \hline
\Cb
\end{array}
\right]\,,
\end{equation}
in such a way that we can rewrite the propagation condition (\ref{propa1}) as
\begin{equation}\label{propa2}
(A-\omega^{2}J)w=0\,.
\end{equation}

We require,  that $A$ be positive-definite for all $\xi>0$ (the semidefiniteness as in \cite{BANE16} being required when $\xi\rightarrow 0$) and then, since $\Ael$ and $\Bel$ are positive-definite by (\ref{pdenergy}) and the definition (\ref{tensorHO1})$_{4}$, the positive-definiteness of $A$ implies that also $\Ab(\mb)$ be positive definite; accordingly the eigenvalue problem (\ref{propa2}) admits the twelve eigencouples with real eigenvalues\footnote{With the notation
\[
<a\,,b>=\sum_{h=1}^{m}a_{h}b^{*}_{h}\,,
\]
we denote the euclidean inner product on a $m$-dimensional complex space $C^{m}$.}
\begin{equation}\label{spectrumdefinition}
(\omega_{k}\,,w_{k})\,,\quad <Jw_{h}\,,w_{k}>=\delta_{hk}\,,\quad h,k=1\,,\ldots\,,12\,.
\end{equation}

The characteristic equation associated with the propagation condition (\ref{propa2}) is
\begin{equation}\label{deltapropa}
\det(A(\xi)-\omega^{2}J)=0\,,
\end{equation}
and since the components of $M$ are functions of the wavenumber $\xi$, then also the eigencouples are:
\begin{equation}\label{dispersion}
\xi\mapsto(\omega_{k}(\xi)\,,w_{k}(\xi))\,,\quad k=1\,,\ldots\,,12\,.
\end{equation}
The functional dependences between $\omega$ and $\xi$ are called the \emph{dispersion relations} and in terms of these relations we can define the \emph{phase} $v^{p}_{k}$ and \emph{group velocity} $v^{g}_{k}$ as:
\begin{equation}\label{velocities}
v_{k}^{p}(\xi)=\frac{\omega_{k}(\xi)}{\xi}\,,\quad v_{k}^{g}(\xi)=\frac{\dif\omega_{k}(\xi)}{\dif\xi}\,,\quad k=1,\ldots, 12\,.
\end{equation}
As pointed out into \cite{MANE18}, waves in micromorphic continua can be classified into:
\begin{itemize}
\item \emph{Acoustic waves}, whose frequencies $\omega_{k}(\xi)$ goes to zero for $\xi\rightarrow 0$\,;
\item \emph{Optic waves} for which the limit for $\xi\rightarrow 0$ is different from zero: the limit $\omega_{k}(0)$ is called the \emph{cut-off frequency} with group velocities $v^{g}_{k}(0)=0$\,;
\item \emph{Standing waves} those associated to immaginary values $\xi=\pm ik$, $k>0$ for some frequencies:
\begin{equation}\label{SW}
\ub(\xb\,,t)=\ab\,e^{\mp k\xb\cdot\mb}e^{-i\omega t}\,,\quad\Lb(\xb\,,t)=\Cb\,e^{\mp k\xb\cdot\mb}e^{-i\omega t}\,;
\end{equation}
these waves do not propagate, but keep oscillating increasing or decreasing in a given, limited, region within the crystal.
\end{itemize}
In view of this classification, we begin to study the behaviour of the eigencouples of (\ref{propa2}) as $\xi\rightarrow 0$  by using some results obtained into \cite{LA64}.  

First of all, provided the components of $A$ are analytic functions of $\xi\geq 0$, there exists a neighbourhood of $\xi=\xi_{o}$ where the eigenvalues $\omega_{k}(\xi)$ are regular and where their derivatives are well-defined. We use this result, in the case of simple eigenvalues, to give an explicit formula for the group velocities (\cite{LA64}, Thm. 5):
\begin{equation}\label{groupvelo1}
v_{k}^{g}(\xi_{o})=\frac{1}{2\omega_{k}(\xi_{o})}<\frac{\dif A}{\dif\xi}\Big|_{\xi=\xi_{o}}w_{k}(\xi_{o})\,,w_{k}(\xi_{o})>\,;
\end{equation}
moreover the eigenvectors $w_{k}(\xi)$ are differentiable functions of the wavenumber.

Now, since for $\xi\rightarrow 0$ the matrix $A$ is real with three multiple eigenvalues $\omega=0$, nine non-zero (simple) eigenvalues $\hat{\omega}_{j}$ and twelve real eigenvectors $\hat{w}_{k}$, then we can use Theorem 2 of \cite{LA64} to show that (\ref{deltapropa}) admits three and only three zero eigenvalues as $\xi$ approaches zero: moreover by the differentiability of eigenvalues and eigenvectors we have that:
\begin{equation}
\hat{\omega}_{k}=\lim_{\xi\rightarrow 0}\omega_{k}(\xi)\,,\quad\hat{w}_{k}=\lim_{\xi\rightarrow 0}w_{k}(\xi)\,,
\end{equation}
and for the non zero eigenvalues we have from (\ref{groupvelo1})
\begin{equation}\label{groupvelo2}
v_{j}^{g}(0)=\frac{1}{2\hat{\omega}_{j}}<\frac{\dif A}{\dif\xi}\Big|_{\xi=0}\hat{w}_{j}\,,\hat{w}_{j}>=0\,.
\end{equation}
Therefore, by the application of the results obtained in \cite{LA64} to our case, we obtain that in any anisotropic crystal :
\begin{itemize}
\item There exist three Acoustic waves and nine Optic waves;
\item The cut-off frequencies for the Optic waves are the limit as $\xi\rightarrow 0$ of the eigenvalues of (\ref{propa2});
\item The frequencies for the Acoustic waves are the eigenvalues of (\ref{propa2}) which goes to zero in the limit for $\xi\rightarrow 0$;
\item The eigenvectors for $\xi\rightarrow 0$ are real.
\end{itemize}

As far as the standing waves are concerned, we notice that by (\ref{SW}) it is easy to show that $A$ would be not positive-definite and therefore we can conclude that no standing waves are possible within this anisotropic model. Indeed as it was observed into \cite{NE13}, \cite{MANE16}, \cite{MANE17} and \cite{MANE18} for the isotropic case, standing waves are associated with band-gap material and are not possible within the classical micromorphic model: they appears instead in the relaxed micromorphic model proposed into \cite{NE13}.

The solutions of the eigenvalues problem (\ref{propa2}) depend, besides the parameter $\xi$, also on the three parameters $L_{m}$, $\zeta_{1}$ and $\zeta_{2}$ whose limiting values, as we already remarked, corresponds to different physical scales: therefore, besides the complete condition given by (\ref{propa1}), we shall study into some details these two limit cases.

\subsubsection{The long wavelength approximation: macroscopic waves}

If we let $\zeta_{1}\rightarrow 0$ and $\zeta_{2}\rightarrow 0$ we are at the same time ''zooming-out'' from the crystal and disregarding the microinertia; such an approximation is called the \emph{long wavelength approximation} \cite{BANE17}. The propagation conditions (\ref{propa1}) then reduce, to within higher-order terms in $\zeta_{1}$ and $\zeta_{2}$, to:
\begin{eqnarray}\label{propa3}
&\xi^{2}\Ab(\mb)\ab_{o}+i\xi L_{m}^{-1}\Qbm(\mb))[\Cb]=\omega^{2}\ab_{o}\,,\nonumber\\
&\\
&-i\xi\Qbm^{T}(\mb)\ab_{o}+L_{m}^{-1}\bar{\Bel}[\Cb]=\bzero\,;\nonumber
\end{eqnarray}
since $\Bel$ is positive definite, then from (\ref{propa3})$_{2}$ we have:
\begin{equation}\label{microlongwave}
\Cb=iL_{m}\xi\Bel^{-1}\Qbm^{T}(\mb)\ab_{o}\,,
\end{equation}
and therefore from (\ref{propa3})$_{1}$ we obtain the classical continuum propagation condition
\begin{equation}\label{macrolongwave}
\hat{\Ab}(\mb)\ab_{o}=c^{2}\ab_{o}\,,\quad \omega=c\xi\,,\quad c=v^{p}=v^{g}\,,
\end{equation}
where the acoustic tensor $\hat{\Ab}(\mb)$ (which is independent on the macroscopic length $L_{m}$) is defined as:
\begin{equation}\label{Cmacromicro}
\hat{\Ab}(\mb)\ab_{o}=\Ab(\mb)\ab_{o}-\Qbm(\mb)\Bel^{-1}\Qbm^{T}(\mb)\ab_{o}=(\Cel-\Cel_{micro})[\ab_{o}\otimes\mb]\mb\,.
\end{equation}
In  this definition $\Cel$ is the elasticity tensor from (\ref{constitutive}), whereas by using (\ref{tensorHO1})$_{3}$ we can represent the positive-definite \emph{microelasticity tensor} $\Cel_{micro}$ as:
\begin{equation}
\Cel_{micro}=\Del\Bel^{-1}\Del^{T}\,,\quad\Cel_{micro}=\Cel^{T}_{micro}\,,\quad\Cel_{micro}:\Sym\rightarrow\Sym\,;
\end{equation}
by (\ref{PDenergy})$_{2}$ $\Cel-\Cel_{micro}$ is positive-definite and hence the acoustic tensor $\hat{\Ab}(\mb)$ is positive-definite too. 

The three eigencouples of (\ref{macrolongwave}) represents acoustic waves; however in this approximation $\hat{\Ab}(\mb)$ is not the acoustic tensor of the linear elasticity and the presence of the microstructure makes the propagation velocities smaller than in linearly elastic bodies: moreover we have also three microdistortions associated to the eigenvectors of (\ref{macrolongwave}) by the means of (\ref{microlongwave}); these microdistorsions are purely immaginary and depend on the ratio $L_{m}\xi=L_{m}/\lambda$ between the macroscopic scale and the wavelength.\footnote{For the solution of (\ref{macrolongwave}) for the various symmetry groups one can refer to \cite{SA41} or \cite{AC73}.}

\subsubsection{The limit $L_{c}\rightarrow 0$: microvibrations}

If we let $L_{c}\rightarrow\infty$, for fixed $L_{m}\approx L_{l}$, then we are zooming into the crystal; into the limit the constitutive relations (\ref{constitutive})$_{1,2}$ remain finite for any choice of material only if $\nabla\Lb=\bzero$. In this case, from (\ref{wavedispla})$_{2}$ we have that $\xi=0$ and the propagation condition (\ref{propa1})$_{1}$ leads to a solution $\omega=0$ with multiplicity three. Since by (\ref{wavedispla})$_{1}$ $\ub$ reduces to a rigid motion, then without loss of generality we can set:
\begin{equation}\label{microvibra1}
\ub(\xb\,,t)=\bzero\,,\quad \Lb(t)=\Cb\,e^{i\omega t}\,,
\end{equation}
and from the propagation condition (\ref{propa1})$_{2}$ then we are led to the characteristic equation
\begin{equation}\label{microvibra2}
\det(\Bel-\rho\omega^{2}\Jel)=\bzero\,.
\end{equation}
For $L_{c}\rightarrow\infty$ we therefore recover the Microvibration solutions of (\ref{propa1}), which was studied in detail into \cite{MI64} for isotropic materials (\emph{vid. also} \cite{MANE18}); the propagation condition (\ref{microvibra2}) admits nine eigencouples $(\omega_{k}\,,\Cb_{k})$, $k=1\,,\ldots\,,9$, with real eigentensors and whose eigenvalues represent the cut-off frequencies for the propagation condition (\ref{propa2}), as it is show into \S. 2.3.

\section{Wave propagation in Tetragonal crystals}

As we already remarked an explicit solution for the propagation condition (\ref{propa2}) and an explicit determination for the dispersion relations (\ref{dispersion}) is not possible in the general anisotropic case, nor it would be particularly useful, since the associated kinematics would be fully coupled. However many of the components of both the acoustic tensors $\Ab$ and $\Ael$, as well as of $\Pbm$ and $\Qbm$ may vanish according to both the crystal symmetry group and the propagation direction $\mb$ and it could make sense to obtain explicit solutions for special cases of symmetry and directions of propagation. 

The simplest case of isotropic material (which depends only on 18 constitutive parameters) was studied in full-length into \cite{MI64} and further analized and extended to a relaxed micromorphic model into \cite{NE13}, \cite{BANE16}). In this section we shall study the wave propagation condition (\ref{propa2}) for crystals belonging to the Tetragonal symmetry group.\footnote{The propagation problme  in Tetragonal materials was previuosly studied  into \cite{ABNE18}: however their analysis, which concerns a relaxed micromorphic model rather then the classical one, was limited to two-dimensional plane strain; their main focus was indeed the parameter identification by means of numerical  homogeneization (see the comments at the end of \S.\ref{conclusions}).}

We took $\eb_{3}$ directed as the tetragonal $c$-axis, hence the acoustic tensors have the representation given into the \S.\ref{quarta} of the Appendix. At a glance, by looking at (\ref{acoustic1LS}), (\ref{acoustic1HS}), (\ref{QBMLS1}), (\ref{QbmHS}), (\ref{PBMLS}), (\ref{PBMHS}), (\ref{AELLS}) and (\ref{AELHS}) for a generic propagation direction $\mb$, the matrix $A$ does not simplify enough and the problem maintains the same complexity as for crystal of the Triclinic group. 

However, for the two relevant cases of propagation direction either parallel ($\mb\times\eb_{3}=\bzero$) or orthogonal ($\mb\cdot\eb_{3}=0$) to the tetragonal $c$-axis, many of the components of $A$ vanish and the number of the independent components reduces too, allowing for an explicit solution of (\ref{propa2}) whose associated kinematics can be understood more easily. Therefore we study these two propagation direction and we begin with the two limit propagation problems we obtained into \S. 2.3.1 (long-wavelength approximation) and \S. 2.3.2 (microvibrations).

\subsection{Long-wavelength approximation}

For the propagation direction $\mb=\eb_{3}$ (\emph{i.e.\/} along the tetragonal $c-$axis: henceforth we shall use $c$ and $\eb_{3}$ as synonimus when we describe the material symmetry) the tensor (\ref{Btensor}) reduces for all classes to the isotropic-like representation:
\begin{equation}\label{wavecaxis}
\hat{\Ab}(\eb_{3})=\frac{1}{\rho}\Big(\hat{\Cel}_{44}(\Ib-\Wb_{3})+\hat{\Cel}_{33}\Wb_{3}\Big)\,,
\end{equation}
and we have have a longitudinal and two transverse acoustic waves with frequencies:
\begin{eqnarray}\label{freqlongLWA}
&\omega_{1}(\xi)=\omega_{2}(\xi)={\displaystyle \xi\sqrt{\frac{\hat{\Cel}_{44}}{\rho}}}\,,&\ab_{1}=\cos\beta\eb_{1}+\sin\beta\eb_{2}\,,\nonumber\\
&&\ab_{2}=-\sin\beta\eb_{1}+\cos\beta\eb_{2}\,,\\
&\omega_{3}(\xi)={\displaystyle \xi\sqrt{\frac{\hat{\Cel}_{33}}{\rho}}}\,,&\ab_{3}=\eb_{3}\,,\nonumber
\end{eqnarray}

By (\ref{microlongwave}), these macroscopic displacements are accompained by a microdistortion associated with the longitudinal wave of frequency (\ref{freqlongLWA})$_{1}$
\begin{equation}\label{microlongwaveLS}
\Cb_{3}=iL_{m}\xi\Bel^{-1}\Qbm^{T}(\eb_{3})\eb_{3}\,;
\end{equation}
since the tensor $\Qbm(\eb_{3})$ for the classes $4$, $\bar{4}$ and $4/m$ has the tabular representation (\ref{QBMLS}) with the non-null components given by (\ref{QBMLS1}) and since $\Bel^{-1}$ has the same non-null components of (\ref{BelLS}), then the tensor $\Cb_{3}$ can be represented as
\begin{equation}
\Cb_{3}=\alpha_{3}(\Ib-\Wb_{3})+\beta_{3}\Wb_{3}+\gamma_{3}\bar{\Wb}_{6}\,,
\end{equation}
where:
\begin{eqnarray}\label{longwavelongi}
\alpha_{3}&=&\Qbm_{311}(\Bel^{-1}_{11}+\Bel^{-1}_{12})+\Qbm_{333}\Bel^{-1}_{13}+\Qbm_{312}(\Bel^{-1}_{16}-\Bel^{-1}_{26})\,,\nonumber\\
\beta_{3}&=&\Qbm_{333}\Bel^{-1}_{33}+2\Qbm_{311}\Bel^{-1}_{13}+2\Qbm_{312}\Bel^{-1}_{36}\,,\\
\gamma_{3}&=&\Qbm_{311}(\Bel^{-1}_{16}+\Bel^{-1}_{26})+\Qbm_{333}\Bel^{-1}_{36}+\Qbm_{312}(\Bel^{-1}_{66}-\Bel^{-1}_{69})\,,\nonumber
\end{eqnarray}
and the components $\Qbm_{311}$, $\Qbm_{333}$ and $\Qbm_{312}$ are obtained from (\ref{QBMLS1}) evaluted for $m_{1}=m_{2}=0$ and $m_{3}=1$. 

The microdistortions associated to the longitudinal waves are therefore a combination of the modes $\Db_{2}$ and $\Rb_{1}$. 

The microdistortions accompanied to the transverse waves which propagates along $\mb=\eb_{3}$ are instead given by
\begin{equation}\label{microtranswaveLS}
\Cb_{k}=iL_{m}\xi\Bel^{-1}\Qbm^{T}(\eb_{3})\ab_{k}\,,\quad k=1\,,2\,,
\end{equation}
which can be represented as
\begin{equation}\label{formula100}
\Cb_{k}=\alpha_{k}\hat{\Wb}_{4}+\beta_{k}\hat{\Wb}_{5}+\gamma_{k}\bar{\Wb}_{4}+\delta_{k}\bar{\Wb}_{5}\,,\quad k=1,2\,,
\end{equation}
with, for instance
\begin{align}
2\alpha_{1}&=(\Qbm_{113}C-\Qbm_{123}S)(\Bel_{44}^{-1}+\Bel_{47}^{-1})+(\Qbm_{132}C-\Qbm_{131}S)(\Bel_{55}^{-1}+\Bel_{47}^{-1})\nonumber\\
&+\Bel_{45}^{-1}((\Qbm_{131}-\Qbm_{123})C-(\Qbm_{132}+\Qbm_{113})S)\,,\nonumber\\
2\beta_{1}&=(\Qbm_{131}C-\Qbm_{132}C)(\Bel_{55}^{-1}+\Bel_{47}^{-1})+(\Qbm_{123}C+\Qbm_{113}S)(\Bel_{44}^{-1}+\Bel_{47}^{-1})\nonumber\\
&+\Bel_{45}^{-1}((\Qbm_{113}+\Qbm_{131})C-(\Qbm_{123}+\Qbm_{132})S)\,,\nonumber\\
2\gamma_{1}&=(\Qbm_{131}C-\Qbm_{132}C)(\Bel_{55}^{-1}-\Bel_{47}^{-1})+(\Qbm_{123}C+\Qbm_{113}S)(\Bel_{44}^{-1}-\Bel_{47}^{-1})\nonumber\\
&+\Bel_{45}^{-1}((\Qbm_{131}-\Qbm_{113})C+(\Qbm_{123}-\Qbm_{132})S)\,,\nonumber\\
2\delta_{1}&=(\Qbm_{113}C-\Qbm_{123}S)(\Bel_{44}^{-1}-\Bel_{47}^{-1})+(\Qbm_{132}C-\Qbm_{131}S)(\Bel_{55}^{-1}-\Bel_{47}^{-1})\nonumber\\
&+\Bel_{45}^{-1}((\Qbm_{131}+\Qbm_{123})C-(\Qbm_{132}-\Qbm_{113})S)\,,\nonumber
\end{align}
where $C=\cos\beta$, $S=\sin\beta$ and with $\Qbm_{123}$, $\Qbm_{131}$, $\Qbm_{132}$ and $\Qbm_{113}$ given by (\ref{QBMLS1}) with $m_{3}=1$ and $m_{1}=m_{2}=0$. 

Therefore each transverse wave $\ab_{k}$ which propagates along $\mb=\eb_{3}$ generates a combination of the modes $\Sbb_{2}$ and $\Rb_{2}$.
\begin{rmk}[Classes $4mmm$, $422$, $4/mm$, $\bar{4}2m$]
For these classes $\Bel^{-1}$ has the same non-null components of (\ref{BelLS}) and we also have $\Qbm_{312}=\Qbm_{123}=\Qbm_{132}=0$. The only noticeable difference with the results thus far obtained is that $\gamma_{3}=0$ into (\ref{longwavelongi}) and hence $\Cb_{3}$ reduces to the mode $\Db_{2}$.
\end{rmk}

Whenever the propagation direction is orthogonal to the $c-$axis, say $\mb=\cos\theta\eb_{1}+\sin\theta\eb_{2}$ then we have, for all $\theta$, a transverse wave which is directed as the $c-$axis:
\begin{equation}\label{transalpha1}
\omega_{1}(\xi)=\xi\sqrt{\frac{\hat{\Cel}_{44}}{\rho}}\,,\quad\ab_{1}=\eb_{3}
\end{equation}
and two waves which in general are neither transverse nor longitudinal: 
\begin{eqnarray}\label{speedneither}
\omega_{2,3}(\xi)&=&\xi\sqrt{\frac{1}{2\rho}\Big(a\pm\sqrt{b\cos^{2}2\theta+c\sin^{2}2\theta+2d\sin 2\theta\cos 2\theta}\Big)}\,,\nonumber\\
\ab_2&=&\cos\beta\eb_1+\sin\beta\eb_2\,,\\
\ab_3&=&-\sin\beta\eb_1+\cos\beta\eb_2\,,\nonumber
\end{eqnarray}
with
\begin{equation}\label{cond1}
\tan\beta=\frac{a-\sqrt{b\cos^{2}2\theta+c\sin^{2}2\theta+2d\sin 2\theta\cos 2\theta}}{2\hat{\Cel}_{16}\cos 2\theta+(\hat{\Cel}_{66}+2\hat{\Cel}_{12})\sin 2\theta}\,,
\end{equation}
where, for the classes $4$, $\bar{4}$ and $4/m$:
\begin{eqnarray}\label{epsilonLSA}
&a=\hat{\Cel}_{11}+\hat{\Cel}_{66}\,,\quad b=(\hat{\Cel}_{11}-\hat{\Cel}_{66})^{2}+4\hat{\Cel}_{16}^{2}\,,\\
&c=\hat{\Cel}^{2}_{16}+(\hat{\Cel}_{66}+2\hat{\Cel}_{12})^{2}\,,\quad d=\hat{\Cel}_{16}(\hat{\Cel}_{11}+\hat{\Cel}_{66}+4\hat{\Cel}_{12})\,.\nonumber
\end{eqnarray}

We may search for the angle $\theta$ such that $\ab_{2}$ is a longitudinal wave with frequency $\omega_{2}$ and $\ab_{3}$ is a transverse wave with frequency $\omega_{3}$, which is equivalent to require either that $\ab_{2}\times\mb=\bzero$ and hence $\theta=\beta$, or that $A_{12}=0$ which gives
\begin{equation}\label{cond2}
\tan 2\theta=-\frac{2\hat{\Cel}_{16}}{\hat{\Cel}_{12}+2\hat{\Cel}_{66}}\,.
\end{equation}

For the classes $4mm\,,422\,,4/mm$ and $\bar{4}2m$ with $\hat{\Cel}_{16}=0$ then
\begin{equation}\label{epsilonHSA}
b=(\hat{\Cel}_{11}-\hat{\Cel}_{66})^{2}\,,\quad c=(\hat{\Cel}_{66}+2\hat{\Cel}_{12})^{2}\,,\quad d=0\,,
\end{equation} 
and then from (\ref{cond2}) and (\ref{cond1}) with $\beta=\theta$  it is easy to see that for $\theta\in\{0\,,\pi/4\,,\pi/2\}$ we have a longitudinal ($\ab_{2}=\mb$) and a transverse ($\ab_{3}=\eb_{3}\times\mb$) waves whose frequencies are given by (\ref{speedneither})$_{1}$ when we use (\ref{epsilonHSA}) in place of (\ref{epsilonLSA}).

The microdistortion which corresponds to the transverse wave (\ref{transalpha1}) is:
\begin{equation}\label{microtransalpha2}
\Cb_{1}=iL_{m}\xi\Bel^{-1}\Qbm^{T}(\theta)\eb_{3}\,,
\end{equation}
and by (\ref{QBMLS}) and (\ref{BelLS}) we obtain again (\ref{formula100}) with for instance, when $k=1$:
\begin{align}\label{formula200}
2\alpha_{1}&=\Qbm_{323}(\Bel_{44}^{-1}+\Bel_{47}^{-1})+\Qbm_{332}(\Bel_{47}^{-1}-\Bel_{45}^{-1})+\Qbm_{331}\Bel_{45}^{-1}+\Qbm_{313}\Bel_{55}^{-1}\,,\nonumber\\
2\beta_{1}&=\Qbm_{331}(\Bel_{47}^{-1}+\Bel_{55}^{-1})+\Qbm_{313}(\Bel_{44}^{-1}+\Bel_{47}^{-1})-\Qbm_{332}\Bel_{45}^{-1}+\Qbm_{323}\Bel_{47}^{-1}\,,\\
2\gamma_{1}&=\Qbm_{323}(\Bel_{44}^{-1}-\Bel_{47}^{-1})+\Qbm_{332}(\Bel_{47}^{-1}+\Bel_{45}^{-1})+\Qbm_{331}\Bel_{45}^{-1}-\Qbm_{313}\Bel_{55}^{-1}\,,\nonumber\\
2\delta_{1}&=\Qbm_{331}(\Bel_{47}^{-1}-\Bel_{55}^{-1})+\Qbm_{313}(\Bel_{44}^{-1}-\Bel_{47}^{-1})-\Qbm_{332}\Bel_{45}^{-1}-\Qbm_{323}\Bel_{47}^{-1}\,;\nonumber
\end{align}
here the components  $\Qbm_{33\alpha}(\theta)$ and $\Qbm_{3\alpha3}(\theta)$, $\alpha=1,2$ are given by (\ref{QBMLS1}) with $m_{1}=\cos\theta$, $m_{2}=\sin\theta$ and $m_{3}=0$. 

Accordingly this transverse wave is associated to a shear between $\mb$ and $\eb_{3}$ and a rigid rotation about $\eb_{3}\times\mb$.

When we turn our attention to the other two waves, which are neither transverse nor longitudinal, we have that the two associated microdistortions
\begin{equation}\label{microtransalpha3}
\Cb_{\gamma}=iL_{m}\xi\Bel^{-1}\Qbm^{T}(\theta)\ab_{\gamma}\,,\quad\gamma=2,3\,,
\end{equation}
in view of (\ref{BelLS}) and (\ref{QBMLS}) can be represented as:
\begin{eqnarray}
\Cb_{2}&=&iL_{m}\xi(\cos\beta\,\Bb_{1}+\sin\beta\,\Bb_{2})\,,\\
\Cb_{3}&=&iL_{m}\xi(-\sin\beta\,\Bb_{1}+\cos\beta\,\Bb_{2})\,;\nonumber
\end{eqnarray}
the two tensors $\Bb_{k}=\Bel^{-1}\Qbm^{T}(\theta)\eb_{k}$, $k=1,2$ have the common structure
\begin{equation}\label{formula102}
\Bb_{k}=\alpha_{k}\Wb_{1}+\beta_{k}\Wb_{2}+\gamma_{k}\Wb_{3}+\delta_{k}\hat{\Wb}_{6}+\epsilon_{k}\bar{\Wb}_{6}\,,
\end{equation}
where
\begin{eqnarray}
\alpha_{k}&=&\Bel_{11}^{-1}\Qbm_{k 11}+\Bel_{12}^{-1}\Qbm_{k 22}+\Bel_{13}^{-1}\Qbm_{k 33}+\Bel_{16}^{-1}\Qbm_{k 12}+\Bel_{26}^{-1}\Qbm_{k 21}\,,\nonumber\\
\beta_{k}&=&\Bel_{12}^{-1}\Qbm_{k 11}+\Bel_{11}^{-1}\Qbm_{k 22}+\Bel_{13}^{-1}\Qbm_{k 33}+\Bel_{26}^{-1}\Qbm_{k 12}+\Bel_{16}^{-1}\Qbm_{k 21}\,,\nonumber\\
\gamma_{k}&=&\Bel_{13}^{-1}(\Qbm_{k 11}+\Qbm_{k 22})+\Bel_{33}^{-1}\Qbm_{k 33}+\Bel_{36}^{-1}(\Qbm_{k 12}-\Qbm_{k 21})\,,\quad k=1,2\,,\\
\delta_{k}&=&\frac{1}{2}(\Bel_{16}^{-1}-\Bel_{26}^{-1})(\Qbm_{k 11}+\Qbm_{k 22})+\frac{1}{2}(\Bel_{66}^{-1}-\Bel_{69}^{-1})(\Qbm_{k 12}+\Qbm_{k 21})\,,\nonumber\\
\epsilon_{k}&=&\frac{1}{2}(\Bel_{16}^{-1}+\Bel_{26}^{-1})(\Qbm_{k 11}+\Qbm_{k 22})+\frac{1}{2}(\Bel_{66}^{-1}-\Bel_{69}^{-1})(\Qbm_{k 12}-\Qbm_{k 21})+\Bel_{36}^{-1}\Qbm_{k 33}\,;\nonumber
\end{eqnarray}
here the components of $\Qbm(\theta)$ are obtained again from (\ref{QBMLS1}) with $m_{1}=\cos\theta$, $m_{2}=\sin\theta$ and $m_{3}=0$. 

Interestingly enough for both the waves with amplitudes $\ab_{2}$ and $\ab_{3}$, the corresponding microdistortions are a combination of the modes $\Db_{1}$, $\Sbb_{1}$ and $\Rb_{1}$. We remark that this situation is maintained even when the propagation direction is given by (\ref{cond2}) and the two waves becomes one transverse and the other longitudinal.

\begin{rmk}[Classes $4mmm$, $422$, $4/mm$, $\bar{4}2m$]
For these classes, by (\ref{QbmHS})  we have that formula (\ref{formula102}) holds with $\alpha_{k}=\beta_{k}$ and $\delta_{k}=\epsilon_{k}$ and therefore the mode $\Db_{1}$ changes into $\Db_{2}$.
\end{rmk}

\subsection{Microvibrations}\label{sectionmicrovibe}

\subsubsection{Classes $4$, $\bar{4}$ and $4/m$}

We begin with the lower-symmetry tetragonal classes: when we  consider the propagation condition (\ref{microvibra2}) with $\Bel$ and $\Jel$ given respectively by (\ref{BelLS}) and (\ref{Jtetra}), we notice that both tensors are reduced by the two subspaces of $\Lin$, $\caU_{1}$ and $\caU_{2}$ with $\Lin\equiv\caU_{1}\oplus\caU_{2}$: 
\begin{eqnarray}\label{setreductionmicro}
&\caU_{1}\equiv\spn\{\Wb_{1}\,,\Wb_{2}\,,\Wb_{3}\,,\Wb_{6}\,,\Wb_{9}\}\,,\\
&\caU_{2}\equiv\spn\{\Wb_{4}\,,\Wb_{5}\,,\Wb_{7}\,,\Wb_{8}\}\,,\nonumber
\end{eqnarray}
that is:
\begin{equation}
\Bel[\Cb_{\alpha}]\in\caU_{\alpha}\,,\quad\Jel[\Cb_{\alpha}]\in\caU_{\alpha}\,,\quad\forall\Cb_{\alpha}\in\caU_{\alpha}\,,\quad\alpha=1,2\,.
\end{equation}
The eigencouples split accordingly into two group: a first one $(\omega_{k}\,,\Cb_{k})$, $k=1,\ldots\,5$ with $\Cb_{k}\in\caU_{1}$ and whose eigentensors are a combination of the modes $\Db_{k}$, $\Sbb_{1}$ and $\Rb_{1}$; the second one $(\omega_{j}\,,\Cb_{j})$, $j=6,\ldots\,9$ with $\Cb_{j}\in\caU_{2}$ and whose eigentensors are a combination of the modes $\Sbb_{2}$ and $\Rb_{2}$.

We define the normalized components of $\Bel$ as follow:
\begin{align}\label{compoU1micro}
&a=\frac{\Bel_{11}}{\rho J_{11}}\,,\quad b=\frac{\Bel_{33}}{\rho J_{33}}\,,\quad d=\frac{\Bel_{12}}{\rho\sqrt{J_{11}J_{33}}}\,,\quad e=\frac{\Bel_{13}}{\rho\sqrt{J_{11}J_{33}}}\,,\nonumber\\
&c=\frac{\Bel_{66}}{\rho J_{11}}\,,\quad f=\frac{\Bel_{69}}{\rho J_{11}}\,,\quad g=\frac{\Bel_{16}}{\rho J_{11}}\,,\quad h=\frac{\Bel_{26}}{\rho J_{11}}\,,\quad l=\frac{\Bel_{36}}{\rho\sqrt{J_{11}J_{33}}}\,,\\
&m=\frac{\Bel_{44}}{\rho J_{33}}\,,\quad n=\frac{\Bel_{55}}{\rho J_{11}}\,,\quad p=\frac{\Bel_{45}}{\rho\sqrt{J_{11}J_{33}}}\,,\quad q=\frac{\Bel_{47}}{\rho\sqrt{J_{11}J_{33}}}\,,\nonumber
\end{align}
and we begin our analysis with the subspace $\caU_{1}$. We notice that the algebraic fifth-order characteristic equation in $\omega^{2}$ can be factorized into
\begin{equation}
(\omega^{4}-B\omega^{2}+C)(\omega^{6}+D\omega^{4}+E\omega^{2}+F)=0\,,
\end{equation}
where
\begin{align}
B&=a+c+f-d\,,\nonumber\\
C&=(a-d)(c+f)-(g-h)^{2}\,,\nonumber\\
D&=a+b+c+d-f\,,\\
E&=2a(b+c)+bc-f(a+b+d)-2(e^{2}+l^{2})-(g+h)^{2}\,,\nonumber\\
F&=(c-b)^2-(g+h)^2-2d^2-9e^2+f^2+18l^2\,,\nonumber\\
&+3d(b+c)-2f(2b+c+d)-3gh-4bc\,.\nonumber
\end{align}
The eigenvalues are thus given by
\begin{equation}\label{RAI1}
\omega_{1,2}^{2}=\frac{1}{2}(B\mp\sqrt{B^{2}-4C})\,,
\end{equation}
with $\omega_{1}<\omega_{2}$ and, by the means of Cardano's formulae \cite{CRC}, by:
\begin{align}\label{RAI2}
\omega_{3}^{2}&=\frac{1}{3}(D+2\sqrt{P}\cos\frac{\theta}{3})\,,\nonumber\\
\omega_{4}^{2}&=\frac{1}{3}(D+2\sqrt{P}\cos\frac{\theta+2\pi}{3})\,,\\
\omega_{5}^{2}&=\frac{1}{3}(D+2\sqrt{P}\cos\frac{\theta-2\pi}{3})\,,\nonumber
\end{align}
with either $\omega_{4}<\omega_{3}<\omega_{5}$ or $\omega_{4}>\omega_{3}>\omega_{5}$ and where:
\begin{equation}
P=D^{2}-3E\,,\quad Q=D^{3}-9DE-27F\,,\quad\theta=\cos^{-1}\frac{Q}{2\sqrt[3]{P}}\,.
\end{equation}
The associated eigentensors combine the modes $\Db_{1}\,,\Sbb_{1}\,,\Rb_{1}$ in all but one case, when it combines $\Db_{2}$ and $\Rb_{1}$:
\begin{align}\label{tensorsU1}
\Cb_{1}&=\alpha_{1}\Wb_{1}+\beta_{1}\Wb_{2}+\gamma_{1}\Wb_{3}+\delta_{1}\hat{\Wb}_{6}+\epsilon_{1}\bar{\Wb}_{6}\,,\nonumber\\
\Cb_{2}&=\beta_{1}\Wb_{1}+\alpha_{1}\Wb_{2}+\gamma_{1}\Wb_{3}-\delta_{1}\hat{\Wb}_{6}+\epsilon_{1}\bar{\Wb}_{6}\,,\nonumber\\
\Cb_{3}&=\alpha_{3}(\Ib-\Wb_{3})+\gamma_{3}\Wb_{3}+2\delta_{3}\bar{\Wb}_{6}\,,\\
\Cb_{4}&=\alpha_{4}\Wb_{1}+\beta_{4}\Wb_{2}+\gamma_{4}\Wb_{3}+\delta_{4}\hat{\Wb}_{6}+\epsilon_{4}\bar{\Wb}_{6}\,,\nonumber\\
\Cb_{5}&=\beta_{4}\Wb_{1}+\alpha_{4}\Wb_{2}+\gamma_{4}\Wb_{3}-\delta_{4}\hat{\Wb}_{6}+\epsilon_{4}\bar{\Wb}_{6}\,;\nonumber
\end{align}
here the real coefficients $\alpha_{k}\,,\beta_{k}\,,\gamma_{k}\,,\delta_{k}$ and $\epsilon_{k}$, which depend on the components of $\Bel$, are given explicitly into the Appendix, \S.\ref{eigencompU1}.

When we turn our attention to the subspace $\caU_{2}$, then we get two eigenvalues of multiplicity 2:
\begin{align}\label{RAI3}
&\omega^{2}_{6}=\omega_{7}^{2}=\frac{1}{2}(m+n)-\sqrt{(\frac{m-n}{2})^{2}+p^{2}+q^{2}}\,,\\
&\omega_{8}^{2}=\omega_{9}^{2}=\frac{1}{2}(m+n)+\sqrt{(\frac{m-n}{2})^{2}+p^{2}+q^{2}}\,,\nonumber
\end{align}
whose (non-normalized ) eigentensors are
\begin{align}\label{EIGEN3}
\Cb_{6}&=p\hat{\Wb}_{4}-p\bar{\Wb}_{4}+(n-q-\omega_{6}^{2})\hat{\Wb}_{5}-(q+n-\omega_{6}^{2})\bar{\Wb}_{5}\,,\nonumber\\
\Cb_{7}&=(m-q-\omega_{6}^{2})\hat{\Wb}_{4}-(q+m-\omega_{6}^{2})\bar{\Wb}_{4}+p\hat{\Wb}_{5}-p\bar{\Wb}_{5}\,,\\
\Cb_{8}&=-p\hat{\Wb}_{4}-p\bar{\Wb}_{4}+(m-q-\omega_{8}^{2})\hat{\Wb}_{5}+(m+q-\omega_{8}^{2})\bar{\Wb}_{5}\,,\nonumber\\
\Cb_{9}&=(n-q-\omega_{8}^{2})\hat{\Wb}_{4}+(n+q-\omega_{8}^{2})\bar{\Wb}_{4}-p\hat{\Wb}_{5}-p\bar{\Wb}_{5}\,,\nonumber
\end{align}
each one representing a combination of the modes $\Sbb_{2}$ and $\Rb_{2}$; from (\ref{RAI3}) we have
\begin{equation}
\omega_{6}=\omega_{7}<\omega_{8}=\omega_{9}\,.
\end{equation}
All together  we have
\begin{align}
\omega_{1}<\omega_{2}\,,\quad\omega_{4}<\omega_{3}<\omega_{5}\,,\mbox{ (or }\omega_{4}>\omega_{3}>\omega_{5})\,,\quad\omega_{6}=\omega_{7}<\omega_{8}=\omega_{9}\,,
\end{align}
and besides this we cannot give a complete ordering between these frequencies without the knowledge of the numerical values of components of $\Bel$.

\subsubsection{Classes $4mm$, $422$, $\bar{4}2m$ and $4/mm$}
For these classes the matrix $\Bel$ is reduced by the three subspaces:
\begin{equation}
\caZ_{1}\equiv\spn\{\Wb_{1}\,,\Wb_{2}\,,\Wb_{3}\}\,,\quad\caZ_{2}\equiv\spn\{\Wb_{6}\,,\Wb_{9}\}\,,\quad\caZ_{3}=\caU_{2},,
\end{equation}
with $\Lin\equiv\caZ_{1}\oplus\caZ_{2}\oplus\caZ_{3}$. 

To find the solution in $\caZ_{1}$ we notice that since $\omega^{2}=a-d$ is a root for the cubic characteristic equation, then we can easily obtain:
\begin{align}\label{cutoffHS}
\omega_{1}^{2}&=a-d\,,\nonumber\\
\omega_{2}^{2}&=\frac{a+b+d}{2}-\sqrt{(\frac{a+d-b}{2})^{2}+2e^{2}}\,,\\
\omega_{3}^{2}&=\frac{a+b+d}{2}+\sqrt{(\frac{a+d-b}{2})^{2}+2e^{2}}\,,\nonumber
\end{align}
with
\begin{equation}\label{microZ1}
\omega_{2}<\omega_{3}\,,\quad\omega_{1}<\omega_{3}\,.
\end{equation}
The corresponding (non-normalized) eigentensors are:
\begin{eqnarray}\label{treeigenHS}
\Cb_{1}&=&\Wb_{1}-\Wb_{2}\,,\nonumber\\
\Cb_{2}&=&\alpha_{2}\Wb_{1}+\beta_{2}\Wb_{2}+\gamma_{2}\Wb_{3}\,,\\
\Cb_{3}&=&\beta_{2}\Wb_{1}+\alpha_{2}\Wb_{2}+\gamma_{2}\Wb_{3}\,,\nonumber
\end{eqnarray}
with:
\begin{equation}
\alpha_{2}=(b-\omega_{2}^{2})d-e^2\,,\quad\beta_{2}=e^2-b(a-\omega_{2}^{2})\,,\quad\gamma_{2}=e(a-d-\omega_{2}^{2})\,.
\end{equation}

In the subspace $\caZ_{2}$ we have two eigencouples associated a shear in the plane orthogonal to the $c$-axis ($\Cb_{4}$) and a rotation about the propagation direction ($\Cb_{5}$):
\begin{align}\label{microZ2}
&(\omega^{2}_{4}=c-f\,,\quad\Cb_{4}=\bar{\Wb}_{6})\,,\\
&(\omega^{2}_{5}=c+f\,,\quad\Cb_{5}=\hat{\Wb}_{6})\,,\nonumber
\end{align}
whereas the solutions on $\caZ_{3}$ are given by (\ref{RAI3}) and (\ref{EIGEN3}) with $p=0$; again we have
\begin{equation}\label{microZ31}
\omega_{6}=\omega_{7}<\omega_{8}=\omega_{9}\,.
\end{equation}
and for $\Bel_{69}>0$ also
\begin{equation}\label{microZ32}
\omega_{4}<\omega_{5}\,,
\end{equation}
the inequality being reversed when $\Bel_{69}<0$.

The kinematics of microdistortions for these classes is represents in $\caZ_{1}$ either $\Db_{1}$, or the traceless plane strain $\Db_{3}$; in $\caZ_{2}$ we have instead a shear in the plane orthogonal to the $c$-axis and a rigid microrotation about the same direction whereas in $\caZ_{3}$ we have the same kinematics as in $\caU_{2}$.

Also for these classes it is not possible to give a complete ordering between all the nine eigenvalues (\ref{microZ1}), (\ref{microZ31}) and (\ref{microZ32}), in absence of the numerical values for the components of $\Bel$.

\subsection{Micromorphic continua}

\subsubsection{Propagation along the tetragonal $c$-axis}

\subparagraph{Classes $4$, $\bar{4}$ and $4/m$.} We begin with the lower-symmetry classes; in this case the blocks of the matrix $A$ have the following non-null components

\begin{equation}\label{matrixMLSx}
\renewcommand{\arraystretch}{1.5} 
A\equiv {\tiny
\left[\begin{array}{@{}ccc|ccccccccc@{}}
\bullet&\bullet&\cdot&\cdot&\cdot&\cdot&\bullet&\bullet&\cdot&\bullet&\bullet&\cdot \\
\bullet&\bullet&\cdot&\cdot&\cdot&\cdot&\bullet&\bullet&\cdot&\bullet&\bullet&\cdot \\
\cdot&\cdot&\bullet&\bullet&\bullet&\bullet&\cdot&\cdot&\bullet&\cdot&\cdot&\bullet \\ \hline
\cdot&\cdot&\bullet&\bullet&\bullet&\bullet&\cdot&\cdot&\bullet&\cdot&\cdot&\bullet \\
\cdot&\cdot&\bullet&\bullet&\bullet&\bullet&\cdot&\cdot&\bullet&\cdot&\cdot&\bullet \\
\cdot&\cdot&\bullet&\bullet&\bullet&\bullet&\cdot&\cdot&\bullet&\cdot&\cdot&\bullet\\
\bullet&\bullet&\cdot&\cdot&\cdot&\cdot&\bullet&\bullet&\cdot&\bullet&\bullet&\cdot \\
\bullet&\bullet&\cdot&\cdot&\cdot&\cdot&\bullet&\bullet&\cdot&\bullet&\bullet&\cdot \\
\cdot&\cdot&\bullet&\bullet&\bullet&\bullet&\cdot&\cdot&\bullet&\cdot&\cdot&\bullet \\
\bullet&\bullet&\cdot&\cdot&\cdot&\cdot&\bullet&\bullet&\cdot&\bullet&\bullet&\cdot \\
\bullet&\bullet&\cdot&\cdot&\cdot&\cdot&\bullet&\bullet&\cdot&\bullet&\bullet&\cdot \\
\cdot&\cdot&\bullet&\bullet&\bullet&\bullet&\cdot&\cdot&\bullet&\cdot&\cdot&\bullet
\end{array}\right]}\,,
\end{equation}

with the independent components given by (\ref{acoustic1LS}), (\ref{PBMLS}), (\ref{QBMLS1}) and (\ref{AELLS}) evaluated for $m_{1}=m_{2}=0$ and $m_{3}=1$.

The matrix (\ref{matrixMLSx}) is reduced by the pairs $\caM_{1}$ and $\caM_{2}$: 
\begin{equation}
\caM_{1}\equiv\spn\{\eb_{3}\}\oplus\caU_{1}\,,\quad\caM_{2}\equiv\spn\{\eb_{1}\,,\eb_{2}\}\oplus\caU_{2}\,,
\end{equation}
were $\caU_{1}$ and $\caU_{2}$ are defined by (\ref{setreductionmicro}).

We begin with the subspace $\caM_{1}$ and define the normalized components:
\begin{align}\label{compoM1}
a(\xi)&=\xi^{2}A_{33}\,,\nonumber\\
b(\xi)&=\xi^{2}\frac{\Pbm_{311}L_{c}^{2}}{L_{m}\sqrt{J_{11}}}+i\xi\frac{\Qbm_{311}}{\sqrt{J_{11}}}\,,\quad c(\xi)=\xi^{2}\frac{\Pbm_{333}L_{c}^{2}}{L_{m}\sqrt{J_{33}}}+i\xi\frac{\Qbm_{333}}{\sqrt{J_{33}}}\,,\nonumber\\
d(\xi)&=\xi^{2}\frac{\Ael_{66}L_{c}^{2}}{J_{11}}+\frac{\Bel_{66}}{\rho J_{11}}\,,\quad e(\xi)=\xi^{2}\frac{\Ael_{69}L_{c}^{2}}{J_{11}}+\frac{\Bel_{69}}{\rho J_{11}}\,,\nonumber\\
f(\xi)&=\xi^{2}\frac{\Ael_{11}L_{c}^{2}}{J_{11}}+\frac{\Bel_{11}}{\rho J_{11}}\,,\quad g(\xi)=\xi^{2}\frac{\Ael_{33}L_{c}^{2}}{J_{33}}+\frac{\Bel_{33}}{\rho J_{33}}\,,\\
h(\xi)&=\xi^{2}\frac{\Ael_{12}L_{c}^{2}}{\sqrt{J_{11}J_{33}}}+\frac{\Bel_{12}}{\rho\sqrt{J_{11}J_{33}}}\,,\quad l(\xi)=\xi^{2}\frac{\Ael_{13}L_{c}^{2}}{\sqrt{J_{11}J_{33}}}+\frac{\Bel_{13}}{\rho\sqrt{J_{11}J_{33}}}\,,\nonumber\\
n(\xi)&=\xi^{2}\frac{\Ael_{16}L_{c}^{2}}{J_{11}}+\frac{\Bel_{16}}{\rho J_{11}}\,,\quad p(\xi)=\xi^{2}\frac{\Ael_{26}L_{c}^{2}}{J_{11}}+\frac{\Bel_{26}}{\rho J_{11}}\,,\nonumber\\
m(\xi)&=\xi^{2}\frac{\Ael_{36}L_{c}^{2}}{\sqrt{J_{11}J_{33}}}+\frac{\Bel_{36}}{\rho\sqrt{J_{11}J_{33}}}\,,\quad q(\xi)=\xi^{2}\frac{\Pbm_{312}L_{c}^{2}}{L_{m}\sqrt{J_{11}}}+i\xi\frac{\Qbm_{312}}{\sqrt{J_{11}}}\,;\nonumber
\end{align}
then, since the characteristic equation can be factorized into a second- and a fourth-grade algebraic equations, we obtain by the means of the Cardano's formulae for the fourth-grade algebraic equations \cite{CRC}, the explicit representation of the six eigenvalues
\begin{align}\label{eigenvaluesM1}
\omega_{1}^{2}&=\frac{1}{2} \left(F-\sqrt{F^2-4G}\right),\nonumber\\
\omega_{2}^{2}&=\frac{1}{2} \left(F+\sqrt{F^2-4G}\right),\nonumber\\
\omega_{3}^{2}&=\frac{1}{4}A -\frac{1}{2} \sqrt{\frac{1}{4} A^2+\frac{2}{3}B+W}\nonumber\\
&-\frac{1}{2} \sqrt{\frac{1}{2} A^2+\frac{4}{3}B-W-\frac{S}{4 \sqrt{\frac{1}{4} A^2+\frac{2}{3}B+W}}}\,,\nonumber\\
\omega_{4}^{2}&=\frac{1}{4}A -\frac{1}{2} \sqrt{\frac{1}{4} A^2+\frac{2}{3}B+W}\\
&+\frac{1}{2} \sqrt{\frac{1}{2} A^2+\frac{4}{3}B-W-\frac{S}{4 \sqrt{\frac{1}{4} A^2+\frac{2}{3}B+W}}}\,,\nonumber\\
\omega_{5}^{2}&=\frac{1}{4}A +\frac{1}{2} \sqrt{\frac{1}{4} A^2+\frac{2}{3}B+W}\nonumber\\
&-\frac{1}{2} \sqrt{\frac{1}{2} A^2+\frac{4}{3}B-W+\frac{S}{4 \sqrt{\frac{1}{4} A^2+\frac{2}{3} B+W}}}\,,\nonumber\\
\omega_{6}^{2}&=\frac{1}{4}A+\frac{1}{2} \sqrt{\frac{1}{4} A^2+\frac{2}{3}B+W}\nonumber\\
&+\frac{1}{2} \sqrt{\frac{1}{2} A^2+\frac{4}{3}B-W+\frac{S}{4 \sqrt{\frac{1}{4} A^2+\frac{2}{3} B+W}}}\,,\nonumber
\end{align}
where:
\begin{align}\label{charaM1}
A&=a+d-e+f+g+h\,,\nonumber\\
B&=cc^{\star}+2(bb^{\star}+l^2+ m^2+ qq^{\star})+(n+p)^2-(d-e)(a+f+g+h)\nonumber\\
&-(f+h)(a+g)-ag\,,\nonumber\\
C&=2bb^{\star}( d -e +g)+2 l^2(a+d-e)+2 m^2(a+f+h)+2qq^{\star}(f+g+h)\nonumber\\
&+(a+g)((n+p)^{2}-(f+h)(d-e))+(cc^{\star}-ag)(d-e+f+h)\\
&-4((lm+qb^{\star})(n+p)+(lb+mq)c^{\star})\,,\nonumber\\
D&=2bb^{\star}(m^{2}-g(d-e))+2qq^{\star}(2l^2-g(f+h))+2l(2bc^{\star}-al)(d-e)\nonumber\\
&+2m(2cq^{\star}-am)(f+h)+((n+p)^2-(f+h)(d-e))(cc^{\star} -a g)\nonumber\\
&+4(n+p)(m(cb^{\star}+al)+q(gb^{\star}-c^{\star}l))-8 l m q b^{\star}\nonumber\\
F&=d+e+f-h\,,\nonumber\\
G&=(d+e)( f-h)-(n-p)^2\,,\nonumber
\end{align}
and
\begin{align}
P&=-2B^3-9ACB+72DB+27 C^2+27 A^2 D\,,\nonumber\\
Q&=B^2+3AC+12 D\,,\nonumber\\
S&=A^3+4AB-8 C\,,\nonumber\\
W&=\frac{\sqrt[3]{P+\sqrt{P^2-4Q^3}}}{3 \sqrt[3]{2}}+\frac{\sqrt[3]{2} Q}{3 \sqrt[3]{P+\sqrt{P^2-4Q^3}}}\,.\nonumber
\end{align}

By looking at (\ref{eigenvaluesM1}), by (\ref{compoM1}) and (\ref{charaM1}) we notice that the frequencies $\omega_{1,2}$ depend solely on the components of $\Ael$ and $\Bel$, whereas the others depend also on the acoustic tensors $\Ab$, $\Pbm$ and $\Qbm$.

The associated eigenvectors:
\begin{align}\label{eigentensorwM1}
w_{1}&=\{\ab_{o}^{1}=\alpha_{1}\eb_{3}\,;\Vb_{1}=\beta_{1}\Wb_{1}+\gamma_{1}\Wb_{2}+\delta_{1}\Wb_{3}+(\theta_{1}+\epsilon_{1})\hat{\Wb}_{6}+(\epsilon_{1}-\theta_{1})\bar{\Wb}_{6}\}\,,\nonumber\\
w_{2}&=\{\ab_{o}^{2}=\alpha_{1}\eb_{3}\,;\Vb_{2}=\gamma_{1}\Wb_{1}+\beta_{1}\Wb_{2}+\delta_{1}\Wb_{3}-(\theta_{1}+\epsilon_{1})\hat{\Wb}_{6}+(\epsilon_{1}-\theta_{1})\bar{\Wb}_{6}\}\,,\nonumber\\
w_{3}&=\{\ab_{o}^{3}=\alpha_{3}\eb_{3}\,;\Vb_{3}=\beta_{3}(\Ib-\Wb_{3})+\delta_{3}\Wb_{3}+2\epsilon_{3}\bar{\Wb}_{6}\}\,,\\
w_{4}&=\{\ab_{o}^{4}=\alpha_{4}\eb_{3}\,;\Vb_{4}=\beta_{4}\Wb_{1}+\gamma_{4}\Wb_{2}+\delta_{4}\Wb_{3}+(\beta_{1}-\gamma_{1})\hat{\Wb}_{6}+(\beta_{1}+\gamma_{1})\bar{\Wb}_{6}\}\,,\nonumber\\
w_{5}&=\{\ab_{o}^{5}=\alpha_{4}\eb_{3}\,;\Vb_{5}=\gamma_{4}\Wb_{1}+\beta_{4}\Wb_{2}+\delta_{4}\Wb_{3}-(\beta_{1}-\gamma_{1})\hat{\Wb}_{6}+(\beta_{1}+\gamma_{1})\bar{\Wb}_{6}\}\,,\nonumber\\
w_{6}&=\{\ab_{o}^{6}=\alpha_{6}\eb_{3}\,;\Vb_{6}=\beta_{6}(\Ib-\Wb_{3})+\delta_{6}\Wb_{3}+2\epsilon_{6}\bar{\Wb}_{6}\}\,,\nonumber
\end{align}
represents a longitudinal wave combined either with the modes $\Db_{1}$, $\Sbb_{1}$, $\Rb_{1}$ or with the modes $\Db_{2}$ and $\Rb_{1}$: the coefficients of (\ref{eigentensorwM1}) are given in \S.\ref{eigencompM1} of the Appendix.

For $\xi\rightarrow 0$ since $a\,,b\,,c$ and $q$ vanish, then from (\ref{eigenvaluesM1}) and the results of \S.\ref{eigencompM1} we have that:
\begin{align}\label{cutoffM1}
&(\omega_{1,2}(\xi)\,,w_{1,2})\rightarrow(\omega_{1,2}\,,\{\bzero\,,\Cb_{1,2}\})\,,\nonumber\\
&(\omega_{6}(\xi)\,,w_{6})\rightarrow(\omega_{3}\,,\{\bzero\,,\Cb_{3}\})\,,\\
&(\omega_{4,5}(\xi)\,,w_{4,5})\rightarrow(\omega_{4,5}\,,\{\bzero\,,\Cb_{4,5}\})\,,\nonumber\\
&(\omega_{3}(\xi)\,,w_{3})\rightarrow(0\,,\{\eb_{3}\,,\bzero\})\,,\nonumber
\end{align}
with the cut-off frequencies $\omega_{1,2}$ and $\omega_{3,4,5}$ given respectively by (\ref{RAI1}) and (\ref{RAI2}) and where the microdistortions $\Cb_{1-5}$ are given by (\ref{tensorsU1}); therefore in the subspace $\caM_{1}$ there are one acoustic longitudinal wave and five optic waves with cut-off frequencies (\ref{RAI1}) and (\ref{RAI2}).

A qualitative graph of the dispersion relations $\omega=\omega(\xi)$ for the solutions in the subspace $\caM_{1}$, which can be obtained for arbitrary values of the components, is given in Fig. \ref{fig1}.

\begin{center}
\begin{picture}(200,180)\label{fig1}
\put(-40,0){\scalebox{.20}{\includegraphics{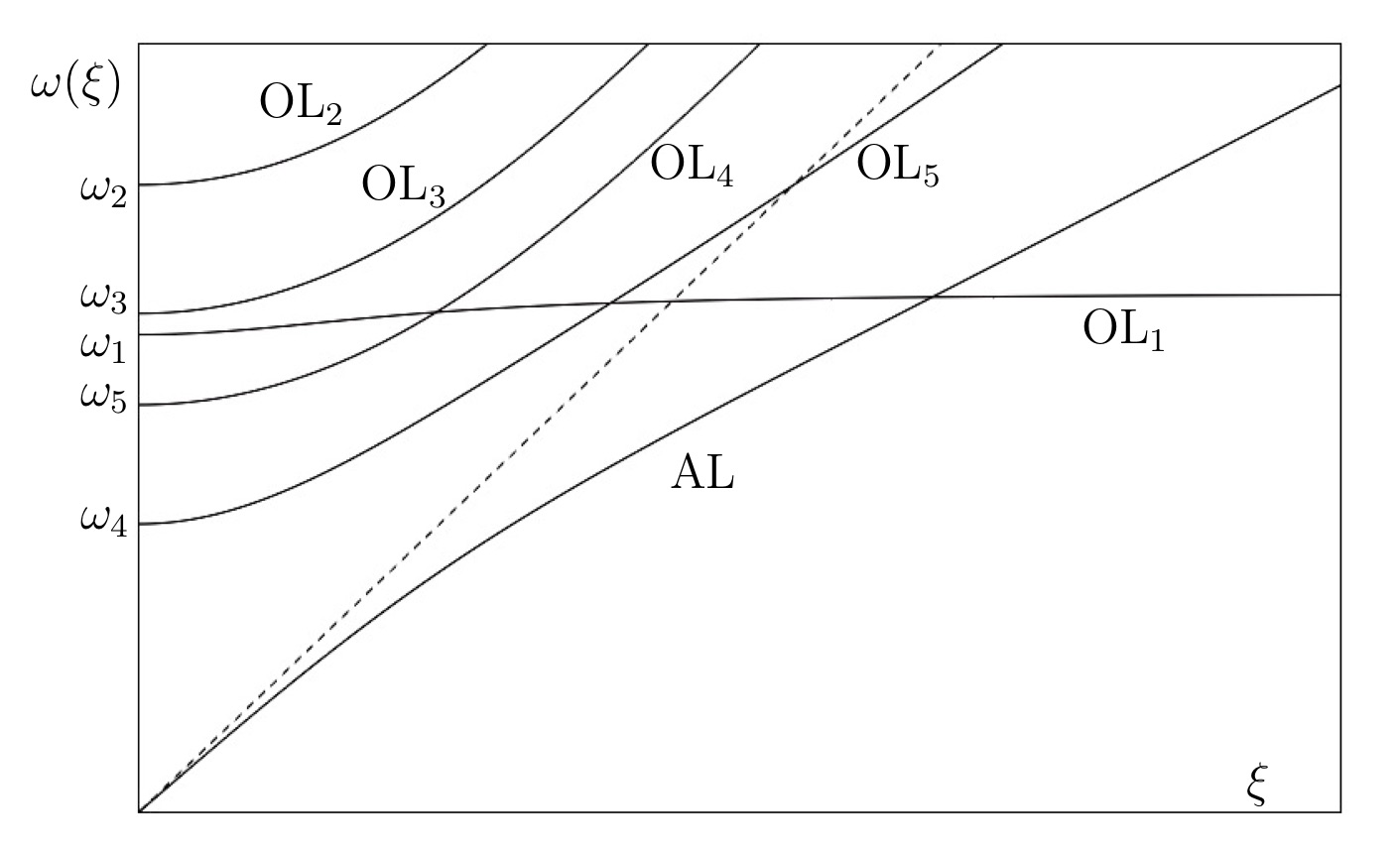}}}
\end{picture}
\end{center}
Fig. \ref{fig1}. \emph{Schematic of the dispersion relations in $\caM_{1}$. The dotted line represent the linearly elastic longitudinal wave. AL=Acoustic longitudinal wave; OL$_{1,2,4,5}$=optic longitudinal waves, modes $\Db_{1}\,,\Sbb_{1}\,,\Rb_{1}$; OL$_{3}$=optic longitudinal wave, modes $\Db_{2}\,,\Rb_{1}$. The frequencies on the $\omega$-axis are the cut-off frequencies (\ref{RAI1}), (\ref{RAI2}).}


In the subspace $\caM_{2}$, provided we define the normalized components:
\begin{align}\label{compoM2}
&a(\xi)=\xi^{2}A_{11}\,,\quad b(\xi)=\xi^{2}A_{12}\,,\nonumber\\
&c(\xi)=\xi^{2}\frac{\Pbm_{113}L_{c}^{2}}{L_{m}\sqrt{J_{11}}}+i\xi\frac{\Qbm_{113}}{\sqrt{J_{11}}}\,,\quad d(\xi)=\xi^{2}\frac{\Pbm_{131}L_{c}^{2}}{L_{m}\sqrt{J_{11}}}+i\xi\frac{\Qbm_{131}}{\sqrt{J_{11}}}\,,\nonumber\\
&e(\xi)=\xi^{2}\frac{\Ael_{44}L^{2}_{c}}{J_{33}}+\frac{\Bel_{44}}{\rho J_{33}}\,,\quad f(\xi)=\xi^{2}\frac{\Ael_{55}L^{2}_{c}}{J_{11}}+\frac{\Bel_{55}}{\rho J_{11}}\,,\nonumber\\
&g(\xi)=\xi^{2}\frac{\Ael_{47}L_{c}^{2}}{\sqrt{J_{11}J_{33}}}+\frac{\Bel_{47}}{\rho\sqrt{J_{11}J_{33}}}\,,\quad h(\xi)=\xi^{2}\frac{\Ael_{45}L_{c}^{2}}{\sqrt{J_{11}J_{33}}}+\frac{\Bel_{45}}{\rho\sqrt{J_{11}J_{33}}}\,,\\
&m(\xi)=\xi^{2}\frac{\Pbm_{123}L_{c}^{2}}{L_{m}\sqrt{J_{11}}}+i\xi\frac{\Qbm_{123}}{\sqrt{J_{11}}}\,,\quad n(\xi)=\xi^{2}\frac{\Pbm_{132}L_{c}^{2}}{L_{m}\sqrt{J_{11}}}+i\xi\frac{\Qbm_{132}}{\sqrt{J_{11}}}\,,\nonumber
\end{align}
we notice that the characteristic equation can be factorized into two cubic equations in $\omega^{2}$. Accordingly the eigenvalues can be obtained by using twice the Cardano's formula:
\begin{align}\label{eigenLSM2}
\omega_{7}^{2}&=\frac{1}{3}(A_{1}+2\sqrt{P_{1}}\cos\frac{\theta_{1}}{3})\,,\nonumber\\
\omega_{8}^{2}&=\frac{1}{3} (A_{1}+2\sqrt{P_{1}}\cos\frac{\theta_{1}-2\pi}{3})\,,\nonumber\\
\omega_{9}^{2}&=\frac{1}{3} (A_{1}+2\sqrt{P_{1}}\cos\frac{\theta_{1}+2\pi}{3})\,,\nonumber\\
\omega_{10}^{2}&=\frac{1}{3} (A_{2}+2\sqrt{P_{2}}\cos\frac{\theta_{2}}{3})\,,\\
\omega_{11}^{2}&=\frac{1}{3} (A_{2}+2\sqrt{P_{2}}\cos\frac{\theta_{2}-2\pi}{3})\,,\nonumber\\
\omega_{12}^{2}&=\frac{1}{3} (A_{2}+2\sqrt{P_{2}}\cos\frac{\theta_{2}+2\pi}{3})\,,\nonumber
\end{align}
where
\begin{align}\label{charaM2}
A_{1,2}&=a\pm b+e+f\,,\nonumber\\
B_{1,2}&=2 a^3\pm 3 a^2(b\mp e\mp f)\nonumber\\
&+3a(2b^2+3(cc^{*}+dd^{*})-(e -f)^2 -6(g^2+h^2) +3(mm^{*} +nn^{*})\nonumber\\
&\mp 2b(e+f)+4ef)\pm 2 (b^3\pm e^3\pm f^3)\\
&\pm 3b(cc^{*}+dd^{*}- e^2-f^2)+9(g^2+h^2)(\mp 2b+e+f)\nonumber\\
&\pm 9mm^{*}(b\pm e\mp 3f)\pm 9nn^{*}(b\pm f\mp 3e)-3e(b^2+f^2-3cc^{*}+6dd^{*})\nonumber\\
&-3f(b^2+6cc^{*}-3dd^{*}+e^2+4b e)+54(g(cd^{*}+mn^{*})+h(dm^{*}-cn^{*}))\nonumber\\
C_{1,2}&=\mp cc^{*}-(dd^{*}+g^2+h^2+mm^{*}+nn^{*})+(a\pm b)(e+f)+e f\,,\nonumber
\end{align}
and
\begin{equation}
\theta_{\alpha}=\cos^{-1}\frac{B_{\alpha}}{2\sqrt[3]{P_{\alpha}}}\,,\quad P_{\alpha}=A_{\alpha}^{2}-3C_{\alpha}\,,\quad \alpha=1,2\,.
\end{equation}
In this case all the frequencies depend on all the components of the block matrix $A$; as far as the corresponding eigenvectors are concerned, they are
\begin{align}\label{vectorsLSM2}
w_{k}&=\{\ab_{o}^{k}=\alpha_{k}\eb_{1}+\beta_{k}\eb_{2}\,;\Vb_{k}=(\gamma_{k}+\epsilon_{k})\hat{\Wb}_{4}+(\delta_{k}+\theta_{k})\hat{\Wb}_{5}\\
&+(\gamma_{k}-\epsilon_{k})\bar{\Wb}_{4}+(\delta_{k}-\theta_{k})\bar{\Wb}_{5}\}\,,\quad k=7\,,\ldots 12\,,\nonumber
\end{align}
the coefficients being given in detail into the Appendix, \S.\ref{eigencompM2}. The kinematics described by these eigenvectors is formed by a macroscopic transverse wave coupled with a combination of the modes $\Sbb_{2}$ and $\Rb_{2}$.

The behavior of the eigencouples (\ref{eigenLSM2}), (\ref{vectorsLSM2}) for $\xi\rightarrow 0$, since in such a case $a=b=c=d=m=n=0$, yields four optic waves (with multiplicity 2) with cut-off frequencies (\ref{RAI3}) and eigenvectors given by (\ref{EIGEN3}):
\begin{equation}
w_{7}(0)=w_{8}(0)=\{\bzero\,;\Cb_{6,7}\}\,,\quad w_{10}(0)=w_{11}(0)=\{\bzero\,;\Cb_{8,9}\}\,,
\end{equation}
and two acoustic waves with eigenvectors
\begin{equation}
w_{9}(0)=\{\eb_{1}-\eb_{2}\,;\bzero\}\,,\quad w_{12}(0)=\{\eb_{1}+\eb_{2}\,;\bzero\}\,.
\end{equation}

As we did for $\caM_{1}$ we give in Fig.\ref{figu2} a representative graph for the dispersion relations in $\caM_{2}$. 

\begin{center}
\begin{picture}(200,190) \label{figu2}
\put(-40,0){\scalebox{.22}{\includegraphics{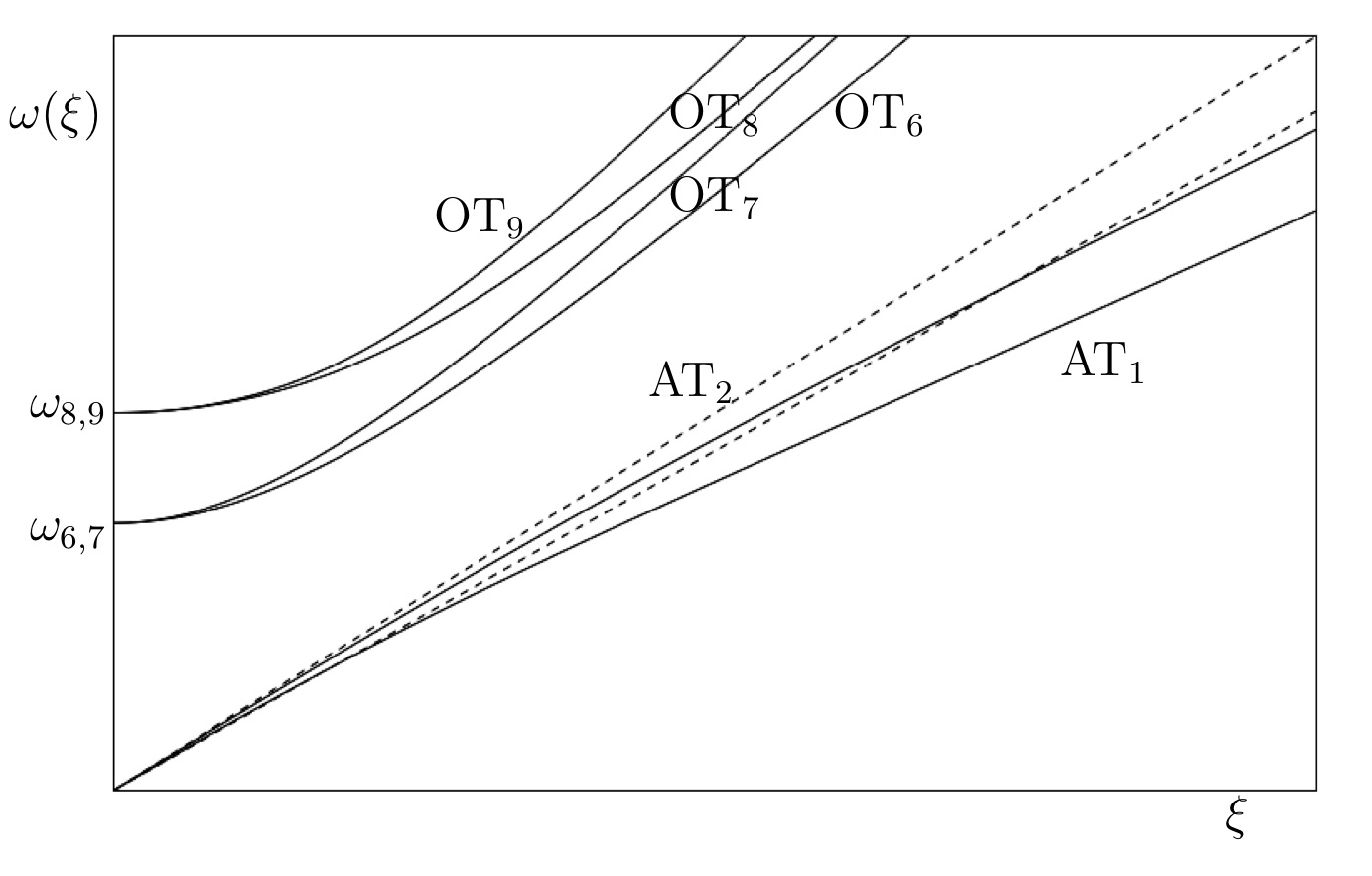}}}
\end{picture}
\end{center}
Fig. \ref{figu2}. \emph{Schematic of the dispersion relations in $\caM_{2}$. The dotted line represent the linearly elastic transverse waves. AT$_{1,2}$=Acoustic transverse wave; OT$_{7,8,10,11}$=optic transverse waves, modes $\Sbb_{2}\,,\Rb_{2}$. The frequencies on the $\omega$-axis are the cut-off frequencies (\ref{RAI3}).}


To summarize, for the Tetragonal classes $4\,,\bar{4}\,,4/m$, for a propagation direction $\mb$ along the tetragonal $c$-axis, we have three Acoustic and nine Optic waves which depend on 43 independent components of $A$ (three components of $\Ab$, thirteen for both $\Ael$ and $\Bel$, seven for both $\Pbm$ and $\Qbm$):
\begin{itemize}
\item[(AL)] One Acoustic wave associated with a macroscopic displacement along $c$ and a combination of the modes $\Db_{1}$, $\Sbb_{1}$ and $\Rb_{1}$, which for $\xi=0$ reduced to a macroscopic longitudinal wave;
\item[(AT$_{1,2}$)] Two Acoustic waves associated with a macroscopic displacement orthogonal to $c$, coupled a combination of the modes $\Sbb_{2}$ and $\Rb_{2}$. For $\xi=0$ these waves reduce to two macroscopic orthogonal transverse waves;
\item[(OL$_{1,2,4,5}$)] Four Optic waves associated with a macroscopic displacement along $c$ and with a microdistortion which combines the modes  $\Db_{1}$, $\Sbb_{1}$ and $\Rb_{1}$ which for $\xi=0$ reduces to the pure microdistortions (\ref{tensorsU1})$_{1,2,4,5}$;
\item[(OL$_{3}$)] One Optic wave associated with a macroscopic displacement along $c$ and with a combination of the modes $\Db_{2}$ and $\Rb_{1}$;
\item[(OT$_{7,8,10,11}$)] Four Optic waves associated with a macroscopic displacement orthogonal to $c$ coupled with a combination of the $\Sbb_{2}$ and $\Rb_{2}$ modes, which for $\xi=0$ reduce to the shear microdistortion (\ref{EIGEN3})$_{2}$ and to the rigid rotation (\ref{EIGEN3})$_{1}$.
\end{itemize}

\subparagraph{Classes $4mm\,,422\,,\bar{4}2m$ and $4/mm$.} When  we deal with the high-symmetric tetragonal classes, for a propagation direction along the tetragonal $c$-axis with $\mb=\eb_{3}$, the matrix $A$ has the following non-null components, the independent ones given by relations (\ref{acoustic1HS}), (\ref{QBMHS}), (\ref{PBMHS}) and (\ref{AELHS}) of the Appendix, evaluted for $m_{1}=m_{2}=0$ and $m_{3}=1$.
\begin{equation}
\renewcommand{\arraystretch}{1.5} 
A\equiv{\tiny 
\left[\begin{array}{@{}ccc|ccccccccc@{}}
\bullet&\bullet&\cdot&\cdot&\cdot&\cdot&\cdot&\bullet&\cdot&\cdot&\bullet&\cdot \\
\bullet&\bullet&\cdot&\cdot&\cdot&\cdot&\bullet&\cdot&\cdot&\bullet&\cdot&\cdot \\
\cdot&\cdot&\bullet&\bullet&\bullet&\bullet&\cdot&\cdot&\cdot&\cdot&\cdot&\cdot \\ \hline
\cdot&\cdot&\bullet&\bullet&\bullet&\bullet&\cdot&\cdot&\cdot&\cdot&\cdot&\cdot \\
\cdot&\cdot&\bullet&\bullet&\bullet&\bullet&\cdot&\cdot&\cdot&\cdot&\cdot&\cdot \\
\cdot&\cdot&\bullet&\bullet&\bullet&\bullet&\cdot&\cdot&\cdot&\cdot&\cdot&\cdot \\
\cdot&\bullet&\cdot&\cdot&\cdot&\cdot&\bullet&\cdot&\cdot&\bullet&\cdot&\cdot \\
\bullet&\cdot&\cdot&\cdot&\cdot&\cdot&\cdot&\bullet&\cdot&\cdot&\bullet&\cdot \\
\cdot&\cdot&\cdot&\cdot&\cdot&\cdot&\cdot&\cdot&\bullet&\cdot&\cdot&\bullet \\
\cdot&\bullet&\cdot&\cdot&\cdot&\cdot&\bullet&\cdot&\cdot&\bullet&\cdot&\cdot \\
\bullet&\cdot&\cdot&\cdot&\cdot&\cdot&\cdot&\bullet&\cdot&\cdot&\bullet&\cdot \\
\cdot&\cdot&\cdot&\cdot&\cdot&\cdot&\cdot&\cdot&\bullet&\cdot&\cdot&\bullet
\end{array}\right]\,:}
\end{equation}
accordingly $A$ is reduced by the three subspaces:
\begin{equation}
\caN_{1}\equiv\spn\{\eb_{3}\}\oplus\caZ_{1}\,,\quad\caN_{2}\equiv\caZ_{2}\,,\quad\caN_{3}\equiv\spn\{\eb_{1}\,,\eb_{2}\}\oplus\caZ_{3}\,;
\end{equation}

We begin our analysis with the subspace $\caN_{1}$ and notice that the eigenvalues can be obtained from those in $\caM_{1}$ when we set $d=e=0$ and $m=n=p=q=0$ into (\ref{compoM1}). Then the characteristic equation admits the root $\omega^{2}=f-h$ and thus, by the means of the Cardano's formulae (\emph{vid. e,g.\/} \cite{CRC}) we obtain the four eigenvalues (for once we write one of them in terms components and characteristic length):
\begin{align}\label{eigen4HS1}
\omega_{1}^{2}(\xi)&=\xi^{2}(\frac{\Ael_{11}}{J_{11}}+\frac{\Ael_{12}}{\sqrt{J_{11}J_{33}}})L_{c}^{2}+\frac{\Bel_{11}}{\rho J_{11}}+\frac{\Bel_{12}}{\rho\sqrt{J_{11}J_{33}}}\,,\nonumber\\
\omega_{2}^{2}(\xi)&=\frac{1}{3}(A+2\sqrt{P}\cos\frac{\theta}{3})\,,\\
\omega_{3}^{2}(\xi)&=\frac{1}{3}(A+2\sqrt{P}\cos\frac{\theta+2\pi}{3})\,,\nonumber\\
\omega_{4}^{2}(\xi)&=\frac{1}{3}(A+2\sqrt{P}\cos\frac{\theta-2\pi}{3})\,,\nonumber
\end{align}
where
\begin{equation}
P=A^{2}-3B\,,\quad Q=2A^{3}-9AB-27C\,,\quad\theta=\cos^{-1}\frac{Q}{2\sqrt[3]{P}}\,,
\end{equation}
and $A, B$ and $C$ are obtained by setting $d=e=m=n=p=q=0$ into (\ref{charaM1}).
The corresponding (non-normalized) eigenvectors are
\begin{align}\label{vector4HS1}
w_{1}(\xi)&=\{\bzero\,;\Vb_{1}=\Wb_{1}-\Wb_{2}\}\,,\\
w_{2}(\xi)&=\{\alpha_{2}\eb_{3}\,;\Vb_{2}=\beta_{2}\Wb_{1}+\gamma_{2}\Wb_{2}+\delta_{2}\Wb_{3}\}\,,\nonumber\\
w_{3}(\xi)&=\{\alpha_{2}\eb_{3}\,;\Vb_{3}=\gamma_{2}\Wb_{1}+\beta_{2}\Wb_{2}+\delta_{2}\Wb_{3}\}\,,\nonumber\\
w_{4}(\xi)&=\{\alpha_{4}\eb_{3}\,;\Vb_{4}=\beta_{4}(\Ib-\Wb_{3})+\delta_{4}\Wb_{3}\}\,,\nonumber
\end{align}
where the complex components $\alpha_{i}\,,\beta_{j},\gamma_{k}$ and $\delta_{k}$ are given in \S.\ref{eigencompN1} of the Appendix.

Looking at (\ref{eigen4HS1}) and (\ref{vector4HS1}) we notice first of all that the frequencies $\omega_{2,3,4}(\xi)$ are associated with the mode $\Db_{2}$ coupled with a macroscopic longitudinal wave; the frequency $\omega_{1}(\xi)$ is instead associated uniquely to the traceless real microdistortion $\Db_{3}$.

In the limit $\xi\rightarrow 0$, since $a\,,b$ and $c$ vanish, then the three frequencies $\omega_{1}(\xi)$ and $\omega_{2,3}(\xi)$ reduce to (\ref{cutoffHS}) with eigenvectors
\begin{equation}\label{vector3HS1}
w_{1}(0)=\{\bzero\,;\Cb_{1}\}\,,\quad w_{2}(0)=\{\bzero\,;\Cb_{2}\}\,,\quad w_{3}=\{\bzero\,;\Cb_{3}\}\,,
\end{equation}
with the three microdistortions given by (\ref{treeigenHS}): these are optic frequencies, the values (\ref{cutoffHS}) being the associated cut-off values; the frequency $\omega_{4}(\xi)$ vanishes instead for $\xi\rightarrow 0$ with 
\begin{equation}
w_{4}(0)=\{\eb_{3}\,;\bzero\}\,,
\end{equation}
which represents a purely macroscopic acoustic longitudinal wave.

The solutions in the subspace $\caN_{2}$ are are the same as those in $\caZ_{2}$ with frequencies
\begin{align}\label{microN2}
\omega^{2}_{5}(\xi)&=\xi^{2}L^{2}_{c}\frac{\Ael_{66}-\Ael_{69}}{J_{11}}+\frac{\Bel_{66}-\Bel_{69}}{\rho J_{11}}\,,\\
\omega^{2}_{6}(\xi)&=\xi^{2}L^{2}_{c}\frac{\Ael_{66}+\Ael_{69}}{J_{11}}+\frac{\Bel_{66}+\Bel_{69}}{\rho J_{11}}\,,\nonumber
\end{align}
and accordingly describe optic waves with purely microdistortion amplitudes and whose cut-off frequencies are given by (\ref{microZ2}); the corresponding real eigenvectors are:
\begin{equation}
w_{5}(0)=\{\bzero\,;\Vb_{5}=\Cb_{4}\}\,,\quad w_{6}(0)=\{\bzero\,;\Vb_{6}=\Cb_{5}\}\,,
\end{equation}
with the microdistortions given by (\ref{microZ2}).

We finish with the subspace $\caN_{3}$ where the solutions are obtained by setting $m=n=h=0$ into (\ref{compoM2}) which yield the six eigenvalues
\begin{align}\label{eigenHSN3}
\omega_{7}^{2}&=\frac{1}{3}(A_{1}+2\sqrt{P_{1}}\cos\frac{\theta_{1}}{3})\,,\nonumber\\
\omega_{8}^{2}&=\frac{1}{3} (A_{1}+2\sqrt{P_{1}}\cos\frac{\theta_{1}+2\pi}{3})\,,\nonumber\\
\omega_{9}^{2}&=\frac{1}{3} (A_{1}+2\sqrt{P_{1}}\cos\frac{\theta_{1}-2\pi}{3})\,,\nonumber\\
\omega_{10}^{2}&=\frac{1}{3} (A_{2}+2\sqrt{P_{2}}\cos\frac{\theta_{2}}{3})\,,\\
\omega_{11}^{2}&=\frac{1}{3} (A_{2}+2\sqrt{P_{2}}\cos\frac{\theta_{2}+2\pi}{3})\,,\nonumber\\
\omega_{12}^{2}&=\frac{1}{3} (A_{2}+2\sqrt{P_{2}}\cos\frac{\theta_{2}-2\pi}{3})\,,\nonumber
\end{align}
where $A_{\alpha}$ and $P_{\alpha}$ for $\alpha=1,2$ are obtained when we put $m=n=h=0$ into (\ref{charaM2}). The eigenvector have the same representation  (\ref{vectorsLSM2}) with the component once again obtained by putting $m=n=h=0$ into those given in \S.\ref{eigencompM2}:
\begin{align}\label{vectorsHSN3}
w_{7}=\{\ab_{o}^{7}&=\alpha_{7}\eb_{1}+\beta_{7}\eb_{2}\,;\nonumber\\
&\Vb_{7}=(\gamma_{7}+\epsilon_{7})\hat{\Wb}_{4}+(\delta_{7}+\theta_{7})\hat{\Wb}_{5}+(\gamma_{7}-\epsilon_{7})\bar{\Wb}_{4}+(\delta_{7}-\theta_{7})\bar{\Wb}_{5}\}\,,\nonumber\\
w_{8}=\{\ab_{o}^{8}&=\alpha_{8}\eb_{1}+\beta_{8}\eb_{2}\,;\nonumber\\
&\Vb_{8}=(\gamma_{8}+\epsilon_{8})\hat{\Wb}_{4}+(\delta_{8}+\theta_{8})\hat{\Wb}_{5}+(\gamma_{8}-\epsilon_{8})\bar{\Wb}_{4}+(\delta_{8}-\theta_{8})\bar{\Wb}_{5}\}\,,\nonumber\\
w_{9}=\{\ab_{o}^{9}&=\beta_{8}\eb_{1}+\alpha_{8}\eb_{2}\,;\nonumber\\
&\Vb_{9}=(\delta_{8}+\theta_{8})\hat{\Wb}_{4}+(\gamma_{8}+\epsilon_{8})\hat{\Wb}_{5}-(\delta_{8}-\theta_{8})\bar{\Wb}_{4}-(\gamma_{8}-\epsilon_{8})\bar{\Wb}_{5}\}\,,\nonumber\\
w_{10}=\{\ab_{o}^{10}&=\beta_{7}\eb_{1}+\alpha_{7}\eb_{2}\,;\\
&\Vb_{10}=(\delta_{7}+\theta_{7})\hat{\Wb}_{4}+(\gamma_{7}+\epsilon_{7})\hat{\Wb}_{5}-(\delta_{7}-\theta_{7})\bar{\Wb}_{4}-(\gamma_{7}-\epsilon_{7})\bar{\Wb}_{5}\}\,,\nonumber\\
w_{11}=\{\ab_{o}^{11}&=\alpha_{11}\eb_{1}+\beta_{11}\eb_{2}\,;\nonumber\\
&\Vb_{11}=(\gamma_{11}+\epsilon_{11})\hat{\Wb}_{4}+(\delta_{11}+\theta_{11})\hat{\Wb}_{5}+(\gamma_{11}-\epsilon_{11})\bar{\Wb}_{4}+(\delta_{11}-\theta_{11})\bar{\Wb}_{5}\}\,,\nonumber\\
w_{12}=\{\ab_{o}^{12}&=\beta_{11}\eb_{1}+\alpha_{11}\eb_{2}\,;\nonumber\\
&\Vb_{12}=(\delta_{11}+\theta_{11})\hat{\Wb}_{4}+(\gamma_{11}+\epsilon_{11})\hat{\Wb}_{5}-(\delta_{11}-\theta_{11})\bar{\Wb}_{4}-(\gamma_{11}-\epsilon_{11})\bar{\Wb}_{5}\}\,,\nonumber
\end{align}
the coefficients being given in detail into the Appendix, \S.\ref{eigencompN3}. These eigenvectors describe a kinematics formed by a macroscopic transverse wave coupled a combination of the modes $\Sbb_{2}$ and $\Rb_{2}$.

The eigenvalues (\ref{eigenHSN3}) and their associated eigenvectors becomes, since for $\xi\rightarrow 0$ it is $a=b=c=d=0$:
\begin{equation}
w_{7,8,9,10}(0)=\{\bzero\,;\Cb_{6,7,8,9}\}\,,
\end{equation}
which correspond to four optic waves (with multiplicity 2) with cut-off frequencies (\ref{RAI3}) with $\Bel_{45}=0$ and eigenvectors given by (\ref{EIGEN3}) and two acoustic waves with eigenvectors:
\begin{equation}
w_{11}(0)=\{\eb_{1}-\eb_{2}\,;\bzero\}\,,\quad w_{12}(0)=\{\eb_{1}+\eb_{2}\,;\bzero\}\,.
\end{equation}

To summarize, in the high-symmetric tetragonal classes $4mm\,,422\,,\bar{4}2m$ and $4/mm$, for the propagation direction along the tetragonal $c$-axis we have the three Acoustic and nine Optic waves, which depend on the 25 independent components of $A$ (three of $\Ab$, seven each of $\Ael$ and $\Bel$ and four each of $\Pbm$ and $\Qbm$):
\begin{itemize}
\item[(AL)] One Acoustic wave associated with a macroscopic displacement along $c$ and the mode $\Db_{2}$ which for $\xi=0$ yields a macroscopic longitudinal wave;
\item[(AT$_{1,2}$)] Two Acoustic waves associated with a macroscopic displacemente orthogonal to $c$, coupled with the $\Sbb_{2}$ and $\Db_{2}$ modes. For $\xi=0$ they reduce to two macroscopic orthogonal transverse waves;
\item[(OL$_{1,2,3}$)] Two Optic waves associated to a macrodisplacement along $\eb_{3}$ coupled with the mode $\Db_{2}$, which for $\xi=0$ reduces to the pure microdistortions (\ref{treeigenHS})$_{2,3}$;
\item[(OD)] One Optic wave associated to the traceless distortion $\Db_{3}$ which is independent on the macroscopic displacement and which reduces, for $\xi\rightarrow 0$, to (\ref{treeigenHS})$_{1}$;
\item[(OS)] One Optic wave associated to the the mode $\Sbb_{1}$ which is independent on the macroscopic displacement and reduces for $\xi\rightarrow 0$ to (\ref{microZ2})$_{2}$;
\item[(OR)] One Optic wave associated to the mode $\Rb_{1}$, which is independent on the macroscopic displacement and reduces for $\xi\rightarrow 0$ to (\ref{microZ2})$_{1}$;
\item[(OT$_{1,2,3,4}$)] Four Optic waves associated with a macroscopic displacement orthogonal to $c$ coupled with a combination of the modes $\Sbb_{2}$ and $\Rb_{2}$, which for $\xi=0$ reduce to the shear microdistortions (\ref{EIGEN3})$_{2}$ and to the rigid rotations (\ref{EIGEN3})$_{1}$ with $\Bel_{45}=0$.
\end{itemize}

As far as the eigenvalues ordering is concerned, we can only say that:
\begin{align}
&\omega_{5}<\omega_{6}\,,\nonumber\\
&\omega_{3}<\omega_{2}<\omega_{4}\,,(\mbox{ or }\omega_{3}>\omega_{2}>\omega_{4})\,,\\
&\omega_{8}<\omega_{7}<\omega_{9}\,,(\mbox{ or }\omega_{8}>\omega_{7}>\omega_{9})\,,\nonumber\\
&\omega_{11}<\omega_{10}<\omega_{12}\,,(\mbox{ or }\omega_{11}>\omega_{10}>\omega_{12})\,.\nonumber
\end{align}

\subsubsection{Propagation orthogonal to the tetragonal $c$-axis}

When the propagation direction is orthogonal to the tetragonal $c$-axis, we may assume into the constitutive relations (\ref{acoustic1LS}), (\ref{acoustic1HS}), (\ref{QBMLS1}), (\ref{QbmHS}), (\ref{PBMLS}), (\ref{PBMHS}), (\ref{AELLS}) and (\ref{AELHS}) that $m_{1}=\cos\theta$, $m_{2}=\sin\theta$ and $m_{3}=0$. An inspection of these relations however shows that the tensor $\Ab(\theta)$ has six independent components, the tensor $\Ael(\theta)$, twenty-six independent components and that the tensor $\xi^{2}\Pbm+i\xi\Qbm$ has twenty-seven independent components for a generic value of $\theta$. It is indeed this last block of $A$ which makes the problem a fully-coupled one: further, for all classes but two, there is no great difference between the classes and the propagation condition is coupled as in the case of any cristal of the Triclinic group. Hence, the best we can say is that there will be three acoustic and nine optic waves with cut-off frequencies given by the solution of the microvibrations problem of \S.\ref{sectionmicrovibe} and that, despite the fact that we can still write the explicit expressions for the eigencouples, these relations will be nearly useless given the total kinematical coupling.  Therefore we do not find useful to pursue the matter any further.

However, for the two centrosymmetric classes $4/m$ and $4/mmm$ the tensor $\Pbm$ vanishes altogether since it depends on the components of an odd-tensor and for the class $4/mm$ the matrix $A$ has only to forty-one independent components:
\begin{equation}
\renewcommand{\arraystretch}{1.5} 
M\equiv{\tiny 
\left[\begin{array}{@{}ccc|ccccccccc@{}}
\bullet&\bullet&\cdot&\bullet&\bullet&\bullet&\cdot&\cdot&\bullet&\cdot&\cdot&\bullet\\
\bullet&\bullet&\cdot&\bullet&\bullet&\bullet&\cdot&\cdot&\bullet&\cdot&\cdot&\bullet\\
\cdot&\cdot&\bullet&\cdot&\cdot&\cdot&\bullet&\bullet&\cdot&\bullet&\bullet&\cdot \\ \hline
\bullet&\bullet&\cdot&\bullet&\bullet&\bullet&\cdot&\cdot&\bullet&\cdot&\cdot&\bullet \\
\bullet&\bullet&\cdot&\bullet&\bullet&\bullet&\cdot&\cdot&\bullet&\cdot&\cdot&\bullet\\
\bullet&\bullet&\cdot&\bullet&\bullet&\bullet&\cdot&\cdot&\bullet&\cdot&\cdot&\bullet \\
\cdot&\cdot&\bullet&\cdot&\cdot&\cdot&\bullet&\bullet&\cdot&\bullet&\cdot&\cdot \\
\cdot&\cdot&\bullet&\cdot&\cdot&\cdot&\bullet&\bullet&\cdot&\cdot&\bullet&\cdot \\
\bullet&\bullet&\cdot&\bullet&\bullet&\bullet&\cdot&\cdot&\bullet&\cdot&\cdot&\bullet \\
\cdot&\cdot&\bullet&\cdot&\cdot&\cdot&\bullet&\cdot&\cdot&\bullet&\bullet&\cdot \\
\cdot&\cdot&\bullet&\cdot&\cdot&\cdot&\cdot&\bullet&\cdot&\bullet&\bullet&\cdot \\
\bullet&\bullet&\cdot&\bullet&\bullet&\bullet&\cdot&\cdot&\bullet&\cdot&\cdot&\bullet
\end{array}\right]\,,}
\end{equation}
and more important it is reduced by the pairs
\begin{equation}
\caL_{1}\equiv\spn\{\eb_{3}\}\oplus\caU_{2}\,,\quad\caL_{2}\equiv\{\eb_{1}\,,\eb_{2}\}\oplus\caU_{1}\,.
\end{equation}

In the subspace $\caL_{1}$ the fifth-order characteristc equation in $\omega^{2}$ depends on 15 independent components (which may reduce to 13 when either $\theta=0$ or $\theta=\pi/2$): therefore, rather than give the explicit solutions (which shall depend on relations more cumbersome than the ones we already obtained before), we limit in this case to a qualitative analysis.

The kinematics described by the five eigenvectors is composed by a macroscopic displacement directed as $\eb_{3}$ and a combination of the modes $\Sbb_{2}$ and $\Rb_{2}$. In the limit $\xi\rightarrow 0$ we have a transverse acoustic wave and the microdeformations (\ref{EIGEN3}).

Likewise, in the subspace $\caL_{2}$ we have seven eigenvectors representing a macroscopic displacement orthogonal to $\eb_{3}$ coupled with a combination of the $\Db_{k}$, $\Sbb_{1}$ and $\Rb_{1}$ modes, which in the limit $\xi\rightarrow 0$ reduce to two acoustic waves (nor longitudinal, neither transverse) and to the microdeformations (\ref{tensorsU1}) (or (\ref{treeigenHS}) and (\ref{microZ2}) for the high-symmetric classes). In fact, in the propagation orthogonal to the $c$-axis the role of the subspaces $\caU_{1}$ and $\caU_{2}$ associated with the microvibrations is exchanged with respect of the case $\mb=\eb_{3}$.

\section{Conclusions}\label{conclusions}

For a bulk three-dimensional crystalline body we modeled the interaction between macroscopic and lattice waves by means of a continuum with affine-structure, or micromorphic. The propagation condition we obtained is described by a twelve-dimensional hermitian acoustic matrix $A$, the eigenvector being elements of $\caV\oplus\Lin$ which represent three macroscopic displacements and nine lattice microdistortions. We showed that for such a propagation condition admits three acoustic and nine optic waves for any crystalline symmetry; moreover it depends on three length scale representing the crystal lattice inhomogeneity, the size of the bulk crystal and the effect of lattice inertia. In terms of these length and their ratios we obtained two limit problems which recover the \emph{long wavelength approximation} and the \emph{microvibration} problems, well-known and studied into details for isotropic materials.

Then we investigated in full detail the propagation condition for crystals of the Tetragonal point group: the the matrix $A$ is reduced by two or three subspaces of $\caV\oplus\Lin$, depending on the classes, whereas in the microvibrations limit case the propagation condition is determined by $\Bel$  which is in turn reduced by two or three subspaces of $\Lin$. The relations between these subspaces and the Tetragonal classes is represented schematically in Fig.\ref{fig3}.
\begin{center}
\begin{picture}(200,180)\label{fig3}
\put(25,120){$\caU_{1}$}
\put(30,123){\oval(30.0,15.0)}
\put(25,30){$\caU_{2}$}
\put(30,33){\oval(30.0,15.0)}
\put(75,120){$\caM_{1}$}
\put(75,30){$\caM_{2}$}
\put(125,30){$\caN_{3}$}
\put(125,90){$\caN_{2}$}
\put(125,150){$\caN_{1}$}
\put(175,90){$\caZ_{2}$}
\put(180,93){\oval(30.0,15.0)}
\put(175,150){$\caZ_{1}$}
\put(180,153){\oval(30.0,15.0)}
\put(175,30){$\caZ_{3}$}
\put(180,33){\oval(30.0,15.0)}
\dottedline{1}(110,0)(110,180)
\put(145,150){$\longrightarrow$}
\put(145,90){$\longrightarrow$}
\put(145,30){$\longrightarrow$}
\put(45,120){$\longleftarrow$}
\put(45,30){$\longleftarrow$}
\put(40,60){$4\,,\bar{4}\,,4/m$}
\put(120,60){$4mm, 422, \bar{4}2m, 4/mm$}
\dashline{2}(95,124)(122,151)
\dashline{2}(95,124)(122,91)
\dashline{2}(97,33)(120,33)
\end{picture}
\end{center}
Fig. \ref{fig3}. \emph{The subspaces $\caM_{j}\subset\caV\oplus\Lin$ and $\caN_{k}\subset\caV\oplus\Lin$ reduced by $A$ and their relation with the subspaces $\caU_{j}\subset\Lin$ and $\caZ_{k}\subset\Lin$ reduced by $\Bel$.}

The results we obtained for the two limit problems are represented in Table \ref{T1}: it is remarkable that in the long-wavelength approximation the macroscopic displacements are real whereas the lattice microdistortions are purely immaginary; conversely, in the microvibration problem the matrix $A$ reduces to a real one and the lattice microdistortions are real. In both cases we find that the lattice modes can be described by the means of three kind of dilatations, two shear deformations and two rigid rotations.

\begin{table}[htbp]
\begin{center}
\caption{Modes in the Long-wavelength (LW) and Microvibration (M) approximations.}\label{T1}

\begin{tabular}{|c|c|c|c|c|c|c|c|c|c|c|c|}
\hline
&$\mb$&$\eb_{1}$&$\eb_{2}$&$\eb_{3}$&$\Db_{1}$&$\Db_{2}$&$\Db_{3}$&$\Sbb_{1}$&$\Sbb_{2}$&$\Rb_{1}$&$\Rb_{2}$\\
\hline
&&$\cdot$&$\cdot$&$\Re_{L}$&$\cdot$&$\Im$&$\cdot$&$\cdot$&$\cdot$&$\Im\dag$&$\cdot$\\
LW&$\parallel\eb_{3}$&$\Re_{T}$&$\cdot$&$\cdot$&$\cdot$&$\cdot$&$\cdot$&$\cdot$&$\Im$&$\cdot$&$\Im$\\
& & $\cdot$ & $\Re_{T}$ & $\cdot$ & $\cdot$ & $\cdot$ & $\cdot$ & $\cdot$ & $\Im$ & $\cdot$ & $\Im$  \\
\hline
& &$\cdot$&$\cdot$&$\Re_{T}$&$\cdot$&$\cdot$&$\cdot$&$\cdot$&$\Im$&$\cdot$&$\Im$\\
LW&$\perp\eb_{3}$&$\Re$&$\cdot$&$\cdot$&$[\Im]$&$(\Im)$&$\cdot$&$\Im$&$\cdot$&$\Im$&$\cdot$\\
&&$\cdot$&$\Re$&$\cdot$&$[\Im]$&$(\Im)$&$\cdot$&$\Im$&$\cdot$&$\Im$&$\cdot$\\
\hline
\hline
&$(2)$&$\cdot$&$\cdot$&$\cdot$&$\Re$&$\cdot$&$\cdot$&$\Re\circ$&$\cdot$&$\Re\bullet$&$\cdot$\\
&$(2)$&$\cdot$&$\cdot$&$\cdot$&$\Re$&$\cdot$&$\cdot$&$\Re\circ$&$\cdot$&$\Re\bullet$&$\cdot$\\
M&&$\cdot$&$\cdot$&$\cdot$&$\cdot$&$\Re$&$\cdot$&$\cdot$&$\cdot$&$\Re$&$\cdot$\\
LS&$(2),\flat$&$\cdot$&$\cdot$&$\cdot$&$\cdot$&$\cdot$&$\cdot$&$\cdot$&$\Re$&$\cdot$&$\Re$\\
&$(2),\flat$&$\cdot$&$\cdot$&$\cdot$&$\cdot$&$\cdot$&$\cdot$&$\cdot$&$\Re$&$\cdot$&$\Re$\\
\hline
&&$\cdot$&$\cdot$&$\cdot$&$\cdot$&$\cdot$&$\Re$&$\cdot$&$\cdot$&$\cdot$&$\cdot$\\
&$(2)$&$\cdot$&$\cdot$&$\cdot$&$\Re$&$\cdot$&$\cdot$&$\cdot$&$\cdot$&$\cdot$&$\cdot$\\
M&&$\cdot$&$\cdot$&$\cdot$&$\cdot$&$\cdot$&$\cdot$&$\Re$&$\cdot$&$\cdot$&$\cdot$\\
HS&&$\cdot$&$\cdot$&$\cdot$&$\cdot$&$\cdot$&$\cdot$&$\cdot$&$\cdot$&$\Re$&$\cdot$\\
&$(2), \flat$&$\cdot$&$\cdot$&$\cdot$&$\cdot$&$\cdot$&$\cdot$&$\cdot$&$\Re$&$\cdot$&$\Re$\\
&$(2), \flat$&$\cdot$&$\cdot$&$\cdot$&$\cdot$&$\cdot$&$\cdot$&$\cdot$&$\Re$&$\cdot$&$\Re$\\
\hline
\end{tabular}
\end{center}
{\scriptsize
HS=classes $4mmm$, $422$, $4/mm$, $\bar{4}2m$, LS=classes $4$, $\bar{4}$, $4/m$; $\Re$=Real component, $\Im$=Immaginary component; $L$=longitudinal wave, $T$=transverse wave; $\bullet$=equal values components; $\circ$=equal value, opposite sign components; $(\sharp)$=number of independent eigentensors with same structure; $\flat$=multiple eigenvalues; $\dag$ vanishes for HS classes; $[\Im]=$ only LS classes; $(\Im)=$ only HS classes.
}
\end{table}%

When we deal with the complete propagation condition in the case of waves propagating along the tetragonal $c$-axis, in all cases but three the modes are complex and fully-couples macroscopic displacements with lattice microdistortions: only for the highly-symmetric classes we have three fully optic modes with real microdistortions without macroscopic displacements. These results are represented in Table \ref{T2}:

\begin{table}[htbp]

\caption{Modes, wave propagation along the direction $c$.}\label{T2}

\begin{center}
\begin{tabular}{|c|c|c|c|c|c|c|c|c|c|c|c|c|}
\hline
&  & $\eb_{1}$ &  $\eb_{2}$ &  $\eb_{3}$ & $\Db_{1}$ & $\Db_{2}$ & $\Db_{3}$ & $\Sbb_{1}$ & $\Sbb_{2}$ & $\Rb_{1}$ & $\Rb_{2}$ &   $\xi\rightarrow 0$  \\
\hline
&$(2)$ & $\cdot$ & $\cdot$ & $\bullet\fraC_{L}$ & $\fraC$ & $\cdot$ & $\cdot$ & $\fraC\circ$ & $\cdot$ & $\fraC\bullet$ & $\cdot$ &   OL$_{1,4}$ \\
 & $(2)$ & $\cdot$ & $\cdot$ & $\bullet\fraC_{L}$ & $\fraC$ & $\cdot$ & $\cdot$ & $\fraC\circ$ & $\cdot$ & $\fraC\bullet$ & $\cdot$ &   OL$_{2,5}$ \\
LS &  & $\cdot$ & $\cdot$ & $\fraC_{L}$ & $\cdot$ & $\fraC$ & $\cdot$ & $\cdot$ &$\cdot$ & $\fraC$ & $\cdot$ &   OL$_{3}$  \\
&  & $\cdot$ & $\cdot$ & $\fraC_{L}$ & $\cdot$ & $\fraC$ & $\cdot$ & $\cdot$ & $\cdot$ & $\fraC$ & $\cdot$ &   AL  \\
&$(4)$ & $\fraC_{T}$ & $\fraC_{T}$ & $\cdot$ & $\cdot$ & $\cdot$ & $\cdot$ & $\cdot$ & $\fraC$ & $\cdot$ & $\fraC$ &   OT$_{6-9}$ \\
& $(2)$ & $\fraC_{T}$ & $\fraC_{T}$& $\cdot$ & $\cdot$ & $\cdot$ & $\cdot$ & $\cdot$ & $\fraC$ & $\cdot$ & $\fraC$ &   AT$_{1-2}$ \\
\hline
\hline
&  & $\cdot$ & $\cdot$ & $\cdot$ & $\cdot$ & $\cdot$ & $\Re$ & $\cdot$ & $\cdot$ & $\cdot$ & $\cdot$ &  OL$_{1}$   \\
&  & $\cdot$ & $\cdot$ & $\bullet\fraC_{L}$ & $\fraC$ & $\cdot$ & $\cdot$ & $\cdot$ & $\cdot$ & $\cdot$ & $\cdot$ &  OL$_{2}$   \\
&  & $\cdot$ & $\cdot$ & $\bullet\fraC_{L}$ & $\fraC$ & $\cdot$ & $\cdot$ & $\cdot$ & $\cdot$ & $\cdot$ & $\cdot$ &  OL$_{3}$   \\
&  & $\cdot$ & $\cdot$ & $\fraC_{L}$ & $\cdot$ & $\fraC$ & $\cdot$ & $\cdot$ & $\cdot$ & $\cdot$ & $\cdot$ &  AL   \\
HS& & $\cdot$ & $\cdot$ & $\cdot$ & $\cdot$ & $\cdot$ & $\cdot$ & $\Re$ & $\cdot$ & $\cdot$ & $\cdot$ &  OS   \\
& & $\cdot$ & $\cdot$ & $\cdot$ & $\cdot$ & $\cdot$ & $\cdot$ & $\cdot$ & $\cdot$ & $\Re$ & $\cdot$ &  OR   \\
& $(2)$ & $\fraC_{T}$ & $\fraC_{T}$ & $\cdot$ & $\cdot$ & $\cdot$ & $\cdot$ & $\cdot$ & $\fraC$ & $\cdot$ & $\fraC$ &  OT$_{6,9}$  \\
& $(2)$ & $\fraC_{T}$ & $\fraC_{T}$ & $\cdot$ & $\cdot$ & $\cdot$ & $\cdot$ & $\cdot$ & $\fraC$ & $\cdot$ & $\fraC$ &  OT$_{7,8}$  \\
& $(2)$ & $\fraC_{T}$ & $\fraC_{T}$ & $\cdot$ & $\cdot$ & $\cdot$ & $\cdot$ & $\cdot$ & $\fraC$ & $\cdot$ & $\fraC$ &  AT$_{1,2}$   \\
\hline
\end{tabular}
\end{center}
{\scriptsize
HS=classes $4mmm$, $422$, $4/mm$, $\bar{4}2m$, LS=classes $4$, $\bar{4}$, $4/m$; $\fraC$=Complex components; $L$=longitudinal wave, $T$=transverse wave; $\bullet$=equal values components; $\circ$=equal value, opposite sign components; $(\sharp)$=number of independent eigentensors with same structure.
}
\end{table}%


We finish by giving an insight for waves propagating in directions orthogonal to the $c$-axis which, in a remarkeble difference with the previous case yields a fully-coupled problem as in the fully anisotropic case.

As far as we know, this is the most complete analysis for the wave propagation problem in classical Tetragonal micromorphic continua: of course, to make such an analyisis a predictive tool for experimental applications we need to know a complete set of parameters (43 for the low-symmetric classes $4$, $\bar{4}$ and $4/m$ and 25 for the high-symmetry classes ). These parameters which are very difficult to obtain and evaluate by the means of a set of simple experiment can be obtained, at the present only by two viable  approach, that is the numerical homogeneization approach used into \cite{ABNE18} for relaxed micromorphic tetragonal continua undergoing plane strain, or by homogeneization and identification methods from lattice dymanics, as it was done for cubic Diamond crystals and Silicon into \cite{MS19}. Such a technique shall be applied in a future paper to evaluate the components of the acoustic matrix $A$ for the Tetragonal $4/m$ crystals of PbWO$_{4}$ (PWO), widely used in high-energy physics as ionizing radiation detectors.

\section*{Acknowledgments}         
The research leading to these results is within the scope of CERN R\&D Experiment 18 "Crystal Clear Collaboration" and the PANDA Collaboration at GSI-Darmstadt. The author declares that he would never have dared to use repeatedly Cardano's formulae for the third- and fourth-order algebraic equations, if he hadn't been locked down by the COVID-19.

\section{Appendix}
{\tiny
In the first part of this Appendix (\S\S.\ref{prima}-\ref{terza}) we shall list, in tabular form when it is possible, the independent components of the fourth-order tensors $\Cel$, $\Del$ and $\Bel$, of the fifth-order tensors $\caF$ and $\caG$ and of the sixth-order tensor $\caH$ which appear into the constitutive relations (\ref{constitutive}), for all the classes of the Tetragonal point group. For the fourth-order tensors we refer to \cite{AU03} whereas for the other tensors we refer to the results obtained into \cite{FIFU53} and \cite{BOU64} (The most recent results given \emph{e.g.\/} into \cite{OLA13a}, \cite{OLA13b}, \cite{ALH18} and \cite{QHE10} cannot be applied to the present case since they were obtained for tensors endowed with some minor symmetries which in our case are missing). We shall list also, for convenience, the independent components of the microinertia fourth-order tensor $\Jel$.

In the second part of this Appendix, \S.\ref{quarta}, we shall list, for the Tetragonal point group, the independent components of the acoustic tensors $\Ab(\mb)$, $\hat{\Ab}(\mb)$, $\Pbm(\mb)$, $\Qbm(\mb)$ and $\Ael(\mb)$ showing, by the means of (\ref{componentsHO}),  the explicit dependence of these components on the propagation direction $\mb$ and on the tensorial quantities listed in the first part.

In the third part, \S.\ref{quinta}, we shall list the coefficients of the linear combinations between the elements of the bases $\{\eb_{k}\}$ and $\{\Wb_{h}\}$, $k=1,2,3$ and $h=1\,,\ldots\,,9$ which describes the eigenvectors of (\ref{microvibra2}) and (\ref{propa2}) for the various crystal symmetries.

\subsection{The fourth-order tensors}\label{prima}

The non-zero components for all the classes of the Tetragonal point group are given in tabular form into \cite{AU03} for the tensors $\Cel$ and $\Bel$ whereas those of $\Del$ can be obtained with the additional conditions induced by the symmetries of the first two components. 

We list the tabular form of these tensors in the Voigt's notation $1=11\,,2=22\,,3=33$, $4=23\,,5=31\,,6=12$, $7=32\,,8=13\,,9=21$, which for tensors mapping from and into $\Sym$ reduces to $4=23=32$, $5=13=31$ and $6=12=21$.

\subsubsection{The elasticity tensor $\Cel$}

All classes (6 independent components):
\begin{equation}\label{tetraC}
[\Cel]\equiv
\left[
\begin{array}{cccccc}
\Cel_{11}  & \Cel_{12}  & \Cel_{13} & 0 & 0  & 0 \\
\cdot  & \Cel_{11}  & \Cel_{13} & 0 & 0  & 0 \\
\cdot  & \cdot & \Cel_{33} & 0 & 0  & 0\\
\cdot &  \cdot & \cdot & \Cel_{44} & 0 & 0 \\
\cdot &  \cdot & \cdot & \cdot & \Cel_{44} & 0 \\
\cdot & \cdot  & \cdot & \cdot & \cdot  & \Cel_{66}   
 \end{array}
\right]\,.
\end{equation}
The tensor $\Cel_{micro}$ has the same non-null components as (\ref{tetraC}).

\subsubsection{The tensor $\Del$}

Classes $4$, $\bar{4}$ and $4/m$ (14 independent components):
\begin{equation}\label{DelLS}
[\Del]\equiv
\left[
\begin{array}{ccccccccc}
\Del_{11}  & \Del_{12}  & \Del_{13} & 0 & 0  & \Del_{16} &0&0&-\Del_{26}\\
\Del_{12}  & \Del_{11}  & \Del_{13} & 0 & 0  & \Del_{26}&0&0&-\Del_{16} \\
\Del_{31}  & \Del_{31}  & \Del_{33} & 0 & 0  & \Del_{36}&0&0&-\Del_{36}\\
0  &  0 & 0 & \Del_{44} & \Del_{45} & 0 &\Del_{55}&-\Del_{54}&0\\
0  &  0 & 0 & \Del_{54} & \Del_{55} & 0&-\Del_{45}&\Del_{44}&0 \\
\Del_{61}  & -\Del_{61}  & 0& 0 & 0  & \Del_{66}&0&0&\Del_{66}  \\ 
0  &  0 & 0 & \Del_{44} & \Del_{45} & 0 &\Del_{55}&-\Del_{54}&0\\
0  &  0 & 0 & \Del_{54} & \Del_{55} & 0&-\Del_{45}&\Del_{44}&0 \\
\Del_{61}  & -\Del_{61}  & 0& 0 & 0  & \Del_{66}&0&0&\Del_{66}
\end{array}
\right]\,;
\end{equation}

Classes $4mmm$, $422$, $4/mm$  and $\bar{4}2m$ (8 independent components): 
\begin{equation}
[\Del]\equiv
\left[
\begin{array}{ccccccccc}
\Del_{11}  & \Del_{12}  & \Del_{13} & 0 & 0  & 0 &0&0&0\\
\Del_{12}  & \Del_{11}  & \Del_{13} & 0 & 0  & 0&0&0&0 \\
\Del_{31}  & \Del_{31}  & \Del_{33} & 0 & 0  & 0&0&0&0\\
0  &  0 & 0 & \Del_{44} & 0 & 0 &\Del_{55}&0&0\\
0  &  0 & 0 & 0 & \Del_{55} & 0&0&\Del_{44}&0 \\
0  &0  & 0 & 0 & 0  & \Del_{66}&0&0&\Del_{66}  \\ 
0  &  0 & 0 & \Del_{44} & 0 & 0 &\Del_{55}&0&0\\
0  &  0 & 0 & 0 & \Del_{55} & 0&0&\Del_{44}&0 \\
0  &0  & 0 & 0 & 0  & \Del_{66}&0&0&\Del_{66} 
\end{array}
\right]\,;
\end{equation}

\subsubsection{The tensor $\Bel$}

Classes $4$, $\bar{4}$ and $4/m$ (13 independent components):
\begin{equation}\label{BelLS}
[\Bel]\equiv
\left[
\begin{array}{ccccccccc}
\Bel_{11}  & \Bel_{12}  & \Bel_{13} & 0 & 0  & \Bel_{16} &0&0&-\Bel_{26}\\
\cdot  & \Bel_{11}  & \Bel_{13} & 0 & 0  & \Bel_{26}&0&0&-\Bel_{16} \\
\cdot   & \cdot   & \Bel_{33} & 0 & 0  & \Bel_{36}&0&0&-\Bel_{36}\\
\cdot   &  \cdot  & \cdot  & \Bel_{44} & \Bel_{45} & 0 &\Bel_{47}&0&0\\
\cdot  &  \cdot  & \cdot  & \cdot  & \Bel_{55} & 0&0&\Bel_{47}&0 \\
\cdot   & \cdot   & \cdot  & \cdot  & \cdot   & \Bel_{66}&0&0&\Bel_{69}  \\
 \cdot &\cdot &\cdot & \cdot  & \cdot &\cdot &\Bel_{55}&-\Bel_{45}&0\\
\cdot  &  \cdot  & \cdot  & \cdot  & \cdot  & \cdot  &\cdot &\Bel_{44}&0\\
\cdot   &  \cdot  & \cdot & \cdot  & \cdot  & \cdot & \cdot & \cdot &\Bel_{66}
  \end{array}
\right]\,;
\end{equation}

Classes $4mmm$, $422$, $4/mm$  and $\bar{4}2m$ (9 independent components):
\begin{equation}\label{BelHS}
[\Bel]\equiv
\left[
\begin{array}{ccccccccc}
\Bel_{11}  & \Bel_{12}  & \Bel_{13} & 0 & 0  & 0 &0&0&0\\
\cdot  & \Bel_{11}  & \Bel_{13} & 0 & 0  &0&0&0&0 \\
\cdot   & \cdot   & \Bel_{33} & 0 & 0  & 0&0&0&0\\
\cdot   &  \cdot  & \cdot  & \Bel_{44} & 0 & 0 &\Bel_{47}&0&0\\
\cdot  &  \cdot  & \cdot  & \cdot  & \Bel_{55} & 0&0&\Bel_{47}&0 \\
\cdot   & \cdot   & \cdot  & \cdot  & \cdot   & \Bel_{66}&0&0&\Bel_{69}  \\
 \cdot &\cdot &\cdot & \cdot  & \cdot &\cdot &\Bel_{55}&0&0\\
\cdot  &  \cdot  & \cdot  & \cdot  & \cdot  & \cdot  &\cdot &\Bel_{44}&0\\
\cdot   &  \cdot  & \cdot & \cdot  & \cdot  & \cdot & \cdot & \cdot &\Bel_{66}
\end{array}
\right]\,;
\end{equation}

\subsubsection{The fourth-order inertia tensor $\Jel$}

Let $\Jb$ be the micro-inertia tensor whose components are $J_{ij}=J_{ji}$, $i,j=1,2,3$: then, by (\ref{Jeldef})$_{2}$, the matrix $\Jel=\Jel^{T}$ is
\begin{equation}\label{Jfourth1}
[\Jel]\equiv
\left[
\begin{array}{ccccccccc}
J_{11}  & 0  & 0 & 0 & 0  & J_{12} &0 & J_{13} & 0\\
\cdot  & J_{22}  & 0 & J_{23} & 0  & 0 & 0 & 0 &J_{12} \\
\cdot    & \cdot   & J_{33} & 0 & J_{13}  & 0&J_{23}&0&0\\
\cdot    &  \cdot   & \cdot  & J_{33} & 0& 0 &0&0&J_{13}\\
\cdot  &  \cdot   & \cdot  & \cdot   & J_{11} & 0&J_{12}&0&0 \\
\cdot   & \cdot    & \cdot   & \cdot   & \cdot   & J_{22} &0&J_{23}&0  \\
\cdot  & \cdot  & \cdot & \cdot  & \cdot  & \cdot  &J_{22}&0&0\\
\cdot  &  \cdot   &\cdot   & \cdot   & \cdot   & \cdot  &\cdot &J_{33}&0\\
\cdot   & \cdot  & \cdot  & \cdot   & \cdot   & \cdot  & \cdot  & \cdot  &J_{11}
 \end{array}
\right]\,;
\end{equation}
for Tetragonal crystals, provided we identify the $c-$axis with the direction $\eb_{3}$, we have $J_{11}=J_{22}$ and $J_{ij}=0\,,i\neq j$: accordingly (\ref{Jfourth1}) reduces to:
\begin{equation}\label{Jtetra}
[\Jel]\equiv
\left[
\begin{array}{ccccccccc}
J_{11}  & 0  & 0 & 0 & 0  &0 &0 & 0 & 0\\
\cdot  & J_{11}  & 0 & 0 & 0  & 0 & 0 & 0 &0 \\
\cdot    & \cdot   & J_{33} & 0 & 0  & 0&0&0&0\\
\cdot    &  \cdot   & \cdot  & J_{33} & 0& 0 &0&0&0\\
\cdot  &  \cdot   & \cdot  & \cdot   & J_{11} & 0&0&0&0 \\
\cdot   & \cdot    & \cdot   & \cdot   & \cdot   & J_{11} &0&0&0  \\
\cdot  & \cdot  & \cdot & \cdot  & \cdot  & \cdot  &J_{11}&0&0\\
\cdot  &  \cdot   &\cdot   & \cdot   & \cdot   & \cdot  &\cdot &J_{33}&0\\
\cdot   & \cdot  & \cdot  & \cdot   & \cdot   & \cdot  & \cdot  & \cdot  &J_{11}
 \end{array}
\right]\,.
\end{equation}

\subsection{The fifth-order tensors}\label{seconda}

A detailed study of the symmetries for fifth- and sixth-order tensor was done into \cite{FIFU53}: for the Tetragonal group the symmetries are different between classes, and accordingly we study them in detail beginning with the fifth-order tensor $\caG$; we also follow \cite{FIFU53} into the use of the notation
\begin{equation}
\caG_{11112}\,[5]\,,
\end{equation}
to denote all the 5 possible combinations of the index, namely: $\caG_{11112}$, $\caG_{11121}$, $\caG_{11211}$, $\caG_{12111}$ and $\caG_{21111}$. 

Since $\caG$ is an odd tensors, then for the two centrosymmetric classes $4/m$ and $4mm$ their components vanishes altogether: for the other classes we have the following restrictions.
\begin{itemize}
\item
Class $4$: for this class we have 61 independent components:
\begin{eqnarray}\label{caG1}
&\caG_{33333}\,[1]\,,\quad\caG_{11113}=\caG_{22223}\,[5]\,,\quad\caG_{11333}=\caG_{22333}\, [10]\,,\\
&\caG_{11123}=-\caG_{22213}\,[20]\,,\quad\caG_{33312}=-\caG_{33321}\,[10]\,,\quad\caG_{11223}=\caG_{22113}\,[15]\,.\nonumber
\end{eqnarray}

\item
Class $\bar{4}$: for this class we have the following restrictions into (\ref{caG1})
\begin{equation}\label{caG2}
\caG_{33333}=0\,,\quad\caG_{33312}=\caG_{33321}=0\,,
\end{equation}
which means that (\ref{caG1})$_{1,4}$ must be zero and the independent components reduce to 50.
\item
Classes $4/mm$, $\bar{4}2m$, $422$: for these classes the tensors $\caG$ splits into polar and axial ones. 

The polar tensors have ha 61 non-zero components:
\begin{eqnarray}
&\caG_{33333}\,[1]\quad\quad\caG_{22223}=\caG_{11113}\,[5]\,,\\
&\quad\caG_{33322}=\caG_{33311}\,[10]\,,\quad\caG_{22113}=\caG_{11223}\,[15]\,,\nonumber
\end{eqnarray}
31 being the independent ones. 

The axial tensors have 60 components with only 30 independent for the class $4/mm$:
\begin{equation}\label{4mmP}
\caG_{22213}=-\caG_{11123}\,[20]\,,\quad\caG_{33312}=-\caG_{33321}\,[10]\,;
\end{equation}
the components for the class $\bar{4}2m$ are obtained by changing the sign of the components (\ref{4mmP}), whereas those for the classes $422$ are obtained by setting to zero (\ref{4mmP}).
\end{itemize}

To obtain the number of the independent components for the tensor $\caF$, we recall that it obeys $\caF_{ijhkm}=\caF_{jihkm}$ and hence the number of independent components reduces to 41 components for the class $4$, 33 components for class $\bar{4}$ and 42 components (21 independent) for the classes  $4/mm$, $\bar{4}2m$ and $422$.

\subsection{The sixth-order tensor $\caH$}\label{terza}

Also for sixth-order tensors the symmetries changes for different classes. Following \cite{FIFU53} we have, by using the same convention we used for the fifth-order tensor, that the for the classes $4$, $\bar{4}$ and $4/m$ the non-null components are:
\begin{eqnarray}\label{4H6nonzero}
&\caH_{111111}=\caH_{222222}\,[1]\quad\caH_{333333}\,[1]\quad\caH_{111112}=-\caH_{222221}\,[6]\,,\nonumber\\
&\caH_{111122}=\caH_{222211}\,[15]\,,\quad\caH_{111133}=\caH_{222233}\,[15]\,,\nonumber\\
&\caH_{333311}=\caH_{333322}\,[15]\,,\quad\caH_{333312}-\caH_{333321}\,[15]\,,\\
&\caH_{111222}\,[20]\,,\quad\caH_{222333}=-\caH_{333222}\,[10]\,,\nonumber\\
&\caH_{111332}=\caH_{222331}\,[60]\,,\quad\caH_{221133}=\caH_{112233}\,[45]\,;\nonumber
\end{eqnarray}
however, the number of non-zero and independent component obtained into \cite{FIFU53} refers to a tensor with no major symmetries, whereas in our case $\caH=\caH^{T}$ and hence there are only 108 independent components into (\ref{4H6nonzero}). 

For the classes $422$, $4mm$, $4/mmm$ and for the polar tensors of the class $\bar{4}2m$ we have the further restrictions into (\ref{4H6nonzero}):
\begin{eqnarray}\label{4H6lower}
&\caH_{122222}=\caH_{211111}=\caH_{333312}=\caH_{333321}=0\,,\\
&\caH_{222331}=\caH_{111332}=\caH_{222333}=\caH_{333222}=0\,,\nonumber
\end{eqnarray}
which further reduce the number of independent components.

\subsection{The acoustic tensors}\label{quarta}

\subsubsection{The generalized acoustic tensor $\Ab(\mb)$}

From relation (\ref{componentsHO})$_{1}$ we have the components of the second-order acoustic tensor $\Ab(\mb)$ for the classes  $4\,,\bar{4}$ and $4/m$ of the Tetragonal group:
\begin{align}\label{acoustic1LS}
A_{11}&=\rho^{-1}\Big((\Cel_{11}+\Bel_{11}+2\Del_{11})m_{1}^{2}+(\Cel_{66}+\Bel_{69}+2\Del_{66})m_{2}^{2}+(\Cel_{44}+\Bel_{47}+\Del_{55}+\Del_{44})m_{3}^{2}\nonumber\\
&+(\Cel_{16}+\Bel_{61}-\Bel_{26}+\Del_{16}+2\Del_{61}-\Del_{26})m_{1}m_{2}\Big)\nonumber\\
A_{22}&=\rho^{-1}\Big((\Cel_{66}+\Bel_{69}+2\Del_{66})m_{1}^{2}+(\Cel_{11}+\Bel_{11}+2\Del_{11})m_{2}^{2}+(\Cel_{44}+\Bel_{47}+\Del_{55}+\Del_{44})m_{3}^{2}\nonumber\\
&-(\Cel_{16}+\Bel_{61}-\Bel_{26}+\Del_{16}+2\Del_{61}-\Del_{26})m_{1}m_{2}\Big)\\
A_{33}&=\rho^{-1}\Big((\Cel_{44}+\Del_{44}+\Del_{55}+\Bel_{47})(m_{1}^{2}+m_{2}^{2})+(\Cel_{33}+2\Del_{33}+\Bel_{33})m_{3}^{2}\Big)\nonumber\\
A_{23}&=\rho^{-1}\Big((-\Bel_{36}+\Del_{45}-\Del_{54})m_{1}m_{3}+(\Cel_{44}+\Bel_{13}+\Bel_{44}+2\Del_{13}+\Del_{44}+\Del_{55})m_{2}m_{3}\Big)\nonumber\\
A_{13}&=\rho^{-1}\Big((\Cel_{44}+\Bel_{13}+\Bel_{55}+2\Del_{13}+\Del_{44}+\Del_{55})m_{1}m_{3}+(\Bel_{36}+\Del_{54}-\Del_{45})m_{2}m_{3}\Big)\nonumber\\
A_{12}&=\rho^{-1}\Big((\Cel_{16}+\Bel_{16}+\Del_{16}-\Del_{26})m_{1}^{2}+(-\Cel_{16}+\Bel_{62}-2\Del_{61})m_{2}^{2}\nonumber\\
&+(-\Bel_{45}+\Del_{54}-\Del_{45})m_{3}^{2}+(\Cel_{66}+2\Cel_{12}+\Bel_{12}+\Bel_{66}+2(\Del_{12}+\Del_{66}))m_{1}m_{2}\Big)\,.\nonumber
\end{align}
For the classes $4mmm\,,422\,,4/mm$ and $\bar{4}2m$ relations (\ref{acoustic1LS}) simplify into:
\begin{align}\label{acoustic1HS}
A_{11}&=\rho^{-1}\Big((\Cel_{11}+\Bel_{11}+2\Del_{11})m_{1}^{2}+(\Cel_{66}+\Bel_{69}+2\Del_{66})m_{2}^{2}+(\Cel_{44}+\Bel_{44}+\Del_{55}+\Del_{44})m_{3}^{2}\Big)\nonumber\\
A_{22}&=\rho^{-1}\Big((\Cel_{66}+\Bel_{69}+2\Del_{66})m_{1}^{2}+(\Cel_{11}+\Bel_{11}+2\Del_{11})m_{2}^{2}+(\Cel_{44}+\Bel_{44}+\Del_{55}+\Del_{44})m_{3}^{2}\Big)\nonumber\\
A_{33}&=\rho^{-1}\Big((\Cel_{44}+\Del_{44}+\Del_{55}+\Bel_{44})(m_{1}^{2}+m_{2}^{2})+(\Cel_{33}+2\Del_{33}+\Bel_{33})m_{3}^{2}\Big)\\
A_{23}&=\rho^{-1}\Big((\Cel_{44}+2\Del_{13}+\Del_{44}+\Del_{55})m_{2}m_{3}+(\Bel_{13}+\Bel_{44})m_{2}m_{3}\Big)\,,\nonumber\\
A_{13}&=\rho^{-1}\Big((\Cel_{44}+2\Del_{13}+\Del_{44}+\Del_{55})m_{1}m_{3}+(\Bel_{13}+\Bel_{55})m_{1}m_{3}\Big)\,,\nonumber\\
A_{12}&=\rho^{-1}\Big(\Cel_{16}(m_1^2-m_2^2)+(\Cel_{66}+2\Cel_{12}+2\Del_{12}+2\Del_{66}+\Bel_{1122}+\Bel_{66})m_{1}m_{2}\Big)\,.\nonumber
\end{align}

\subsubsection{The long wavelength approximation acoustic tensor $\hat{\Ab}(\mb)$}

The definition of the acoustic tensor $\hat{\Ab}(\mb)$ leads to the same result as in the classical linearly elastic case with the components of $\hat{\Cel}=\Cel-\Cel_{micro}$ in place of those of $\Cel$, accordingly, since $A_{ij}=\rho^{-1}\Cel_{iljk}m_{k}m_{l}$, then for the classes $4$, $\bar{4}$ and $4/m$ we have the explicit representation:
\begin{eqnarray}\label{Btensor}
\hat{A}_{11}&=&\rho^{-1}(\hat{\Cel}_{11}m_1^2+\hat{\Cel}_{66}m_2^2+\hat{\Cel}_{16}m_1m_2+\hat{\Cel}_{44}m_3^2)\,,\nonumber\\
\hat{A}_{22}&=&\rho^{-1}(\hat{\Cel}_{11}m_2^2+\hat{\Cel}_{66}m_1^2-\hat{\Cel}_{16}m_1m_2+\hat{\Cel}_{44}m_3^2)\,,\nonumber\\
\hat{A}_{33}&=&\rho^{-1}(m_3^2\hat{\Cel}_{33}+\hat{\Cel}_{44}(m_{1}^{2}+m_{2}^{2}))\,,\\
\hat{A}_{23}&=&\rho^{-1}\hat{\Cel}_{44}m_{2}m_{3}\,,\nonumber\\
\hat{A}_{13}&=&\rho^{-1}\hat{\Cel}_{44}m_{1}m_{3}\,,\nonumber\\
\hat{A}_{12}&=&\rho^{-1}(\hat{\Cel}_{16}(m_1^2-m_2^2)+(\hat{\Cel}_{66}+2\hat{\Cel}_{12})m_1m_2)\,;\nonumber
\end{eqnarray}
the components for the classes $4mmm$, $422$, $4/mm$  and $\bar{4}2m$ can be obtained by taking $\hat{\Cel}_{16}=0$ into (\ref{Btensor}) .

\subsubsection{The third-order tensor $\Qbm(\mb)$}

For the classes  $4\,,\bar{4}$ and $4/m$ the components of the third-order tensor $\Qbm(\mb)$, calculated by the means of (\ref{componentsHO})$_{3}$, obey the  restrictions
\begin{equation}
\Qbm_{232}=\Qbm_{131}\,,\quad\Qbm_{223}=\Qbm_{113}\,,\quad\Qbm_{322}=\Qbm_{311}\,,\quad\Qbm_{213}=-\Qbm_{123}\,,\quad\Qbm_{231}=-\Qbm_{132}\,,\quad\Qbm_{321}=-\Qbm_{312}\,;
\end{equation}
the 21 independent components are given in explicit by:
\begin{equation}
\begin{aligned}\label{QBMLS1}
&\Qbm_{111}=\rho^{-1}((\Del_{11}+\Bel_{11})m_{1}+(\Del_{16}+\Bel_{16})m_{2})\,,&&\Qbm_{123}=\rho^{-1}\Bel_{36}m_{3}\,,\\
&\Qbm_{122}=\rho^{-1}((\Del_{66}+\Bel_{69})m_{1}-(\Del_{61}-\Bel_{26})m_{2})\,,&&\Qbm_{131}=\rho^{-1}(\Del_{44}+\Bel_{44})m_{3}\,,\\
&\Qbm_{133}=\rho^{-1}((\Del_{55}+\Bel_{47})m_{1}-(\Del_{45}+\Bel_{45})m_{2})\,,&&\Qbm_{132}=\rho^{-1}\Del_{54}m_{3}\,,\\
&\Qbm_{112}=\rho^{-1}((\Del_{12}+\Bel_{12})m_{1}-(\Del_{26}+\Bel_{26})m_{2})\,,&&\Qbm_{113}=\rho^{-1}(\Del_{13}+\Bel_{13})m_{3}\,,\\
&\Qbm_{121}=\rho^{-1}((\Del_{16}+\Bel_{16})m_{1}+(\Del_{66}+\Bel_{66})m_{2})\,,&&\Qbm_{311}=\rho^{-1}(\Del_{44}+\Bel_{47})m_{3}\,,\\
&\Qbm_{211}=\rho^{-1}((\Del_{26}-\Bel_{26})m_{1}+(\Del_{66}+\Bel_{69})m_{2})\,,&&\Qbm_{333}=\rho^{-1}(\Del_{33}+\Bel_{33})m_{3}\,,\\
&\Qbm_{222}=\rho^{-1}(-(\Del_{16}+\Bel_{16})m_{1}+(\Del_{11}+\Bel_{11})m_{2})\,,&&\Qbm_{312}=\rho^{-1}(\Del_{54}+\Bel_{45})m_{3}\,,\\
&\Qbm_{233}=\rho^{-1}((\Del_{45}+\Bel_{45})m_{1}+(\Del_{55}+\Bel_{47})m_{2})\,,&&\\
&\Qbm_{212}=\rho^{-1}((\Del_{66}+\Bel_{66})m_{1}-(\Del_{61}+\Bel_{16})m_{2})\,,&&\\
&\Qbm_{221}=\rho^{-1}((\Del_{12}+\Bel_{12})m_{1}+(\Del_{26}+\Bel_{26})m_{2})\,,&&\\
&\Qbm_{323}=\rho^{-1}(\Del_{45}m_{1}+(\Del_{55}+\Bel_{55})m_{2})\,,&&\\
&\Qbm_{331}=\rho^{-1}((\Del_{31}+\Bel_{13})m_{1}+(\Del_{36}+\Bel_{36})m_{2})\,,&&\\
&\Qbm_{332}=\rho^{-1}(-(\Del_{36}+\Bel_{36})m_{1}+(\Del_{31}+\Bel_{13})m_{2})\,,&&\\
&\Qbm_{313}=\rho^{-1}((\Del_{55}+\Bel_{55})m_{1}-\Del_{45}m_{2})\,.&&
\end{aligned}
\end{equation}
The tabular representation is
\begin{equation}\label{QBMLS}
[\Qbm]=
\left[
\begin{array}{ccccccccc}
( \Qbm_{111}) &  (\Qbm_{122})  & (\Qbm_{133})  & \Qbm_{123}& \Qbm_{131}& (\Qbm_{112})& \Qbm_{132}& \Qbm_{113}& (\Qbm_{121})\\
 ( \Qbm_{211}) &  (\Qbm_{222} ) &  (\Qbm_{233})  & \Qbm_{113}& -\Qbm_{132}& (\Qbm_{212})& \Qbm_{131}& -\Qbm_{123}& (\Qbm_{221})\\
  \Qbm_{311} &  \Qbm_{311}  &  \Qbm_{333}  & (\Qbm_{323})& (\Qbm_{331})& \Qbm_{312}& (\Qbm_{332)}& (\Qbm_{313})& -\Qbm_{312}
\end{array}
\right]
\end{equation}
where the 14 components which depend solely $m_{1}\,,m_{2}$ are represented within brackets $(\cdot)$, the other 13 components depending only on $m_{3}$..

For the tetragonal classes $4mmm$, $422$, $4/mm$  and $\bar{4}2m$ the 6 components $\Qbm_{123}$, $\Qbm_{132}$, $\Qbm_{231}$, $\Qbm_{213}$, $\Qbm_{312}$ and $\Qbm_{321}$ vanishes and we have the three conditions $\Qbm_{113}=\Qbm_{223}$, $\Qbm_{311}=\Qbm_{322}$ and $\Qbm_{331}=\Qbm_{332}$. There are 18 independent components:
\begin{equation}\label{QbmHS}
\begin{array}{ccc}
\Qbm_{111}=\rho^{-1}(\Del_{11}+\Bel_{11})m_{1}&\Qbm_{211}=\rho^{-1}(\Del_{66}+\Bel_{69})m_{2}&\Qbm_{311}=\rho^{-1}(\Del_{44}+\Bel_{47})m_{3}\\
\Qbm_{122}=\rho^{-1}(\Del_{66}+\Bel_{69})m_{1}&\Qbm_{222}=\rho^{-1}(\Del_{11}+\Bel_{11})m_{2}&\Qbm_{333}=\rho^{-1}(\Del_{33}+\Bel_{33})m_{3}\\
\Qbm_{133}=\rho^{-1}(\Del_{55}+\Bel_{47})m_{1}&\Qbm_{233}=\rho^{-1}(\Del_{55}+\Bel_{47})m_{2}&\Qbm_{323}=\rho^{-1}(\Del_{55}+\Bel_{45})m_{2}\\
\Qbm_{131}=\rho^{-1}(\Del_{44}+\Bel_{44})m_{3}&\Qbm_{212}=\rho^{-1}(\Del_{66}+\Bel_{66})m_{1}&\Qbm_{331}=\rho^{-1}(\Del_{31}+\Bel_{13})m_{1}\\
\Qbm_{112}=\rho^{-1}(\Del_{12}+\Bel_{12})m_{2}&\Qbm_{232}=\rho^{-1}(\Del_{44}+\Bel_{44})m_{3}&\Qbm_{313}=\rho^{-1}(\Del_{55}+\Bel_{55})m_{1}\\
\Qbm_{113}=\rho^{-1}(\Del_{13}+\Bel_{13})m_{3}&\Qbm_{221}=\rho^{-1}(\Del_{21}+\Bel_{12})m_{1}&\\
\Qbm_{121}=\rho^{-1}(\Del_{66}+\Bel_{66})m_{2}&&
\end{array}
\end{equation}
with the tabular representation, where we put in square brackets $[\cdot]$ those depending on $m_{1}$ and in round brackets $(\cdot)$ those which depends only on $m_{2}$, the remaining depending on $m_{3}$:
\begin{equation}\label{QBMHS}
[\Qbm]=
\left[
\begin{array}{ccccccccc}
 [\Qbm_{111}] &  [\Qbm_{122}]  &  [\Qbm_{133}]  & 0& \Qbm_{131}& (\Qbm_{112})& 0& \Qbm_{113}& (\Qbm_{121})\\
  (\Qbm_{211}) &  (\Qbm_{222} ) &  (\Qbm_{233})  & \Qbm_{113}& 0& [\Qbm_{212}]&\Qbm_{131}& 0& [\Qbm_{221}]\\
 \Qbm_{311} &  \Qbm_{311}  &  \Qbm_{333}  & [\Qbm_{323}]&( \Qbm_{331})& 0&( \Qbm_{331})& (\Qbm_{313})& 0
\end{array}
\right]\,.
\end{equation}

\subsubsection{The third-order tensor $\Pbm(\mb)$}

We recall that for the centrosymmetric classes $4/m$ and $4mm$ since $\caF=\caG=\bzero$ then $\Pbm=\bzero$. The components for the class $4$, evaluated by the means of (\ref{componentsHO})$_{2}$  and (\ref{caG1}) are (here we denote $\caA=\caF+\caG$):
{\allowdisplaybreaks
\begin{align}\label{PBMLS}
\Pbm_{111}&=\rho^{-1}((\caA_{11113}+\caA_{13111})m_{1}m_{3}+(\caA_{12113}+\caA_{13112})m_{2}m_{3})\,,\nonumber\\
\Pbm_{122}&=\rho^{-1}((\caA_{11223}+\caA_{13221})m_{1}m_{3}-(\caA_{21113}+\caA_{23111})m_{2}m_{3})\,,\nonumber\\
\Pbm_{133}&=\rho^{-1}((\caA_{11333}+\caA_{13331})m_{1}m_{3}+(\caA_{12333}+\caA_{13332})m_{2}m_{3}\nonumber)\,,\\
\Pbm_{123}&=\rho^{-1}(\caA_{11231}m_{1}^{2}-\caA_{21131}m_{2}^{2}+\caA_{13233}m_{3}^{2}+(\caA_{11232}+\caA_{12231})m_{1}m_{2})\,,\nonumber\\
\Pbm_{131}&=\rho^{-1}(\caA_{11311}m_{1}^{2}+\caA_{12312}m_{2}^{2}+\caA_{13313}m_{3}^{2}+(\caA_{11312}+\caA_{12311})m_{1}m_{2})\,,\nonumber\\
\Pbm_{112}&=\rho^{-1}((\caA_{11123}+\caA_{13121})m_{1}m_{3}+(\caA_{12123}+\caA_{13122})m_{2}m_{3})\,,\nonumber\\
\Pbm_{132}&=\rho^{-1}(\caA_{11321}m_{1}^{2}-\caA_{21311}m_{2}^{2}+\caA_{13323}m_{3}^{2}+(\caA_{11322}+\caA_{12321})m_{1}m_{2}\nonumber\\
&+(\caA_{12323}+\caA_{13322})m_{2}m_{3})\,,\nonumber\\
\Pbm_{113}&=\rho^{-1}(\caA_{11131}m_{1}^{2}+\caA_{12132}m_{2}^{2}+\caA_{13133}m_{3}^{2}+(\caA_{11132}+\caA_{12131})m_{1}m_{2})\,,\nonumber\\
\Pbm_{121}&=\rho^{-1}((\caA_{11213}+\caA_{13211})m_{1}m_{3}+(\caA_{12213}+\caA_{13212})m_{2}m_{3})\,,\nonumber\\
\Pbm_{211}&=\rho^{-1}((\caA_{21113}+\caA_{23111})m_{1}m_{3}+(\caA_{11223}+\caA_{13221})m_{2}m_{3})\,,\nonumber\\
\Pbm_{222}&=\rho^{-1}(-(\caA_{12113}+\caA_{13112})m_{1}m_{3}+(\caA_{11113}+\caA_{13111})m_{2}m_{3})\,,\nonumber\\
\Pbm_{233}&=\rho^{-1}((\caA_{21332}+\caA_{22331})m_{1}m_{2}-(\caA_{12333}+\caA_{13332})m_{1}m_{3}\nonumber\\
&+(\caA_{11333}+\caA_{13331})m_{2}m_{3})\,,\\
\Pbm_{223}&=\rho^{-1}(\caA_{12132}m_{1}^{2}+\caA_{11131}m_{2}^{2}+\caA_{13133}m_{3}^{2}+(\caA_{21232}+\caA_{22231})m_{1}m_{2})\,,\nonumber\\
\Pbm_{231}&=\rho^{-1}(\caA_{21311}m_{1}^{2}-\caA_{11321}m_{2}^{2}-\caA_{13323}m_{3}^{2}+(\caA_{12321}+\caA_{11322})m_{1}m_{2}\nonumber\\
&+(\caA_{21313}+\caA_{23311})m_{1}m_{3})\,,\nonumber\\
\Pbm_{212}&=\rho^{-1}((\caA_{12213}+\caA_{13212})m_{1}m_{3}-(\caA_{11213}+(\caA_{13211})m_{2}m_{3})\,,\nonumber\\
\Pbm_{232}&=\rho^{-1}(\caA_{12312}m_{1}^{2}+\caA_{11311}m_{2}^{2}+\caA_{13313}m_{3}^{2}-(\caA_{12311}+\caA_{11312})m_{1}m_{2})\,,\nonumber\\
\Pbm_{213}&=\rho^{-1}(\caA_{21131}m_{1}^{2}-\caA_{11231}m_{2}^{2}-\caA_{13233}m_{3}^{2}+(\caA_{12231}+\caA_{11232})m_{1}m_{2}\nonumber\\
&+(\caA_{22133}+\caA_{23132})m_{2}m_{3})\,,\nonumber\\
\Pbm_{221}&=\rho^{-1}((\caA_{12123}+\caA_{13122})m_{1}m_{3}-(\caA_{11123}+\caA_{13121})m_{2}m_{3})\,,\nonumber\\
\Pbm_{311}&=\rho^{-1}(\caA_{31111}m_{1}^{2}+\caA_{32112}m_{2}^{2}+\caA_{33113}m_{3}^{2}\nonumber\\
&+(\caA_{31112}+\caA_{32111})m_{1}m_{2}+(\caA_{31113}+\caA_{33111})m_{1}m_{3})\,,\nonumber\\
\Pbm_{322}&=\rho^{-1}(\caA_{31221}m_{1}^{2}+\caA_{31111}m_{2}^{2}+\caA_{33113}m_{3}^{2}-(\caA_{32111}+\caA_{31112})m_{1}m_{2})\,,\nonumber\\
\Pbm_{333}&=\rho^{-1}(\caA_{31331}m_{1}^{2}+\caA_{31331}m_{2}^{2}+\caA_{33333}m_{3}^{2}+(\caA_{31332}+\caA_{32331})m_{1}m_{2})\,,\nonumber\\
\Pbm_{323}&=\rho^{-1}((\caA_{31233}+\caA_{33231})m_{1}m_{3}+(\caA_{31133}+\caA_{33131})m_{2}m_{3})\,,\nonumber\\
\Pbm_{331}&=\rho^{-1}((\caA_{31313}+\caA_{33311})m_{1}m_{3}-(\caA_{31323}+\caA_{33321})m_{2}m_{3})\,,\nonumber\\
\Pbm_{312}&=\rho^{-1}(\caA_{31121}m_{1}^{2}-\caA_{31211}m_{2}^{2}+\caA_{33123}m_{3}^{2}\nonumber\\
&+(\caA_{31122}+\caA_{32121})m_{1}m_{2}-(\caA_{31213}+\caA_{33212})m_{2}m_{3})\,,\nonumber\\
\Pbm_{332}&=\rho^{-1}((\caA_{31323}+\caA_{33321})m_{1}m_{3}+(\caA_{32323}+\caA_{33322})m_{2}m_{3})\,,\nonumber\\
\Pbm_{313}&=\rho^{-1}((\caA_{31133}+\caA_{33131})m_{1}m_{3}-(\caA_{31233}+\caA_{33231})m_{2}m_{3})\,,\nonumber\\
\Pbm_{321}&=\rho^{-1}(\caA_{31211}m_{1}^{2}+\caA_{32212}m_{2}^{2}-\caA_{33123}m_{3}^{2}+(\caA_{32121}+\caA_{31122})m_{1}m_{2})\,.\nonumber
\end{align}
}
In the tabular form of the tensor $\Pbm(\mb)$ we show the components which are different from zero when either $\mb=\eb_{3}$ or $\mb\cdot\eb_{3}=0$:
\begin{equation}
[\Pbm]=
\left[
\begin{array}{ccccccccc}
0 &  0  &  0  & \Pbm_{123}& \Pbm_{131}& 0 & \Pbm_{132}& \Pbm_{113}& 0\\
0 &  0  &  0  & \Pbm_{223}& \Pbm_{231}& 0& \Pbm_{232}& \Pbm_{213}& 0\\
  \Pbm_{311} &  \Pbm_{322}  &  \Pbm_{333}  & 0 & 0 & \Pbm_{312}& 0 & 0 & \Pbm_{321}\\
\end{array}
\right]\,.
\end{equation}
In the case $\mb=\eb_{3}$ we also have that for the class $4$ 
\begin{equation}
\Pbm_{223}=\Pbm_{113}\,,\quad\Pbm_{231}=-\Pbm_{132}\,,\quad\Pbm_{232}=-\Pbm_{131}\,,\quad\Pbm_{213}=-\Pbm_{123}\,,\quad\Pbm_{322}=\Pbm_{311}\,,\quad\Pbm_{321}=-\Pbm_{312}\,,
\end{equation}
whereas for the class $\bar{4}$ the following components vanish:
\begin{equation}
\Pbm_{123}=\Pbm_{132}=\Pbm_{231}=\Pbm_{213}=\Pbm_{333}=\Pbm_{312}=\Pbm_{321}=0\,.
\end{equation}

For the remaining classes $\bar{4}2m$, $422$, $4/mm$ of the Tetragonal group we have that:
{\allowdisplaybreaks
\begin{align}\label{PBMHS}
\Pbm_{111}&=\rho^{-1}(\caA_{11113}+\caA_{13111})m_{1}m_{3}\,,\nonumber\\
\Pbm_{122}&=\rho^{-1}(\caA_{11223}+\caA_{13221})m_{1}m_{3}\,,\nonumber\\
\Pbm_{133}&=\rho^{-1}(\caA_{11333}+\caA_{13331})m_{1}m_{3}\,,\nonumber\\
\Pbm_{123}&=\rho^{-1}(\caA_{11232}+\caA_{12231})m_{1}m_{2}\,,\nonumber\\
\Pbm_{131}&=\rho^{-1}(\caA_{11311}m_{1}^{2}+\caA_{12312}m_{2}^{2}+\caA_{13313}m_{3}^{2})\,,\nonumber\\
\Pbm_{112}&=\rho^{-1}(\caA_{12123}+\caA_{13122})m_{2}m_{3}\,,\nonumber\\
\Pbm_{132}&=\rho^{-1}((\caA_{11322}+\caA_{12321})m_{1}m_{2}+(\caA_{12323}+\caA_{13322})m_{2}m_{3})\,,\nonumber\\
\Pbm_{113}&=\rho^{-1}(\caA_{11131}m_{1}^{2}+\caA_{12132}m_{2}^{2}+\caA_{13133}m_{3}^{2})\,,\nonumber\\
\Pbm_{121}&=\rho^{-1}(\caA_{12213}+\caA_{13212})m_{2}m_{3}\,,\nonumber\\
\Pbm_{211}&=\rho^{-1}(\caA_{11223}+\caA_{13221})m_{2}m_{3}\,,\nonumber\\
\Pbm_{222}&=\rho^{-1}(\caA_{11113}+\caA_{13111})m_{2}m_{3}\,,\nonumber\\
\Pbm_{233}&=\rho^{-1}((\caA_{21332}+\caA_{22331})m_{1}m_{2}+(\caA_{11333}+\caA_{13331})m_{2}m_{3})\,,\\
\Pbm_{223}&=\rho^{-1}(\caA_{12132}m_{1}^{2}+\caA_{11131}m_{2}^{2}+\caA_{13133}m_{3}^{2}+(\caA_{21232}+\caA_{22231})m_{1}m_{2})\,,\nonumber\\
\Pbm_{231}&=\rho^{-1}((\caA_{12321}+\caA_{11322})m_{1}m_{2}+(\caA_{21313}+\caA_{23311})m_{1}m_{3})\,,\nonumber\\
\Pbm_{212}&=\rho^{-1}(\caA_{12213}+\caA_{13212})m_{1}m_{3}\,,\nonumber\\
\Pbm_{232}&=\rho^{-1}(\caA_{12312}m_{1}^{2}+\caA_{11311}m_{2}^{2}+\caA_{13313}m_{3}^{2})\,,\nonumber\\
\Pbm_{213}&=\rho^{-1}((\caA_{12231}+\caA_{11232})m_{1}m_{2}+(\caA_{22133}+\caA_{23132})m_{2}m_{3})\,,\nonumber\\
\Pbm_{221}&=\rho^{-1}(\caA_{12123}+\caA_{13122})m_{1}m_{3}\,,\nonumber\\
\Pbm_{311}&=\rho^{-1}(\caA_{31111}m_{1}^{2}+\caA_{32112}m_{2}^{2}+\caA_{33113}m_{3}^{2}+(\caA_{31113}+\caA_{33111})m_{1}m_{3})\,,\nonumber\\
\Pbm_{322}&=\rho^{-1}(\caA_{31221}m_{1}^{2}+\caA_{31111}m_{2}^{2}+\caA_{33113}m_{3}^{2})\,,\nonumber\\
\Pbm_{333}&=\rho^{-1}(\caA_{31331}m_{1}^{2}+\caA_{31331}m_{2}^{2}+\caA_{33333}m_{3}^{2})\,,\nonumber\\
\Pbm_{323}&=\rho^{-1}(\caA_{31133}+\caA_{33131})m_{2}m_{3}\,,\nonumber\\
\Pbm_{331}&=\rho^{-1}(\caA_{31313}+\caA_{33311})m_{1}m_{3}\,,\nonumber\\
\Pbm_{312}&=\rho^{-1}((\caA_{31122}+\caA_{32121})m_{1}m_{2}-(\caA_{31213}+\caA_{33212})m_{2}m_{3})\,,\nonumber\\
\Pbm_{332}&=\rho^{-1}(\caA_{32323}+\caA_{33322})m_{2}m_{3}\,,\nonumber\\
\Pbm_{313}&=\rho^{-1}(\caA_{31133}+\caA_{33131})m_{1}m_{3}\,,\nonumber\\
\Pbm_{321}&=\rho^{-1}(\caA_{32212}m_{2}^{2}+(\caA_{32121}+\caA_{31122})m_{1}m_{2})\,.\nonumber
\end{align}
}
We show in the tabular form the components which are different from zero when $\mb=\eb_{3}$
\begin{equation}
[\Pbm]=
\left[
\begin{array}{ccccccccc}
0 &  0  &  0  &0& \Pbm_{131}& 0 & 0& \Pbm_{113}& 0\\
0 &  0  &  0  & \Pbm_{113}& 0& 0& \Pbm_{131}& 0& 0\\
  \Pbm_{311} &  \Pbm_{311}  &  \Pbm_{333}  & 0 & 0 & 0 & 0 & 0 & 0
\end{array}
\right]\,;
\end{equation}
whereas when $\mb\cdot\eb_{3}=0$ we have
\begin{equation}
[\Pbm]=
\left[
\begin{array}{ccccccccc}
0 &  0  &  0  &\Pbm_{123}& \Pbm_{131}& 0 & \Pbm_{132}& \Pbm_{113}& 0\\
0 &  0  &  \Pbm_{233}  & \Pbm_{223}& \Pbm_{231}& 0& \Pbm_{232}& \Pbm_{213}& 0\\
  \Pbm_{311} &  \Pbm_{322}  &  \Pbm_{333}  & 0 & 0 & \Pbm_{312} & 0 & 0 &\Pbm_{321}
\end{array}
\right]\,.
\end{equation}

\subsubsection{The fourth-order tensor $\Ael(\mb)$}

For the classes $4$, $\bar{4}$ and $4/m$ by (\ref{componentsHO})$_{4}$ and (\ref{4H6nonzero}) we have:
{\allowdisplaybreaks
\begin{align}\label{AELLS}
\Ael_{11}&=\rho^{-1}(\caH_{111111}m^{2}_{1}+\caH_{112112}m_{2}^{2}+\caH_{113113}m_{3}^{2}+2\caH_{111112}m_{1}m_{2})\,,\nonumber\\
\Ael_{22}&=\rho^{-1}(\caH_{112112}m_{1}^{2}+\caH_{111111}m_{2}^{2}+\caH_{113113}m_{3}^{2}-2\caH_{111112}m_{1}m_{2})\,,\nonumber\\
\Ael_{33}&=\rho^{-1}(\caH_{331331}(m_{1}^{2}+m_{2}^{2})+\caH_{333333}m_{3}^{2}+2\caH_{331332}m_{1}m_{2})\,,\nonumber\\
\Ael_{44}&=\rho^{-1}(\caH_{231231}m_{1}^{2}+\caH_{131131}m_{2}^{2}+\caH_{133133}m_{3}^{2}-2\caH_{131132}m_{1}m_{2})\,,\nonumber\\
\Ael_{55}&=\rho^{-1}(\caH_{311311}m_{1}^{2}+\caH_{312312}m_{2}^{2}+\caH_{313313}m_{3}^{2}+2\caH_{311312}m_{1}m_{2})\,,\nonumber\\
\Ael_{66}&=\rho^{-1}(\caH_{121121}m_{1}^{2}+\caH_{211211}m_{2}^{2}+\caH_{123123}m_{3}^{2}+2\caH_{121122}m_{1}m_{2})\,,\nonumber\\
\Ael_{77}&=\rho^{-1}(\caH_{312312}m_{1}^{2}+\caH_{311311}m_{2}^{2}+\caH_{313313}m_{3}^{2}-2\caH_{311312}m_{1}m_{2})\,,\nonumber\\
\Ael_{88}&=\rho^{-1}(\caH_{131131}m_{1}^{2}+\caH_{231231}m_{2}^{2}+\caH_{133133}m_{3}^{2}+2\caH_{131132}m_{1}m_{2})\,,\nonumber\\
\Ael_{99}&=\rho^{-1}(\caH_{211211}m_{1}^{2}+\caH_{212212}m_{2}^{2}+\caH_{123123}m_{3}^{2}+2\caH_{211212}m_{1}m_{2})\,,\nonumber\\
\Ael_{12}&=\rho^{-1}(\caH_{111221}m_{1}^{2}+\caH_{221111}m_{2}^{2}+\caH_{113223}m_{3}^{2}+2\caH_{111222}m_{1}m_{2})\,,\nonumber\\
\Ael_{13}&=\rho^{-1}(\caH_{111331}m_{1}^{2}+\caH_{112332}m_{2}^{2}+\caH_{113333}m_{3}^{2}+2\caH_{111332}m_{1}m_{2})\,,\nonumber\\
\Ael_{14}&=\rho^{-1}(2\caH_{111233}m_{1}m_{3}+2\caH_{112233}m_{2}m_{3})\,,\nonumber\\
\Ael_{15}&=\rho^{-1}(2\caH_{111313}m_{1}m_{3}+2\caH_{112313}m_{2}m_{3})\,,\nonumber\\
\Ael_{16}&=\rho^{-1}(\caH_{111121}m_{1}^{2}+\caH_{112122}m_{2}^{2}+\caH_{113123}m_{3}^{2}+2\caH_{111122}m_{1}m_{2})\,,\nonumber\\
\Ael_{17}&=\rho^{-1}(2\caH_{111323}m_{1}m_{3}+2\caH_{112323}m_{2}m_{3})\,,\nonumber\\
\Ael_{18}&=\rho^{-1}(2\caH_{111133}m_{1}m_{3}+2\caH_{112133}m_{2}m_{3})\,,\nonumber\\
\Ael_{19}&=\rho^{-1}(\caH_{111211}m_{1}^{2}+\caH_{112212}m_{2}^{2}+\caH_{113213}m_{3}^{2}+2\caH_{111212}m_{1}m_{2})\,,\nonumber\\
\Ael_{23}&=\rho^{-1}(\caH_{112332}m_{1}^{2}+\caH_{111331}m_{2}^{2}+\caH_{113333}m_{3}^{2}-2\caH_{112331}m_{1}m_{2})\,,\nonumber\\
\Ael_{24}&=-\rho^{-1}(2\caH_{112133}m_{1}m_{3}+2\caH_{111133}m_{2}m_{3})\,,\nonumber\\
\Ael_{25}&=\rho^{-1}(2\caH_{112323}m_{1}m_{3}-2\caH_{111323}m_{2}m_{3})\,,\nonumber\\
\Ael_{26}&=\rho^{-1}(\caH_{221121}m_{1}^{2}-\caH_{111211}m_{2}^{2}-\caH_{113213}m_{3}^{2}+2\caH_{112211}m_{1}m_{2})\,,\nonumber\\
\Ael_{27}&=-\rho^{-1}(2\caH_{112313}m_{1}m_{3}+2\caH_{111313}m_{2}m_{3})\,,\nonumber\\
\Ael_{28}&=\rho^{-1}(2\caH_{112233}m_{1}m_{3}-2\caH_{111233}m_{2}m_{3})\,,\\
\Ael_{29}&=\rho^{-1}(\caH_{221211}m_{1}^{2}-\caH_{111121}m_{2}^{2}-\caH_{113123}m_{3}^{2}+2\caH_{112121}m_{1}m_{2})\,,\nonumber\\
\Ael_{34}&=\rho^{-1}(2\caH_{331233}m_{1}m_{3}+2\caH_{331133}m_{2}m_{3})\,,\nonumber\\
\Ael_{35}&=\rho^{-1}(2\caH_{331313}m_{1}m_{3}-2\caH_{331323}m_{2}m_{3})\,,\nonumber\\
\Ael_{36}&=\rho^{-1}(\caH_{331121}m_{1}^{2}-\caH_{331211}m_{2}^{2}+\caH_{333123}m_{3}^{2}+2\caH_{331122}m_{1}m_{2})\,,\nonumber\\
\Ael_{37}&=\rho^{-1}(2\caH_{331323}m_{1}m_{3}+2\caH_{331313}m_{2}m_{3})\,,\nonumber\\
\Ael_{38}&=\rho^{-1}(2\caH_{331133}m_{1}m_{3}-2\caH_{331233}m_{2}m_{3})\,,\nonumber\\
\Ael_{39}&=\rho^{-1}(\caH_{331211}m_{1}^{2}-\caH_{331121}m_{2}^{2}-\caH_{333123}m_{3}^{2}+2\caH_{331212}m_{1}m_{2})\,,\nonumber\\
\Ael_{45}&=\rho^{-1}(\caH_{231311}m_{1}^{2}-\caH_{131321}m_{2}^{2}-\caH_{133323}m_{3}^{2}+2\caH_{231312}m_{1}m_{2})\,,\nonumber\\
\Ael_{46}&=\rho^{-1}(2\caH_{132213}m_{1}m_{3}-2\caH_{131213}m_{2}m_{3})\,,\nonumber\\
\Ael_{47}&=\rho^{-1}(\caH_{132312}m_{1}^{2}+\caH_{131311}m_{2}^{2}+\caH_{133313}m_{3}^{2}-2\caH_{132311}m_{1}m_{2})\,,\nonumber\\
\Ael_{48}&=\rho^{-1}(\caH_{231131}m_{1}^{2}-\caH_{131231}m_{2}^{2}-2\caH_{132231}m_{1}m_{2})\,,\nonumber\\
\Ael_{49}&=\rho^{-1}(2\caH_{132123}m_{1}m_{3}-2\caH_{131123}m_{2}m_{3})\,,\nonumber\\
\Ael_{56}&=\rho^{-1}(\caH_{311131}m_{1}^{2}+\caH_{132312}m_{2}^{2}+\caH_{313133}m_{3}^{2}+2\caH_{311132}m_{1}m_{2})\,,\nonumber\\
\Ael_{57}&=\rho^{-1}(\caH_{311321}m_{1}^{2}-\caH_{321311}m_{2}^{2}+2\caH_{311322}m_{1}m_{2})\,,\nonumber\\
\Ael_{58}&=\rho^{-1}(\caH_{311131}m_{1}^{2}+\caH_{132312}m_{2}^{2}+\caH_{313133}m_{3}^{2}+2\caH_{311132}m_{1}m_{2})\,,\nonumber\\
\Ael_{59}&=\rho^{-1}(2\caH_{311213}m_{1}m_{3}+2\caH_{312213}m_{2}m_{3})\,,\nonumber\\
\Ael_{67}&=\rho^{-1}(2\caH_{121323}m_{1}m_{3}+2\caH_{122323}m_{2}m_{3})\,,\nonumber\\
\Ael_{68}&=\rho^{-1}(2\caH_{121133}m_{1}m_{3}+2\caH_{122133}m_{2}m_{3})\,,\nonumber\\
\Ael_{69}&=\rho^{-1}(\caH_{121211}m_{1}^{2}+\caH_{211121}m_{2}^{2}+\caH_{123213}m_{3}^{2}+2\caH_{121212}m_{1}m_{2})\,,\nonumber\\
\Ael_{78}&=\rho^{-1}(\caH_{311131}m_{1}^{2}+\caH_{312132}m_{2}^{2}+\caH_{313133}m_{3}^{2}+2\caH_{311132}m_{1}m_{2})\,,\nonumber\\
\Ael_{59}&=\rho^{-1}(2\caH_{311213}m_{1}m_{3}+2\caH_{123321}m_{2}m_{3})\,,\nonumber\\
\Ael_{89}&=\rho^{-1}(2\caH_{131213}m_{1}m_{3}+2\caH_{132213}m_{2}m_{3})\,.\nonumber
\end{align}
}
From (\ref{AELLS}) we have that when $\mb=\eb_{3}$ $\Ael$ has the same 13 independent components of of $\Bel$. For $\mb\cdot\eb_{3}=0$ we have instead that there are  26 independent components.

For the classes $422$, $4mm$, $4/mmm$ and for the polar tensors of the class $\bar{4}2m$, by using (\ref{4H6lower}) into (\ref{AELLS}) we get
{\allowdisplaybreaks
\begin{align}\label{AELHS}
&\Ael_{11}=\rho^{-1}(\caH_{111111}m^{2}_{1}+\caH_{112112}m_{2}^{2}+\caH_{113113}m_{3}^{2})\,,&&\Ael_{14}=\rho^{-1}2\caH_{112233}m_{2}m_{3}\,,\nonumber\\
&\Ael_{22}=\rho^{-1}(\caH_{112112}m_{1}^{2}+\caH_{111111}m_{2}^{2}+\caH_{113113}m_{3}^{2})\,,&&\Ael_{15}=\rho^{-1}2\caH_{111313}m_{1}m_{3}\,,\nonumber\\
&\Ael_{33}=\rho^{-1}(\caH_{331331}(m_{1}^{2}+m_{2}^{2})+\caH_{333333}m_{3}^{2})\,,&&\Ael_{16}=\rho^{-1}2\caH_{111122}m_{1}m_{2}\,,\nonumber\\
&\Ael_{44}=\rho^{-1}(\caH_{231231}m_{1}^{2}+\caH_{131131}m_{2}^{2}+\caH_{133133}m_{3}^{2})\,,&&\Ael_{17}=\rho^{-1}2\caH_{112323}m_{2}m_{3}\,,\nonumber\\
&\Ael_{55}=\rho^{-1}(\caH_{311311}m_{1}^{2}+\caH_{312312}m_{2}^{2}+\caH_{313313}m_{3}^{2})\,,&&\Ael_{18}=\rho^{-1}2\caH_{111133}m_{1}m_{3}\,,\nonumber\\
&\Ael_{66}=\rho^{-1}(\caH_{121121}m_{1}^{2}+\caH_{211211}m_{2}^{2}+\caH_{123123}m_{3}^{2})\,,&&\Ael_{19}=\rho^{-1}2\caH_{111212}m_{1}m_{2}\,,\nonumber\\
&\Ael_{77}=\rho^{-1}(\caH_{312312}m_{1}^{2}+\caH_{311311}m_{2}^{2}+\caH_{313313}m_{3}^{2})\,,&&\Ael_{24}=-\rho^{-1}2\caH_{111133}m_{2}m_{3}\,,\nonumber\\
&\Ael_{88}=\rho^{-1}(\caH_{131131}m_{1}^{2}+\caH_{231231}m_{2}^{2}+\caH_{133133}m_{3}^{2})\,,&&\Ael_{25}=\rho^{-1}2\caH_{112323}m_{1}m_{3}\,,\nonumber\\
&\Ael_{99}=\rho^{-1}(\caH_{211211}m_{1}^{2}+\caH_{212212}m_{2}^{2}+\caH_{123123}m_{3}^{2})\,,&&\Ael_{26}=\rho^{-1}2\caH_{112211}m_{1}m_{2}\,,\\
&\Ael_{12}=\rho^{-1}(\caH_{111221}m_{1}^{2}+\caH_{221111}m_{2}^{2}+\caH_{113223}m_{3}^{2})\,,&&\Ael_{27}=-\rho^{-1}2\caH_{111313}m_{2}m_{3}\,,\nonumber\\
&\Ael_{13}=\rho^{-1}(\caH_{111331}m_{1}^{2}+\caH_{112332}m_{2}^{2}+\caH_{113333}m_{3}^{2})\,,&&\Ael_{28}=\rho^{-1}2\caH_{112233}m_{1}m_{3}\,,\nonumber\\
&\Ael_{23}=\rho^{-1}(\caH_{112332}m_{1}^{2}+\caH_{111331}m_{2}^{2}+\caH_{113333}m_{3}^{2})\,,&&\Ael_{29}=\rho^{-1}2\caH_{112121}m_{1}m_{2}\,,\nonumber\\
&\Ael_{47}=\rho^{-1}(\caH_{132312}m_{1}^{2}+\caH_{131311}m_{2}^{2}+\caH_{133313}m_{3}^{2})\,,&&\Ael_{34}=\rho^{-1}2\caH_{331133}m_{2}m_{3}\,,\nonumber\\
&\Ael_{56}=\rho^{-1}(\caH_{311131}m_{1}^{2}+\caH_{132312}m_{2}^{2}+\caH_{313133}m_{3}^{2})\,,&&\Ael_{35}=\rho^{-1}2\caH_{331313}m_{1}m_{3}\,,\nonumber\\
&\Ael_{58}=\rho^{-1}(\caH_{311131}m_{1}^{2}+\caH_{132312}m_{2}^{2}+\caH_{313133}m_{3}^{2})\,,&&\Ael_{36}=\rho^{-1}2\caH_{331122}m_{1}m_{2}\,,\nonumber\\
&\Ael_{69}=\rho^{-1}(\caH_{121211}m_{1}^{2}+\caH_{211121}m_{2}^{2}+\caH_{123213}m_{3}^{2})\,,&&\Ael_{37}=\rho^{-1}2\caH_{331313}m_{2}m_{3}\,,\nonumber\\
&\Ael_{78}=\rho^{-1}(\caH_{311131}m_{1}^{2}+\caH_{312132}m_{2}^{2}+\caH_{313133}m_{3}^{2})\,,&&\Ael_{38}=\rho^{-1}2\caH_{331133}m_{1}m_{3}\,,\nonumber\\
&\Ael_{39}=\rho^{-1}2\caH_{331212}m_{1}m_{2}\,,&&\Ael_{45}=\rho^{-1}2\caH_{231312}m_{1}m_{2}\,,\nonumber\\
&\Ael_{46}=\rho^{-1}2\caH_{132213}m_{1}m_{3}\,,&&\Ael_{48}=-\rho^{-1}2\caH_{132231}m_{1}m_{2}\,,\nonumber\\
&\Ael_{49}=\rho^{-1}2\caH_{132123}m_{1}m_{3}\,,&&\Ael_{57}=\rho^{-1}2\caH_{311322}m_{1}m_{2}\,,\nonumber\\
&\Ael_{59}=\rho^{-1}2\caH_{312213}m_{2}m_{3}\,,&&\Ael_{67}=\rho^{-1}2\caH_{121323}m_{1}m_{3}\,,\nonumber\\
&\Ael_{68}=\rho^{-1}2\caH_{122133}m_{2}m_{3}\,,&&\Ael_{59}=\rho^{-1}2\caH_{123321}m_{2}m_{3}\,,\nonumber\\
&\Ael_{89}=\rho^{-1}2\caH_{132213}m_{2}m_{3}\,.&&\nonumber
\end{align}
}
For the direction of propagation $\mb=\eb_{3}$ the tensor $\Ael$ has the same non-null independent components of $\Bel$, whereas in the case $\mb\cdot\eb_{3}=0$ there are 26 independent components.

\subsection{The components of the eigentensors $\Cb_{k}$ and the eigenvectors $w_{k}$}\label{quinta}

These components are calculated by the means of the free web application WIMS (wims.unice.fr).

\subsubsection{Subspace $\caU_{1}$}\label{eigencompU1}

Here $\bar{a}=a-\omega^{2}$, $\bar{b}=b-\omega^{2}$ and $\bar{c}=c-\omega^{2}$, the real coefficients $a, b, c, d, e$ and $f$ being defined by (\ref{compoU1micro}). All the components must be divided by the norm $\|\Cb_{k}\|=(\alpha_{k}^{2}+\beta_{k}^{2}+\gamma_{k}^{2}+\delta_{k}^{2}+\epsilon_{k}^{2})^{1/2}$, $k=1\,,\ldots 5$.
\begin{itemize}
\item $\Cb_{1}$ and $\Cb_{2}$:
\begin{align}
\alpha_{1}&=-\beta_{2}=(hd-\bar{a}g)(l^2+\bar{b}f)+(h\bar{a}-gd)(l^2-\bar{b}\bar{c})+hg(2el-\bar{b}g)+e^2(f-\bar{c})(g+h)+el(g^2-3h^2+(f+\bar{c})(\bar{a}-d))+\bar{b}h^3\,,\nonumber\\
\beta_{1}&=-\alpha_{2}=-(h\bar{a}+gd)(l^2+\bar{b}f)-(hd+\bar{a}g)(l^2-\bar{b}\bar{c})+e^2(g+h)(f-\bar{c})+\bar{b}g(h^2-g^2)+el(2gh-h^2+3g^2+(f+\bar{c})(d-\bar{a}))\,,\nonumber\\
\gamma_{1}&=-\gamma_{2}=-l(\bar{a}+d)(h^2+g^2)-egh(h+g)+e(d-\bar{a})(f-\bar{c})(g-h)+e(h^3+g^3)+l(f+\bar{c})(\bar{a}^2-d^2)\,,\nonumber\\
\delta_{1}&=\delta_{2}=(\bar{a}^2-d^2)(2l^2+\bar{b}(f-c))+(h^2+g^2)(\bar{b}(a+d)-2e^2)+2(\bar{a}-d)(h(g\bar{b}-el)-e^2\bar{c})+2(\bar{a}+d)e^2f\,,\nonumber\\
\epsilon_{1}&=-\epsilon_{2}=\bar{b}((\bar{a}^2-d^2)(f+c)+ (h+g)^2(d-\bar{a}))+2(h(el-g\bar{b})+2e^2\bar{c})(d-\bar{a})-2e^2f(\bar{a}+d)\,.\nonumber
\end{align}
\item $\Cb_{3}$:
\begin{align}
\alpha_{3}&=\beta_{3}=e(\bar{c}-f)-l(h+g)\,,\quad\gamma_{3}=(h+g)^2+(d+\bar{a})(f-\bar{c})\,,\quad\delta_{3}=-\epsilon_{3}=l(d+\bar{a})-e(h+g)\,.\nonumber
\end{align}
\item $\Cb_{4}$ and $\Cb_{5}$:
\begin{align}
\alpha_{4}&=\beta_{5}=(\bar{b}d-e^2)(f^2-\bar{c}^2)+(l^2-\bar{b}f)(h^2+g^2)+2(\bar{b}\bar{c}-l^2)gh+2l(dl-e(g+h))(f+\bar{c})  \,,\nonumber\\
\beta_{4}&=\alpha_{5}=(\bar{a}\bar{b}-e^2)(\bar{c}^2-f^2)+(l^2-\bar{b}\bar{c})(h^2+g^2)+2(\bar{b}f-l^2)gh-2l(\bar{a}l-e(g+h))(f+\bar{c})\,,\nonumber\\
\gamma_{4}&=\gamma_{5}=e(h-g)^2(\bar{c}-f)-l(h^3+g^3-gh(h+g))+e(\bar{a}-d)(f^2-\bar{c}^2)+l(g+h)(\bar{a}-d)(f+\bar{c})\,,\nonumber\\ 
\delta_{4}&=-\delta_{5}= (h-g)((2l^2+\bar{b}(f-\bar{c}))(\bar{a}+d)+2e^2(\bar{c}-f)+\bar{b}(h^2-g^2)+4l(h+g))\,,\nonumber\\ 
\epsilon_{4}&=-\epsilon_{5}=(\bar{b}(h-g)-2el)(h^2-g^2+(\bar{a}-d)(f+\bar{c}))\,,\nonumber 
\end{align}
\end{itemize}

\subsubsection{Subspace $\caM_{1}$}\label{eigencompM1}

The components of $A$ are defined by (\ref{compoM1}); we denote $\bar{a}=a-\omega^{2}$, $\bar{f}=f-\omega^{2}$, $\bar{g}=g-\omega^{2}$ and $\bar{d}=d-\omega^{2}$. All the components must be divided by $\|w_{k}\|=(\alpha_{k}^{2}+\beta_{k}^{2}+\gamma_{k}^{2}+\delta_{k}^{2}+\epsilon_{k}^{2}+\theta_{k}^{2})^{1/2}$, $k=1\,,\ldots 6$.
\begin{itemize}
\item $w_{1}$ and $w_{2}$:
\begin{align}
\alpha_{1}&=-\alpha_{2}=-(((2l^2-\bar{g}h-\bar{f}\bar{g})(p-n)^2+2(\bar{d}+e)(h-\bar{f})l^2+\bar{g}(\bar{d}+e)(\bar{f}^2-h^2))q \nonumber\\
&+c(-lp^3+(ln+hm+\bar{f}m)p^2 +(ln^2-2mn(h+\bar{f})+(\bar{d}+ e)(\bar{f}-h)l)p\nonumber\\
&-ln^3 +(hm+\bar{f}m)n^2+((-\bar{d}- e)h+(\bar{d}+ e)\bar{f})ln+(\bar{d}+ e)h^2m +(-\bar{d}- e)\bar{f}^2m) +b(\bar{g}p^3+(-\bar{g}n-2lm)p^2\nonumber\\
&+(-\bar{g}n^2+4lmn+(\bar{d}+ e)\bar{g}h+(-\bar{d}- e)\bar{f}\bar{g})p +\bar{g}n^3-2lmn^2+((\bar{d}+ e)\bar{g}h+(-\bar{d}- e)\bar{f}\bar{g})n +l((-2\bar{d}-2 e)hm+(2\bar{d}+2 e)\bar{f}m)))\,,\nonumber\\
\beta_{1}&=-\gamma_{2}=-(((2l^2-\bar{g}h-\bar{f}\bar{g})p+(-2l^2+\bar{g}h+\bar{f}\bar{g})n)q^2 +(c(-3lp^2+(2ln+2hm+2\bar{f}m)p+ln^2\nonumber\\
&+(-2hm-2\bar{f}m)n +((-\bar{d}- e)h+(\bar{d}+ e)\bar{f})l) +b(3\bar{g}p^2+(-2\bar{g}n-4lm)p-\bar{g}n^2+4lmn+(\bar{d}+ e)\bar{g}h+(-\bar{d}- e)\bar{f}\bar{g})) q\nonumber\\
& +\bar{a}(-\bar{g}p^3+3lmp^2 +(\bar{g}n^2-2lmn+\bar{f}(\bar{d}\bar{g}-m^2)+h(-m^2- e\bar{g})+( e-\bar{d})l^2)p -lmn^2+(h(m^2-\bar{d}\bar{g})+\bar{f}(m^2+ e\bar{g})\nonumber\\
&+(\bar{d}- e)l^2)n +l((\bar{d}+ e)hm+(-\bar{d}- e)\bar{f}m)) +c^2(p^3+(-n^2+ eh-\bar{d}\bar{f})p+(\bar{d}h- e\bar{f})n) +bc (-3mp^2+(2mn+(2\bar{d}-2 e)l)p\nonumber\\
&+mn^2+(2 e-2\bar{d})ln+(-\bar{d}- e)hm +(\bar{d}+ e)\bar{f}m) +b^2((2m^2+( e-\bar{d})\bar{g})p+((\bar{d}- e)\bar{g}-2m^2)n))\,,\nonumber\\ 
\gamma_{1}&=-\beta_{2}= (((2l^2-\bar{g}h-\bar{f}\bar{g})p+(-2l^2+\bar{g}h+\bar{f}\bar{g})n)q^2 +(c(-lp^2+(-2ln+2hm+2\bar{f}m)p+3ln^2+(-2hm-2\bar{f}m)n \nonumber\\
&+((\bar{d}+ e)h+(-\bar{d}- e)\bar{f})l) +b(\bar{g}p^2+(2\bar{g}n-4lm)p-3\bar{g}n^2+4lmn+(-\bar{d}- e)\bar{g}h+(\bar{d}+ e)\bar{f}\bar{g})) q +\bar{a}((lm-\bar{g}n)p^2\nonumber\\
&+(2lmn+h(\bar{d}\bar{g}-m^2)+\bar{f}(-m^2- e\bar{g})+( e-\bar{d})l^2)p +\bar{g}n^3-3lmn^2 +(\bar{f}(m^2-\bar{d}\bar{g})+h(m^2+ e\bar{g})+(\bar{d}- e)l^2)n\nonumber\\
& +l((-\bar{d}- e)hm+(\bar{d}+ e)\bar{f}m)) +c^2(np^2+( e\bar{f}-\bar{d}h)p-n^3+(\bar{d}\bar{f}- eh)n) +bc (-mp^2+((2\bar{d}-2 e)l-2mn)p+3mn^2\nonumber\\
&+(2 e-2\bar{d})ln+(\bar{d}+ e)hm +(-\bar{d}- e)\bar{f}m)+b^2((2m^2+( e-\bar{d})\bar{g})p+((\bar{d}- e)\bar{g}-2m^2)n))\,,\nonumber\\ 
\delta_{1}&=-\delta_{2}=-((b(-2lp^2+4lnp-2ln^2+((-2\bar{d}-2 e)h+(2\bar{d}+2 e)\bar{f})l) +c((h+\bar{f})p^2+(-2h-2\bar{f})np+(h+\bar{f})n^2+(\bar{d}+ e)h^2\nonumber\\
&+(-\bar{d}- e)\bar{f}^2)) q +\bar{a}(lp^3+(-ln-hm-\bar{f}m)p^2 +(-ln^2+(2hm+2\bar{f}m)n+((\bar{d}+ e)h+(-\bar{d}- e)\bar{f})l)p+ln^3\nonumber\\
& +(-hm-\bar{f}m)n^2+((\bar{d}+ e)h+(-\bar{d}- e)\bar{f})ln+(-\bar{d}- e)h^2m +(\bar{d}+ e)\bar{f}^2m) +bc (-p^3+np^2+(n^2+(-\bar{d}- e)h+(\bar{d}+ e)\bar{f})p\nonumber\\
&-n^3 +((-\bar{d}- e)h+(\bar{d}+ e)\bar{f})n) +b^2(2mp^2-4mnp+2mn^2+(2\bar{d}+2 e)hm+(-2\bar{d}-2 e)\bar{f}m))\,,\nonumber\\ 
\epsilon_{1}&=\theta_{2}= (((2h-2\bar{f})l^2-\bar{g}h^2+\bar{f}^2\bar{g})q^2 +(c((2\bar{f}-2h)lp+(2\bar{f}-2h)ln+2h^2m-2\bar{f}^2m) +b((2\bar{g}h-2\bar{f}\bar{g})p+(2\bar{g}h-2\bar{f}\bar{g})n\nonumber\\
&+l(4\bar{f}m-4hm))) q +\bar{a}((l^2-\bar{g}h)p^2+((2\bar{f}\bar{g}-2l^2)n+l(2hm-2\bar{f}m))p+(l^2-\bar{g}h)n^2 +l(2hm-2\bar{f}m)n+\bar{f}^2(m^2+ e\bar{g})\nonumber\\
&+h^2(-m^2- e\bar{g}) +(2 eh-2 e\bar{f})l^2) +bc (-2lp^2+(4ln-2hm+2\bar{f}m)p-2ln^2+(2\bar{f}m-2hm)n +(4 e\bar{f}-4 eh)l)\nonumber\\
& +c^2(hp^2-2\bar{f}np+hn^2+ eh^2- e\bar{f}^2) +b^2(\bar{g}p^2-2\bar{g}np+\bar{g}n^2+h(2m^2+2 e\bar{g})+\bar{f}(-2m^2-2 e\bar{g})))\,,\nonumber\\ 
\theta_{1}&=\epsilon_{2}=(((2h-2\bar{f})l^2-\bar{g}h^2+\bar{f}^2\bar{g})q^2 +(c((2\bar{f}-2h)lp+(2\bar{f}-2h)ln+2h^2m-2\bar{f}^2m) +b((2\bar{g}h-2\bar{f}\bar{g})p+(2\bar{g}h-2\bar{f}\bar{g})n\nonumber\\
&+l(4\bar{f}m-4hm))) q +\bar{a}((\bar{f}\bar{g}-l^2)p^2+((2l^2-2\bar{g}h)n+l(2hm-2\bar{f}m))p+(\bar{f}\bar{g}-l^2)n^2 +l(2hm-2\bar{f}m)n+\bar{f}^2(m^2-\bar{d}\bar{g})\nonumber\\
&+h^2(\bar{d}\bar{g}-m^2) +(2\bar{d}\bar{f}-2\bar{d}h)l^2) +bc (2lp^2+(-4ln-2hm+2\bar{f}m)p+2ln^2+(2\bar{f}m-2hm)n +(4\bar{d}h-4\bar{d}\bar{f})l)\nonumber\\
&+b^2(-\bar{g}p^2+2\bar{g}np-\bar{g}n^2+h(2m^2-2\bar{d}\bar{g})+\bar{f}(2\bar{d}\bar{g}-2m^2)) +c^2(-\bar{f}p^2+2hnp-\bar{f}n^2-\bar{d}h^2+\bar{d}\bar{f}^2))\,;\nonumber
\end{align}

\item $w_{3}$:
\begin{align}
\alpha_{3}&=2(ql+bm)(p+n)-c(p+n)^2+(c(\bar{d}-e)-2qm)(\bar{f}+h)-2bl(\bar{d}-e)\,,\nonumber\\
\beta_{3}&=\gamma_{3}=(qc-\bar{a}m)(p+n)+2q(bm-lq)+(\bar{a}l+bc)(e-\bar{d})\,,\nonumber\\
\delta_{3}&=2q^2(h+\bar{f})-4bq(p+n)+\bar{a}(p+n)^2+\bar{a}(e-\bar{d})(h+\bar{f}) +2b^2(\bar{d}-e)\,,\nonumber\\
\epsilon_{3}&=-\theta_{3}=(\bar{a}m-qc)(\bar{f}+h)+(bc-\bar{a}l)(p+n)-2b(bm+ql)\,.\nonumber
\end{align}

\item $w_{4}$ and $w_{5}$:
\begin{align}
\alpha_{4}&=\alpha_{5}=-((\bar{g}p^3+(-\bar{g}n-2lm)p^2 +(-\bar{g}n^2+4lmn+(\bar{d}+ e)\bar{g}h+(-\bar{d}- e)\bar{f}\bar{g})p+\bar{g}n^3-2lmn^2 +((\bar{d}+ e)\bar{g}h\nonumber\\
&+(-\bar{d}- e)\bar{f}\bar{g})n +l((-2\bar{d}-2 e)hm+(2\bar{d}+2 e)\bar{f}m)) q +c(-mp^3+(mn+(\bar{d}- e)l)p^2 +(mn^2+(2 e-2\bar{d})ln+(-\bar{d}- e)hm\nonumber\\
&+(\bar{d}+ e)\bar{f}m)p-mn^3 +(\bar{d}- e)ln^2+((-\bar{d}- e)hm+(\bar{d}+ e)\bar{f}m)n +((\bar{d}^2- e^2)h+( e^2-\bar{d}^2)\bar{f})l) +b((2m^2+( e-\bar{d})\bar{g})p^2\nonumber\\
&+((2\bar{d}-2 e)\bar{g}-4m^2)np +(2m^2+( e-\bar{d})\bar{g})n^2 +h((2\bar{d}+2 e)m^2+( e^2-\bar{d}^2)\bar{g}) +\bar{f}((-2\bar{d}-2 e)m^2+(\bar{d}^2- e^2)\bar{g})))\,,\nonumber\\  
&+(\bar{d}- e)ln^2+(\bar{d}+ e)(\bar{f}-h)mn +((\bar{d}^2- e^2)(h-\bar{f}))l) +b((2m^2+( e-\bar{d})\bar{g})p^2+((2\bar{d}-2 e)\bar{g}-4m^2)np\nonumber\\
& +(2m^2+( e-\bar{d})\bar{g})n^2 +h((2\bar{d}+2 e)m^2+( e^2-\bar{d}^2)\bar{g}) +\bar{f}((-2\bar{d}-2 e)m^2+(\bar{d}^2- e^2)\bar{g})))\,,\nonumber\\
\beta_{4}&=\gamma_{5}= (\bar{g}q^2(p-n)^2+2(\bar{d}+e)(\bar{g}h-l^2)q^2 +(c(-2mp^2+(4mn+(2\bar{d}+2 e)l)p-2mn^2+(2\bar{d}+2 e)ln\nonumber\\
& +(-4\bar{d}-4 e)hm) +b((-2\bar{d}-2 e)\bar{g}p+(-2\bar{d}-2 e)\bar{g}n+(4\bar{d}+4 e)lm)) q +\bar{a}((m^2+ e\bar{g})p^2+((2\bar{d}\bar{g}-2m^2)n+(-2\bar{d}-2 e)lm)p \nonumber\\
&+(m^2+ e\bar{g})n^2+(-2\bar{d}-2 e)lmn +h((2\bar{d}+2 e)m^2+( e^2-\bar{d}^2)\bar{g})+(\bar{d}^2- e^2)l^2) +c^2(- ep^2-2\bar{d}np- en^2+(\bar{d}^2- e^2)h)\nonumber\\
& +bc((2\bar{d}+2 e)mp+(2\bar{d}+2 e)mn+(2 e^2-2\bar{d}^2)l) +b^2((-2\bar{d}-2 e)m^2+(\bar{d}^2- e^2)\bar{g})\,,\nonumber\\
\gamma_{4}&=\beta_{5}=(\bar{g}q^2(p-n)^2+2(\bar{d}+e)(l^2-\bar{f}\bar{g})q^2 +(c(-2mp^2+(4mn+(-2\bar{d}-2 e)l)p-2mn^2+(-2\bar{d}-2 e)ln +(4\bar{d}+4 e)\bar{f}m)\nonumber\\
& +b((2\bar{d}+2 e)\bar{g}p+(2\bar{d}+2 e)\bar{g}n+(-4\bar{d}-4 e)lm)) q +\bar{a}((m^2-\bar{d}\bar{g})p^2+((-2m^2-2 e\bar{g})n+(2\bar{d}+2 e)lm)p +(m^2-\bar{d}\bar{g})n^2\nonumber\\
&+(2\bar{d}+2 e)lmn +\bar{f}((-2\bar{d}-2 e)m^2+(\bar{d}^2- e^2)\bar{g})+( e^2-\bar{d}^2)l^2) +c^2(\bar{d}p^2+2 enp+\bar{d}n^2+( e^2-\bar{d}^2)\bar{f})\nonumber\\
& +bc((-2\bar{d}-2 e)mp+(-2\bar{d}-2 e)mn+(2\bar{d}^2-2 e^2)l) +b^2((2\bar{d}+2 e)m^2+( e^2-\bar{d}^2)\bar{g})\,,\nonumber\\ 
\delta_{4}&=\delta_{5} -(2l(p-n)^2q^2+2(\bar{d}+e)(h-\bar{f})lq^2 +(c(-p^3+np^2+(n^2+(-\bar{d}- e)h+(\bar{d}+ e)\bar{f})p-n^3 +((-\bar{d}- e)h\nonumber\\
&+(\bar{d}+ e)\bar{f})n) +b(-2mp^2+4mnp-2mn^2+(-2\bar{d}-2 e)hm+(2\bar{d}+2 e)\bar{f}m)) q +\bar{a}(mp^3+(( e-\bar{d})l-mn)p^2\nonumber\\
& +(-mn^2+(2\bar{d}-2 e)ln+(\bar{d}+ e)hm+(-\bar{d}- e)\bar{f}m)p+mn^3 +( e-\bar{d})ln^2+((\bar{d}+ e)hm+(-\bar{d}- e)\bar{f}m)n\nonumber\\
& +(( e^2-\bar{d}^2)h+(\bar{d}^2- e^2)\bar{f})l) +bc ((\bar{d}- e)p^2+(2 e-2\bar{d})np+(\bar{d}- e)n^2+(\bar{d}^2- e^2)h+( e^2-\bar{d}^2)\bar{f})) \,,\nonumber\\ 
\epsilon_{4}&=-\theta_{5}=\beta_{1} \,,\nonumber\\
\theta_{4}&=-\epsilon_{5}= -\gamma_{1}\,;\nonumber
\end{align}

\item $w_{6}$:
\begin{align}
\alpha_{6}&=\bar{g}(p+n)^2-4lm(p+n)+(h+\bar{f})(2m^2+(e-\bar{d})\bar{g})+2l^2(\bar{d}-e)\,,\nonumber\\
\beta_{6}&=\gamma_{6}=l(2mq+c( e-\bar{d}))+(p+n)(cm-\bar{g}q)+b((\bar{d}- e)\bar{g}-2m^{2})\,,\nonumber\\
\delta_{6}& =-(f+h)(2mq+c( e-\bar{d}))-2(p+n)(lq+bm)+c(p+n)^2+2bl(\bar{d}-e)\,,\nonumber\\
\epsilon_{6}&=-\theta_{6}=(h+\bar{f})(\bar{g}q-cm)+(cl-b\bar{g})(p+n)+2l(bm-lq)\,,\nonumber
\end{align}
\end{itemize}

\subsubsection{Subspace $\caM_{2}$}\label{eigencompM2}

The components of $A$ are defined by (\ref{compoM2}); we denote $\bar{a}=a-\omega^{2}$,  $\bar{e}=e-\omega^{2}$, $\bar{f}=f-\omega^{2}$. All the components must be divided by $\|w_{k}\|=(\alpha_{k}^{2}+\beta_{k}^{2}+\gamma_{k}^{2}+\delta_{k}^{2}+\epsilon_{k}^{2}+\theta_{k}^{2})^{1/2}$, $k=7\,,\ldots 12$.

\begin{itemize}
\item $w_{7}$
\begin{align}
\alpha_{7}&= ( \bar{e}hn^3+(-2ghm+c(2h^2+ \bar{e}\bar{f})- \bar{e}dg)n^2 +(\bar{f}hm^2+(d(2g^2-2h^2)-2c\bar{f}g)m +\bar{a}(h^3+(g^2- \bar{e}\bar{f})h)+b(-gh^2-g^3+ \bar{e}\bar{f}g)\nonumber\\
& -4cdgh+3c^2\bar{f}h+ \bar{e}d^2h) n+(c\bar{f}^2-d\bar{f}g)m^2 +(b(\bar{f}h^2+\bar{f}g^2- \bar{e}\bar{f}^2)+2d^2gh-2cd\bar{f}h)m +d(b(( \bar{e}\bar{f}-g^2)h-h^3)\nonumber\\
&+\bar{a}(-gh^2-g^3+ \bar{e}\bar{f}g)-3c^2\bar{f}g) +\bar{a}c(\bar{f}h^2+\bar{f}g^2- \bar{e}\bar{f}^2)+cd^2(2g^2+ \bar{e}\bar{f})- \bar{e}d^3g +c^3\bar{f}^2) \,,\nonumber\\ 
\beta_{7}&= ( \bar{e}gn^3+((-2g^2- \bar{e}\bar{f})m+2cgh+ \bar{e}dh)n^2 +(3\bar{f}gm^2+(-4dgh-2c\bar{f}h)m+b(( \bar{e}\bar{f}-g^2)h-h^3) +\bar{a}(gh^2+g^3- \bar{e}\bar{f}g)\nonumber\\
&+cd(2h^2-2g^2) +c^2\bar{f}g+ \bar{e}d^2g) n-\bar{f}^2m^3+3d\bar{f}hm^2 +(\bar{a}(-\bar{f}h^2-\bar{f}g^2+ \bar{e}\bar{f}^2)+d^2(-2h^2- \bar{e}\bar{f})+2cd\bar{f}g -c^2\bar{f}^2) m\nonumber\\
&+d(\bar{a}(h^3+(g^2- \bar{e}\bar{f})h)+b(gh^2+g^3- \bar{e}\bar{f}g)+c^2\bar{f}h) +bc(-\bar{f}h^2-\bar{f}g^2+ \bar{e}\bar{f}^2)-2cd^2gh+ \bar{e}d^3h) \,,\nonumber\\ 
\gamma_{7}&= -(b(h^2-g^2)n^2+(2b\bar{f}gm-4bdgh+2bc\bar{f}h)n-b\bar{f}^2m^2 +2bd\bar{f}hm+bd^2(g^2-h^2)-2bcd\bar{f}g +bc^2\bar{f}^2) \,,\nonumber\\ 
\delta_{7}&= -( \bar{e}mn^3+(-2gm^2+2chm- \bar{e}bh- \bar{e}\bar{a}g+ \bar{e}cd)n^2 +(\bar{f}m^3-2dhm^2 +(\bar{a}(h^2+3g^2- \bar{e}\bar{f})-4cdg+c^2\bar{f}+ \bar{e}d^2)m\nonumber\\
& +c(b(-h^2-g^2- \bar{e}\bar{f})-2\bar{a}gh) +d(2c^2h+2 \bar{e}bg)) n+(b\bar{f}h-\bar{a}\bar{f}g+cd\bar{f})m^2 +(d(b(-h^2-g^2- \bar{e}\bar{f})+2\bar{a}gh)-2cd^2h+2bc\bar{f}g)m\nonumber\\
& +d(\bar{a}c(h^2+3g^2- \bar{e}\bar{f})+c^3\bar{f})+b^2(gh^2+g^3- \bar{e}\bar{f}g) +\bar{a}^2(-gh^2-g^3+ \bar{e}\bar{f}g)+c^2(-b\bar{f}h-\bar{a}\bar{f}g) +d^2( \bar{e}bh-2c^2g- \bar{e}\bar{a}g)+ \bar{e}cd^3) \,,\nonumber\\  
\epsilon_{7}&= -( \bar{e}cn^3+((-2cg- \bar{e}d)m+2c^2h+ \bar{e}\bar{a}h+ \bar{e}bg)n^2 +((2dg+c\bar{f})m^2+(b(-h^2-g^2- \bar{e}\bar{f})-2\bar{a}gh-4cdh)m +\bar{a}c(3h^2+g^2- \bar{e}\bar{f})\nonumber\\
& +d(2 \bar{e}bh-2c^2g)+c^3\bar{f}+ \bar{e}cd^2) n-d\bar{f}m^3+(\bar{a}\bar{f}h+2d^2h+b\bar{f}g)m^2 +(d(\bar{a}(-3h^2-g^2+ \bar{e}\bar{f})-c^2\bar{f}) -2bc\bar{f}h+2cd^2g- \bar{e}d^3) m\nonumber\\
&+\bar{a}^2(h^3+(g^2- \bar{e}\bar{f})h)+b^2(( \bar{e}\bar{f}-g^2)h-h^3) +cd(b(h^2+g^2+ \bar{e}\bar{f})-2\bar{a}gh)+c^2(\bar{a}\bar{f}h-b\bar{f}g) +d^2( \bar{e}\bar{a}h- \bar{e}bg))\,,\nonumber\\ 
\theta_{7}&=( \bar{e}n^4+(2ch-2gm)n^3 +(\bar{f}m^2-2dhm+\bar{a}(h^2+g^2-2 \bar{e}\bar{f})-2bgh-2cdg+c^2\bar{f} +2 \bar{e}d^2) n^2 +((2b\bar{f}h+2\bar{a}\bar{f}g-2d^2g)m\nonumber\\
&+bd(2g^2-2h^2) +c(-2\bar{a}\bar{f}h-2b\bar{f}g) +2cd^2h) n+(d^2\bar{f}-\bar{a}\bar{f}^2)m^2 +(d(2\bar{a}\bar{f}h-2b\bar{f}g)-2d^3h+2bc\bar{f}^2)m +d^2(\bar{a}(h^2+g^2-2 \bar{e}\bar{f})\nonumber\\
&+2bgh+c^2\bar{f}) +b^2(\bar{f}h^2+\bar{f}g^2- \bar{e}\bar{f}^2)+\bar{a}^2(-\bar{f}h^2-\bar{f}g^2+ \bar{e}\bar{f}^2) +cd(2\bar{a}\bar{f}g-2b\bar{f}h)-2cd^3g-\bar{a}c^2\bar{f}^2+ \bar{e}d^4)\,;\nonumber
\end{align}

\item $w_{8}$
\begin{align}
\alpha_{8}&= ( \bar{e}^2n^3+( \bar{e}ch-3 \bar{e}gm)n^2 +((-2h^2+2g^2+ \bar{e}\bar{f})m^2-2cghm +\bar{a}( \bar{e}h^2+ \bar{e}g^2- \bar{e}^2\bar{f}) +c^2( \bar{e}\bar{f}-2h^2)-2 \bar{e}bgh -2 \bar{e}cdg+ \bar{e}^2d^2) n\nonumber\\
&-\bar{f}gm^3+(2dgh-c\bar{f}h)m^2 +(b((g^2+ \bar{e}\bar{f})h-h^3)+\bar{a}(gh^2-g^3+ \bar{e}\bar{f}g)+2cdg^2 -c^2\bar{f}g- \bar{e}d^2g) m+c(\bar{a}((g^2+ \bar{e}\bar{f})h-h^3)\nonumber\\
&+b(-gh^2+g^3- \bar{e}\bar{f}g)) +d(b(- \bar{e}h^2- \bar{e}g^2+ \bar{e}^2\bar{f})-2 \bar{e}\bar{a}gh)-c^3\bar{f}h + \bar{e}cd^2h)\,,\nonumber\\ 
\beta_{8}&= -(( \bar{e}hm+ \bar{e}cg- \bar{e}^2d)n^2 +((2 \bar{e}dg-2cg^2)m+b( \bar{e}h^2+ \bar{e}g^2- \bar{e}^2\bar{f})+2c^2gh -2 \bar{e}\bar{a}gh) n-\bar{f}hm^3+(d(2h^2- \bar{e}\bar{f})+c\bar{f}g)m^2\nonumber\\
& +(\bar{a}((g^2+ \bar{e}\bar{f})h-h^3)+b(gh^2-g^3+ \bar{e}\bar{f}g)-2cdgh-c^2\bar{f}h + \bar{e}d^2h) m+c(b((g^2+ \bar{e}\bar{f})h-h^3)+\bar{a}(-gh^2+g^3- \bar{e}\bar{f}g))\nonumber\\
& +d(\bar{a}(- \bar{e}h^2- \bar{e}g^2+ \bar{e}^2\bar{f})+c^2(2h^2-2g^2- \bar{e}\bar{f})-2 \bar{e}bgh) +c^3\bar{f}g+3 \bar{e}cd^2g- \bar{e}^2d^3)\,,\nonumber\\ 
\gamma_{8}&= -( \bar{e}mn^3+(-2gm^2+ \bar{e}bh- \bar{e}\bar{a}g+ \bar{e}cd)n^2 +(\bar{f}m^3+(\bar{a}(-h^2+3g^2- \bar{e}\bar{f})-2bgh-4cdg+c^2\bar{f} + \bar{e}d^2) m+bc(-h^2+g^2+ \bar{e}\bar{f})\nonumber\\
&-2 \bar{e}bdg) n+(b\bar{f}h-\bar{a}\bar{f}g+cd\bar{f})m^2 +(bd(-h^2+g^2+ \bar{e}\bar{f})-2bc\bar{f}g)m +\bar{a}^2(gh^2-g^3+ \bar{e}\bar{f}g)+b^2(-gh^2+g^3- \bar{e}\bar{f}g)\nonumber\\
 &+d(c(\bar{a}(-h^2+3g^2- \bar{e}\bar{f})+2bgh)+c^3\bar{f}) +c^2(-b\bar{f}h-\bar{a}\bar{f}g)+d^2(- \bar{e}bh-2c^2g- \bar{e}\bar{a}g) + \bar{e}cd^3)\,,\nonumber\\ 
\delta_{8}&= ( \bar{e}^2bn^2+(2hm^3+(2c^2h-2 \bar{e}\bar{a}h-2 \bar{e}bg)m)n +(b(h^2+g^2)-2\bar{a}gh+2cdh)m^2 +c^2(b(-h^2-g^2)-2\bar{a}gh)\nonumber\\
& +d(c(2 \bar{e}bg-2 \bar{e}\bar{a}h)+2c^3h)-2 \bar{e}b^2gh +2 \bar{e}\bar{a}^2gh- \bar{e}^2bd^2) \,,\nonumber\\
\epsilon_{8}&=(( \bar{e}m^2+ \bar{e}c^2- \bar{e}^2\bar{a})n^2+(-2gm^3+2chm^2 +(2 \bar{e}bh-2c^2g+2 \bar{e}\bar{a}g)m +c(2 \bar{e}bg-2 \bar{e}\bar{a}h)+2c^3h -2 \bar{e}^2bd) n+\bar{f}m^4-2dhm^3\nonumber\\
& +(\bar{a}(h^2+g^2-2 \bar{e}\bar{f})-2bgh-2cdg +2c^2\bar{f}+ \bar{e}d^2) m^2 +(bc(2h^2-2g^2) +d(-2c^2h+2 \bar{e}\bar{a}h+2 \bar{e}bg)) m+c^2(\bar{a}(h^2+g^2-2 \bar{e}\bar{f})+2bgh)\nonumber\\
& +b^2( \bar{e}h^2+ \bar{e}g^2- \bar{e}^2\bar{f}) +\bar{a}^2(- \bar{e}h^2- \bar{e}g^2+ \bar{e}^2\bar{f}) +d(c(2 \bar{e}\bar{a}g-2 \bar{e}bh)-2c^3g)+c^4\bar{f} +( \bar{e}c^2- \bar{e}^2\bar{a})d^2)\,,\nonumber\\
\theta_{8}&= -( \bar{e}cn^3+(2hm^2+(-2cg- \bar{e}d)m+2c^2h- \bar{e}\bar{a}h+ \bar{e}bg)n^2 +((2dg+c\bar{f})m^2+(b(3h^2-g^2- \bar{e}\bar{f})-2\bar{a}gh)m\nonumber\\
& +\bar{a}c(-h^2+g^2- \bar{e}\bar{f}) +d(-2 \bar{e}bh-2c^2g)+c^3\bar{f}+ \bar{e}cd^2) n-d\bar{f}m^3+(-\bar{a}\bar{f}h+2d^2h+b\bar{f}g)m^2 +(d(\bar{a}(h^2-g^2+ \bar{e}\bar{f})-c^2\bar{f})\nonumber\\
&+2bc\bar{f}h+2cd^2g- \bar{e}d^3) m+b^2(h^3+(-g^2- \bar{e}\bar{f})h)+\bar{a}^2((g^2+ \bar{e}\bar{f})h-h^3) +cd(b(-3h^2+g^2+ \bar{e}\bar{f})-2\bar{a}gh)\nonumber\\
&+c^2(-\bar{a}\bar{f}h-b\bar{f}g) +d^2(2c^2h- \bar{e}\bar{a}h- \bar{e}bg))\,;\nonumber
\end{align}

\item $w_{9}$
\begin{align}
\alpha_{9}&= -(( \bar{e}hm+ \bar{e}cg- \bar{e}^2d)n^2 +(-2ghm^2+(c(2h^2-2g^2)+2 \bar{e}dg)m+b(- \bar{e}h^2- \bar{e}g^2+ \bar{e}^2\bar{f}) +2c^2gh-2 \bar{e}cdh) n+\bar{f}hm^3\nonumber\\
&+(d(-2h^2- \bar{e}\bar{f})+c\bar{f}g)m^2 +(\bar{a}(h^3+(g^2- \bar{e}\bar{f})h)+b(gh^2+g^3- \bar{e}\bar{f}g)-4cdgh+c^2\bar{f}h +3 \bar{e}d^2h) m+c(b(( \bar{e}\bar{f}-g^2)h-h^3)\nonumber\\
&+\bar{a}(gh^2+g^3- \bar{e}\bar{f}g)) +d(\bar{a}(- \bar{e}h^2- \bar{e}g^2+ \bar{e}^2\bar{f})+c^2(-2g^2- \bar{e}\bar{f}))+c^3\bar{f}g +3 \bar{e}cd^2g- \bar{e}^2d^3)\,,\nonumber\\  
\beta_{9}&= -( \bar{e}^2n^3+(3 \bar{e}ch-3 \bar{e}gm)n^2 +((2g^2+ \bar{e}\bar{f})m^2+(-4cgh-2 \bar{e}dh)m +\bar{a}( \bar{e}h^2+ \bar{e}g^2- \bar{e}^2\bar{f}) +c^2(2h^2+ \bar{e}\bar{f})-2 \bar{e}cdg+ \bar{e}^2d^2) n\nonumber\\
&-\bar{f}gm^3+(2dgh+c\bar{f}h)m^2 +(b(( \bar{e}\bar{f}-g^2)h-h^3)+\bar{a}(-gh^2-g^3+ \bar{e}\bar{f}g) +cd(2g^2-2h^2)-c^2\bar{f}g - \bar{e}d^2g) m\nonumber\\
&+c(\bar{a}(h^3+(g^2- \bar{e}\bar{f})h)+b(-gh^2-g^3+ \bar{e}\bar{f}g)) +d(b( \bar{e}h^2+ \bar{e}g^2- \bar{e}^2\bar{f})-2c^2gh)+c^3\bar{f}h + \bar{e}cd^2h)\,,\nonumber\\  
\gamma_{9}&= ( \bar{e}cn^3+((-2cg- \bar{e}d)m+2c^2h+ \bar{e}\bar{a}h- \bar{e}bg)n^2 +((2dg+c\bar{f})m^2+(b(h^2+g^2+ \bar{e}\bar{f})-2\bar{a}gh-4cdh)m +\bar{a}c(3h^2+g^2- \bar{e}\bar{f})\nonumber\\
& +d(-2 \bar{e}bh-2c^2g)+c^3\bar{f}+ \bar{e}cd^2) n-d\bar{f}m^3+(\bar{a}\bar{f}h+2d^2h-b\bar{f}g)m^2 +(d(\bar{a}(-3h^2-g^2+ \bar{e}\bar{f})-c^2\bar{f}) +2bc\bar{f}h+2cd^2g- \bar{e}d^3) m\nonumber\\
&+\bar{a}^2(h^3+(g^2- \bar{e}\bar{f})h)+b^2(( \bar{e}\bar{f}-g^2)h-h^3) +cd(b(-h^2-g^2- \bar{e}\bar{f})-2\bar{a}gh)+c^2(\bar{a}\bar{f}h+b\bar{f}g) +d^2( \bar{e}\bar{a}h+ \bar{e}bg))\,,\nonumber\\  
\delta_{9}&=(( \bar{e}m^2+ \bar{e}c^2- \bar{e}^2\bar{a})n^2+(-2gm^3+2chm^2 +(-2 \bar{e}bh-2c^2g+2 \bar{e}\bar{a}g) m+c(-2 \bar{e}\bar{a}h-2 \bar{e}bg) +2c^3h+2 \bar{e}^2bd) n+\bar{f}m^4\nonumber\\
&-2dhm^3 +(\bar{a}(h^2+g^2-2 \bar{e}\bar{f})+2bgh-2cdg +2c^2\bar{f}+ \bar{e}d^2) m^2 +(bc(2g^2-2h^2) +d(-2c^2h+2 \bar{e}\bar{a}h-2 \bar{e}bg)) m\nonumber\\
&+c^2(\bar{a}(h^2+g^2-2 \bar{e}\bar{f})-2bgh) +b^2( \bar{e}h^2+ \bar{e}g^2- \bar{e}^2\bar{f}) +\bar{a}^2(- \bar{e}h^2- \bar{e}g^2+ \bar{e}^2\bar{f}) +d(c(2 \bar{e}bh+2 \bar{e}\bar{a}g)-2c^3g)+c^4\bar{f} +( \bar{e}c^2- \bar{e}^2\bar{a})d^2)\,,\nonumber\\ 
\epsilon_{9}&= ( \bar{e}^2bn^2+(2 \bar{e}bch-2 \bar{e}bgm)n+b(g^2-h^2)m^2 +(2 \bar{e}bdh-4bcgh)m+bc^2(h^2-g^2)+2 \bar{e}bcdg - \bar{e}^2bd^2)\,,\nonumber\\  
\theta_{9}&= -( \bar{e}mn^3+(-2gm^2+2chm- \bar{e}bh- \bar{e}\bar{a}g+ \bar{e}cd)n^2 +(\bar{f}m^3-2dhm^2 +(\bar{a}(h^2+3g^2- \bar{e}\bar{f})-4cdg+c^2\bar{f}+ \bar{e}d^2)m\nonumber\\
& +c(b(-h^2-g^2- \bar{e}\bar{f})-2\bar{a}gh) +d(2c^2h+2 \bar{e}bg)) n+(b\bar{f}h-\bar{a}\bar{f}g+cd\bar{f})m^2 +(d(b(-h^2-g^2- \bar{e}\bar{f})+2\bar{a}gh)-2cd^2h+2bc\bar{f}g)m\nonumber\\
& +d(\bar{a}c(h^2+3g^2- \bar{e}\bar{f})+c^3\bar{f})+b^2(gh^2+g^3- \bar{e}\bar{f}g) +\bar{a}^2(-gh^2-g^3+ \bar{e}\bar{f}g)+c^2(-b\bar{f}h-\bar{a}\bar{f}g) +d^2( \bar{e}bh-2c^2g- \bar{e}\bar{a}g)+ \bar{e}cd^3)\,;\nonumber 
\end{align}

\item $w_{10}$
\begin{align}
\alpha_{10}&= -( \bar{e}gn^3+((-2h^2-2g^2- \bar{e}\bar{f})m+ \bar{e}dh)n^2 +(3\bar{f}gm^2-2c\bar{f}hm+b((-g^2- \bar{e}\bar{f})h-h^3) +\bar{a}(gh^2+g^3- \bar{e}\bar{f}g)\nonumber\\
&+cd(-2h^2-2g^2) +c^2\bar{f}g+ \bar{e}d^2g) n-\bar{f}^2m^3+d\bar{f}hm^2 +(\bar{a}(\bar{f}h^2-\bar{f}g^2+ \bar{e}\bar{f}^2)+2b\bar{f}gh+2cd\bar{f}g-c^2\bar{f}^2 - \bar{e}d^2\bar{f}) m\nonumber\\
& +d(\bar{a}((-g^2- \bar{e}\bar{f})h-h^3)+b(-gh^2-g^3+ \bar{e}\bar{f}g)-c^2\bar{f}h) +c(b(-\bar{f}h^2+\bar{f}g^2- \bar{e}\bar{f}^2)+2\bar{a}\bar{f}gh)+ \bar{e}d^3h)\,,\nonumber\\  
\beta_{10}&= ( \bar{e}hn^3+( \bar{e}c\bar{f}- \bar{e}dg)n^2 +(-\bar{f}hm^2+(d(2h^2+2g^2)-2c\bar{f}g)m +\bar{a}((-g^2- \bar{e}\bar{f})h-h^3)+b(gh^2+g^3- \bar{e}\bar{f}g)\nonumber\\
& +c^2\bar{f}h+ \bar{e}d^2h) n+(c\bar{f}^2-d\bar{f}g)m^2 +(b(\bar{f}h^2-\bar{f}g^2+ \bar{e}\bar{f}^2)+2\bar{a}\bar{f}gh-2cd\bar{f}h)m +d(b((-g^2- \bar{e}\bar{f})h-h^3)+\bar{a}(-gh^2-g^3+ \bar{e}\bar{f}g)\nonumber\\
& -3c^2\bar{f}g)+c(\bar{a}(-\bar{f}h^2+\bar{f}g^2- \bar{e}\bar{f}^2)+2b\bar{f}gh) +cd^2(2h^2+2g^2+ \bar{e}\bar{f})- \bar{e}d^3g+c^3\bar{f}^2) \,,\nonumber\\
\gamma_{10}&=( \bar{e}n^4-2gmn^3 +(\bar{f}m^2+\bar{a}(-h^2+g^2-2 \bar{e}\bar{f})-2cdg+c^2\bar{f}+2 \bar{e}d^2)n^2 +((2\bar{a}\bar{f}g-2d^2g)m+bd(-2h^2-2g^2)+2bc\bar{f}g)n\nonumber\\
& +(d^2\bar{f}-\bar{a}\bar{f}^2)m^2+(2bd\bar{f}g-2bc\bar{f}^2)m +\bar{a}^2(\bar{f}h^2-\bar{f}g^2+ \bar{e}\bar{f}^2)+b^2(-\bar{f}h^2+\bar{f}g^2- \bar{e}\bar{f}^2) +d^2(\bar{a}(-h^2+g^2-2 \bar{e}\bar{f})+c^2\bar{f})\nonumber\\
&+2\bar{a}cd\bar{f}g-2cd^3g -\bar{a}c^2\bar{f}^2+ \bar{e}d^4)\,,\nonumber\\
\delta_{10}&= ( \bar{e}cn^3+(-2hm^2+(-2cg- \bar{e}d)m+ \bar{e}\bar{a}h- \bar{e}bg)n^2 +((2dg+c\bar{f})m^2+(b(h^2+g^2+ \bar{e}\bar{f})+2\bar{a}gh-4cdh)m\nonumber\\
& +\bar{a}c(-h^2+g^2- \bar{e}\bar{f}) +d(-2 \bar{e}bh-2c^2g)+c^3\bar{f}+ \bar{e}cd^2) n-d\bar{f}m^3+(\bar{a}\bar{f}h-b\bar{f}g)m^2 +(d(\bar{a}(h^2-g^2+ \bar{e}\bar{f})-c^2\bar{f})\nonumber\\
&+2bc\bar{f}h+2cd^2g- \bar{e}d^3) m+b^2(h^3+(g^2+ \bar{e}\bar{f})h)+\bar{a}^2((-g^2- \bar{e}\bar{f})h-h^3) +cd(b(-h^2-g^2- \bar{e}\bar{f})+2\bar{a}gh)\nonumber\\
&+c^2(\bar{a}\bar{f}h+b\bar{f}g) +d^2(-2c^2h+ \bar{e}\bar{a}h+ \bar{e}bg))\,,\nonumber\\ 
\epsilon_{10}&= -( \bar{e}mn^3+(-2gm^2+2chm+ \bar{e}bh- \bar{e}\bar{a}g+ \bar{e}cd)n^2 +(\bar{f}m^3-2dhm^2 +(\bar{a}(h^2+3g^2- \bar{e}\bar{f})-4cdg+c^2\bar{f}+ \bar{e}d^2)m\nonumber\\
& +c(b(h^2+g^2+ \bar{e}\bar{f})-2\bar{a}gh) +d(2c^2h-2 \bar{e}bg)) n+(-b\bar{f}h-\bar{a}\bar{f}g+cd\bar{f})m^2 +(d(b(h^2+g^2+ \bar{e}\bar{f})+2\bar{a}gh)-2cd^2h-2bc\bar{f}g)m\nonumber\\
& +d(\bar{a}c(h^2+3g^2- \bar{e}\bar{f})+c^3\bar{f})+b^2(gh^2+g^3- \bar{e}\bar{f}g) +\bar{a}^2(-gh^2-g^3+ \bar{e}\bar{f}g)+c^2(b\bar{f}h-\bar{a}\bar{f}g) +d^2(- \bar{e}bh-2c^2g- \bar{e}\bar{a}g)+ \bar{e}cd^3) \,,\nonumber\\
\theta_{10}&= (2hmn^3+(b(h^2+g^2)-2\bar{a}gh+2cdh)n^2 +(-2\bar{a}\bar{f}h+2d^2h-2b\bar{f}g)mn+b\bar{f}^2m^2 +d^2(b(-h^2-g^2)-2\bar{a}gh)\nonumber\\
&+cd(2b\bar{f}g-2\bar{a}\bar{f}h) -2b^2\bar{f}gh+2\bar{a}^2\bar{f}gh+2cd^3h-bc^2\bar{f}^2)\,;\nonumber
\end{align}

\item $w_{11}$
\begin{align}
\alpha_{11}&= ((2h^3m+\bar{a}(gh^2+g^3- \bar{e}\bar{f}g)+c(gh^2-g^3+ \bar{e}\bar{f}g)-2 \bar{e}dh^2)n +(c(\bar{f}h^2+\bar{f}g^2- \bar{e}\bar{f}^2)+\bar{a}(\bar{f}h^2-\bar{f}g^2+ \bar{e}\bar{f}^2)-2dgh^2)m\nonumber\\
& +b(h^4-g^4+2 \bar{e}\bar{f}g^2- \bar{e}^2\bar{f}^2) +d(c(h^3+(-g^2- \bar{e}\bar{f})h)+\bar{a}((-g^2- \bar{e}\bar{f})h-h^3))+2\bar{a}c\bar{f}gh +2 \bar{e}d^2gh)\,,\nonumber\\
\beta_{11}&=(( \bar{e}h^2+ \bar{e}g^2- \bar{e}^2\bar{f})n^2+((2 \bar{e}\bar{f}g-2g^3)m+2cg^2h -2 \bar{e}dgh) n+(-\bar{f}h^2+\bar{f}g^2- \bar{e}\bar{f}^2)m^2 +(d(2h^3+2 \bar{e}\bar{f}h)-2c\bar{f}gh)m \nonumber\\
&+\bar{a}(-h^4+g^4-2 \bar{e}\bar{f}g^2+ \bar{e}^2\bar{f}^2) +c^2(\bar{f}h^2+\bar{f}g^2- \bar{e}\bar{f}^2) +d^2(- \bar{e}h^2+ \bar{e}g^2- \bar{e}^2\bar{f}) +cd(2 \bar{e}\bar{f}g-2g^3))\,,\nonumber\\
\gamma_{11}&= ( \bar{e}hn^3+(-2ghm+cg^2+\bar{a}( \bar{e}\bar{f}-g^2)- \bar{e}dg)n^2 +(\bar{f}hm^2+(2dg^2-2c\bar{f}g)m+\bar{a}((g^2- \bar{e}\bar{f})h-h^3) +b(-gh^2+g^3- \bar{e}\bar{f}g)\nonumber\\
&-2cdgh+c^2\bar{f}h + \bar{e}d^2h) n+(c\bar{f}^2-d\bar{f}g)m^2+b(\bar{f}h^2-\bar{f}g^2+ \bar{e}\bar{f}^2)m +d(b((g^2- \bar{e}\bar{f})h-h^3)+\bar{a}(gh^2-g^3+ \bar{e}\bar{f}g)-c^2\bar{f}g\nonumber\\
& -2\bar{a}c\bar{f}g) +d^2(\bar{a}(h^2+ \bar{e}\bar{f})+c(2g^2-h^2)) +\bar{a}^2(-\bar{f}h^2+\bar{f}g^2- \bar{e}\bar{f}^2)- \bar{e}d^3g+\bar{a}c^2\bar{f}^2)\,,\nonumber\\
\delta_{11}&= -( \bar{e}^2n^3+( \bar{e}\bar{a}h-3 \bar{e}gm)n^2 +((2h^2+2g^2+ \bar{e}\bar{f})m^2+(-cgh+\bar{a}gh-2 \bar{e}dh)m +\bar{a}(- \bar{e}h^2+ \bar{e}g^2- \bar{e}^2\bar{f}) -2 \bar{e}bgh+\bar{a}cg^2\nonumber\\
& +c^2( \bar{e}\bar{f}-g^2) +d(- \bar{e}cg- \bar{e}\bar{a}g)+ \bar{e}^2d^2) n-\bar{f}gm^3+c\bar{f}hm^2 +(b(h^3+(g^2+ \bar{e}\bar{f})h) +d(c(h^2+2g^2- \bar{e}\bar{f})+\bar{a}(h^2+ \bar{e}\bar{f}))\nonumber\\
& +\bar{a}(-gh^2-g^3+ \bar{e}\bar{f}g)-\bar{a}c\bar{f}g- \bar{e}d^2g) m+\bar{a}^2((-g^2- \bar{e}\bar{f})h-h^3) +d(b(- \bar{e}h^2+ \bar{e}g^2- \bar{e}^2\bar{f})-c^2gh+\bar{a}cgh +2 \bar{e}\bar{a}gh)\nonumber\\
& +bc(-gh^2-g^3+ \bar{e}\bar{f}g)+\bar{a}c^2\bar{f}h- \bar{e}cd^2h)\,,\nonumber\\ 
\epsilon_{11}&= -(( \bar{e}hm+ \bar{e}cg- \bar{e}^2d)n^2 +(-2ghm^2+(c(2h^2-g^2- \bar{e}\bar{f})+\bar{a}( \bar{e}\bar{f}-g^2)+2 \bar{e}dg)m +b( \bar{e}h^2+ \bar{e}g^2- \bar{e}^2\bar{f})\nonumber\\
&+d(- \bar{e}ch- \bar{e}\bar{a}h)+c^2gh +\bar{a}cgh) n+\bar{f}hm^3+(d(-2h^2- \bar{e}\bar{f})+c\bar{f}g)m^2 +(\bar{a}(h^3+(g^2- \bar{e}\bar{f})h)\nonumber\\
&+b(-gh^2-g^3+ \bar{e}\bar{f}g)+d(-3cgh-\bar{a}gh) +\bar{a}c\bar{f}h+3 \bar{e}d^2h) m+bc(h^3+(g^2- \bar{e}\bar{f})h) +d(\bar{a}(- \bar{e}h^2- \bar{e}g^2+ \bar{e}^2\bar{f})\nonumber\\
&+c^2(h^2- \bar{e}\bar{f})+\bar{a}c(-h^2-2g^2)) +\bar{a}^2(gh^2+g^3- \bar{e}\bar{f}g)+d^2(2 \bar{e}cg+ \bar{e}\bar{a}g)+\bar{a}c^2\bar{f}g- \bar{e}^2d^3)\,,\nonumber\\ 
\theta_{11}&= ( \bar{e}gn^3+((2h^2-2g^2- \bar{e}\bar{f})m+cgh+\bar{a}gh- \bar{e}dh)n^2 +(3\bar{f}gm^2+(-2dgh-c\bar{f}h+\bar{a}\bar{f}h)m +b(h^3+(-g^2- \bar{e}\bar{f})h)\nonumber\\
& +d(c(h^2-g^2)+\bar{a}(-h^2-g^2)) +\bar{a}(-gh^2+g^3- \bar{e}\bar{f}g)+\bar{a}c\bar{f}g+ \bar{e}d^2g) n-\bar{f}^2m^3+d\bar{f}hm^2 +(\bar{a}(-\bar{f}h^2-\bar{f}g^2+ \bar{e}\bar{f}^2)\nonumber\\
&+d^2(2h^2- \bar{e}\bar{f})+2b\bar{f}gh +d(c\bar{f}g+\bar{a}\bar{f}g)-\bar{a}c\bar{f}^2) m +d(\bar{a}((g^2+ \bar{e}\bar{f})h-h^3)+b(-gh^2+g^3- \bar{e}\bar{f}g)+\bar{a}c\bar{f}h)\nonumber\\
& +bc(-\bar{f}h^2-\bar{f}g^2+ \bar{e}\bar{f}^2)+d^2(\bar{a}gh-cgh)-2\bar{a}^2\bar{f}gh - \bar{e}d^3h)\,,\nonumber
\end{align}

\item $w_{12}$
\begin{align}
\alpha_{12}&=(( \bar{e}h^2- \bar{e}g^2+ \bar{e}^2\bar{f})n^2+((2g^3-2 \bar{e}\bar{f}g)m+c(2h^3+2 \bar{e}\bar{f}h) -2 \bar{e}dgh) n+(-\bar{f}h^2-\bar{f}g^2+ \bar{e}\bar{f}^2)m^2 +(2dg^2h-2c\bar{f}gh)m\nonumber\\
& +\bar{a}(h^4-g^4+2 \bar{e}\bar{f}g^2- \bar{e}^2\bar{f}^2) +c^2(\bar{f}h^2-\bar{f}g^2+ \bar{e}\bar{f}^2) +d^2(- \bar{e}h^2- \bar{e}g^2+ \bar{e}^2\bar{f}) +cd(2g^3-2 \bar{e}\bar{f}g))\,,\nonumber\\
\beta_{12} &= (2 \bar{e}ghn^2+((-2g^2-2 \bar{e}\bar{f})hm+2cgh^2+2 \bar{e}dh^2)n +2\bar{f}ghm^2+(-2dgh^2-2c\bar{f}h^2)m +b(-h^4+g^4-2 \bar{e}\bar{f}g^2+ \bar{e}^2\bar{f}^2)+2cdh^3) \,,\nonumber\\
\gamma_{12}&= ( \bar{e}gn^3+((-2g^2- \bar{e}\bar{f})m+ \bar{e}dh)n^2 +(3\bar{f}gm^2-2dghm+b((g^2- \bar{e}\bar{f})h-h^3) +\bar{a}(-gh^2+g^3- \bar{e}\bar{f}g)-2cdg^2+c^2\bar{f}g + \bar{e}d^2g) n\nonumber\\
&-\bar{f}^2m^3+d\bar{f}hm^2 +(\bar{a}(\bar{f}h^2-\bar{f}g^2+ \bar{e}\bar{f}^2)+2cd\bar{f}g-c^2\bar{f}^2- \bar{e}d^2\bar{f})m +d(\bar{a}((g^2- \bar{e}\bar{f})h-h^3)+b(gh^2-g^3+ \bar{e}\bar{f}g)+c^2\bar{f}h)\nonumber\\
& +bc(-\bar{f}h^2+\bar{f}g^2- \bar{e}\bar{f}^2)-2cd^2gh+ \bar{e}d^3h) \,,\nonumber\\
\delta_{12}&  = -(( \bar{e}hm- \bar{e}cg+ \bar{e}^2d)n^2 +((c(2h^2+2g^2)-2 \bar{e}dg)m+b(- \bar{e}h^2+ \bar{e}g^2- \bar{e}^2\bar{f})-2 \bar{e}\bar{a}gh +2 \bar{e}cdh) n-\bar{f}hm^3\nonumber\\
&+( \bar{e}d\bar{f}-c\bar{f}g)m^2 +(\bar{a}(h^3+(g^2+ \bar{e}\bar{f})h)+b(-gh^2-g^3+ \bar{e}\bar{f}g)-c^2\bar{f}h- \bar{e}d^2h)m +c(b((-g^2- \bar{e}\bar{f})h-h^3)\nonumber\\
&+\bar{a}(-gh^2-g^3+ \bar{e}\bar{f}g)) +d(\bar{a}(- \bar{e}h^2+ \bar{e}g^2- \bar{e}^2\bar{f})+c^2(2h^2+2g^2+ \bar{e}\bar{f})+2 \bar{e}bgh) -c^3\bar{f}g-3 \bar{e}cd^2g+ \bar{e}^2d^3)\,,\nonumber\\ 
\epsilon_{12}&= -( \bar{e}^2n^3+(3 \bar{e}ch-3 \bar{e}gm)n^2 +((2g^2+ \bar{e}\bar{f})m^2+(-4cgh-2 \bar{e}dh)m +\bar{a}( \bar{e}h^2+ \bar{e}g^2- \bar{e}^2\bar{f}) +c^2(2h^2+ \bar{e}\bar{f})-2 \bar{e}cdg+ \bar{e}^2d^2) n\nonumber\\
&-\bar{f}gm^3+(2dgh+c\bar{f}h)m^2 +(b(h^3+(g^2- \bar{e}\bar{f})h)+\bar{a}(-gh^2-g^3+ \bar{e}\bar{f}g) +cd(2g^2-2h^2)-c^2\bar{f}g - \bar{e}d^2g) m+c(\bar{a}(h^3\nonumber\\
&+(g^2- \bar{e}\bar{f})h)+b(gh^2+g^3- \bar{e}\bar{f}g)) +d(b(- \bar{e}h^2- \bar{e}g^2+ \bar{e}^2\bar{f})-2c^2gh)+c^3\bar{f}h + \bar{e}cd^2h) \,,\nonumber\\
\theta_{12}& = ( \bar{e}hn^3+(c(2h^2- \bar{e}\bar{f})+ \bar{e}dg)n^2 +(-\bar{f}hm^2+(2c\bar{f}g-2dg^2)m+\bar{a}(h^3+(-g^2- \bar{e}\bar{f})h) +b(-gh^2+g^3- \bar{e}\bar{f}g)+2cdgh\nonumber\\
&-c^2\bar{f}h + \bar{e}d^2h) n+(d\bar{f}g-c\bar{f}^2)m^2 +(b(-\bar{f}h^2-\bar{f}g^2+ \bar{e}\bar{f}^2)+2\bar{a}\bar{f}gh-2d^2gh)m +d(b((g^2+ \bar{e}\bar{f})h-h^3)\nonumber\\
&+\bar{a}(-gh^2+g^3- \bar{e}\bar{f}g)+3c^2\bar{f}g) +c(\bar{a}(-\bar{f}h^2-\bar{f}g^2+ \bar{e}\bar{f}^2)-2b\bar{f}gh) +cd^2(2h^2-2g^2- \bar{e}\bar{f})+ \bar{e}d^3g-c^3\bar{f}^2)\,.\nonumber
\end{align}
\end{itemize}

\subsubsection{Subspace $\caN_{1}$}\label{eigencompN1}

The components, which are obtained from those in the subspace $\caM_{1}$ by setting $d=e=m=n=p=q=0$, are:
\begin{itemize}
\item $w_{1}$
\begin{align}
\alpha_{1}&= c(h+\bar{f})-2bl\,,\quad\beta_{1}=\gamma_{1}= \bar{a}l-bc^{*}\,,\quad\delta_{1}=-\bar{a}(h+\bar{f})+2bb^{*}\,,\nonumber
\end{align}
\item $w_{2}$ and $w_{3}$
\begin{align}
\alpha_{2}&=\alpha_{3}=(b\bar{g}-cl)(\bar{f}-h)\,,\quad\beta_{2}=\gamma_{3}=-(\bar{a}(l^2-\bar{g}h)-2bc^{*}l+cc^{*}h+bb^{*}\bar{g})\,,\nonumber\\
\gamma_{2}&=\beta_{3}=\bar{a}(l^2-\bar{f}\bar{g})-2bc^{*}l+b^2\bar{g}+cc^{*}\bar{f}\,,\quad\delta_{2}=\delta_{3}=(\bar{a}l-bc^{*})(\bar{f}-h)\,.\nonumber
\end{align}
\item $w_{4}$
\begin{align}
\alpha_{4}&=2l^2-\bar{g}(h+\bar{f})\,,\quad \beta_{4}=\gamma_{4}=b\bar{g}-cl\,,\quad\delta_{4}=c(h+\bar{f})-2bl\,;\nonumber
\end{align}
\end{itemize}
they must be divided by $\|w_{k}\|=(\alpha_{k}^{2}+\beta_{k}^{2}+\gamma_{k}^{2}+\delta_{k}^{2})^{1/2}$, $k=1\,,\ldots 4$.

\subsubsection{Subspace $\caN_{3}$}\label{eigencompN3}

The components, which are obtained from those in the subspace $\caM_{2}$ by setting $m=n=h=0$:
\begin{itemize}
\item $w_{7}$ and $w_{10}$:
\begin{align}
\alpha_{7}&=\beta_{10}=(\bar{e}\bar{f}-dg)(\bar{a}(g^2-c^{\star}\bar{f})+d(d^{\star}\bar{e}-c^{\star}g)+c(c^{\star}\bar{f}-d^{\star}g)) \,,\nonumber\\
\beta_{7}&=\alpha_{10}=b(dg-c\bar{f})(g^2- \bar{e}\bar{f})  \,,\nonumber\\
\gamma_{7}&=\theta_{10}=-b(dg-c\bar{f})^2 \,,\nonumber\\
\delta_{7}&=\epsilon_{10}=-((g^2- \bar{e}\bar{f})(g(b^2-\bar{a}^2)+\bar{a}cd^{\star})+(dc-\bar{a}g)(c^{\star}(c\bar{f}-dg)+d(d^{\star}\bar{e}-c^{\star}g))) \,,\nonumber\\
\epsilon_{7}&=\delta_{10}=b(\bar{e}d-cg)(gd^{\star}-\bar{e}\bar{f})\,,\nonumber\\
\theta_{7}&=\gamma_{10}=\bar{f}(b^2-\bar{a}^2)(g^2- \bar{e}\bar{f})+\bar{a}(dg+c\bar{f})^2-2\bar{a}\bar{f}(cc^{\star}bar{f}+d^2\bar{e})+cdd^{\star}(c\bar{f}-dg)+dd^{\star}d(\bar{e}d^{\star}-c^{\star}g)\,,\nonumber
\end{align}
\item $w_{8}$ and $w_{9}$:
\begin{align}
\alpha_{8}&=\beta_{9}=b(cg-d\bar{e})(g^2-\bar{e}\bar{f}) \,,\nonumber\\
\beta_{8}&=\alpha_{9}=(d\bar{e}-cg)(\bar{a}(g^2-\bar{e}\bar{f})+d(d\bar{e}-gc)+c^{\star}(c\bar{f}-dg))\,,\nonumber\\
\gamma_{8}&=\theta_{9}= -((g^2- \bar{e}\bar{f})(g(b^2-\bar{a}^2)+\bar{a}cd^{\star})+(dc-\bar{a}g)(c^{\star}(c\bar{f}-dg)+d(d^{\star}\bar{e}-c^{\star}g))) \,,\nonumber\\
\delta_{8}&=\epsilon_{9}= -b(cg-\bar{e}d)^2 \,,\nonumber\\
\epsilon_{8}&=\delta_{9}=\bar{e}(b^2-\bar{a}^2)( g^2- \bar{e}\bar{f})+\bar{a}(cg+d\bar{e})^2-2\bar{a}\bar{e}(cc^{\star}\bar{f}+dd^{\star}\bar{e})+dc^{\star}c(\bar{e}d^{\star}-c^{\star}g)-cc^{\star}c(d^{\star}g-\bar{e}\bar{f})\,,\nonumber\\
\theta_{8}&=\gamma_{9}=b(dg-c\bar{f})(\bar{e}d^{\star}-c^{\star}g) \,,\nonumber 
\end{align}
\item $w_{11}$ and $w_{12}$:
\begin{align}
\alpha_{11}&=\beta_{12}= -b(g^2- \bar{e}\bar{f})^2\,,\nonumber\\
\beta_{11}&=\alpha_{12}=(g^2- \bar{e}\bar{f})(\bar{a}(g^2- \bar{e}\bar{f})+c(c^{\star}\bar{f}-d^{\star}g)+d(\bar{e}d^{\star}-c^{\star}g))\,,\nonumber\\
\gamma_{11}&=\theta_{12}=(c\bar{f}-dg)(\bar{a}(g^2- \bar{e}\bar{f})+c^{\star}(c\bar{f}-dg)+d^{\star}(\bar{e}d-cg))\,,\nonumber\\ 
\delta_{11}&=\epsilon_{12}= b(g^2- \bar{e}\bar{f})(cg-\bar{e}d)\,,\nonumber\\
\epsilon_{11}&=\delta_{12}=( \bar{e}d-cg)(\bar{a}(g^2- \bar{e}\bar{f})+c^{\star}(c\bar{f}-dg)+d^{\star}(\bar{e}d-cg))\,,\nonumber\\ 
\theta_{11}&=\gamma_{12}= b(g^2- \bar{e}\bar{f})(dg-c\bar{f})\,,\nonumber
\end{align}
\end{itemize}
also these components must be divided by $\|w_{k}\|=(\alpha_{k}^{2}+\beta_{k}^{2}+\gamma_{k}^{2}+\delta_{k}^{2}+\epsilon_{k}^{2}+\theta_{k}^{2})^{1/2}$, $k=7\,,\ldots 12$.


\section*{References}  

\begin{thebibliography}{99}

\bibitem{LEC06} P.~Lecoq, A.~Annekov, A.~Getkin, M.~Korzhik, C.~Pedrini, \emph{Inorganic {S}cintillators for {D}etector
{S}ystems}, Springer, Berlin Heidelberg New York, 2006.

\bibitem{ER13}  W.~Erni, I.~Keshelashvili, B.~Krusche, et al., \emph{Technical design report for the {PANDA} ({A}nti{P}roton {A}nnihilations at {D}armstadt) {S}traw {T}ube {T}racker}, 
Eur. Phys. J. A, \textbf{49}, (2013), 25.

\bibitem{DO05} V.I.~Dormenev,  G.Y.~Drobyshev, M.V.~Korzhik, et al. \emph{Studying the Kinetics of Radiation Damage in PWO Crystals for the CMS Electromagnetic Calorimeter (CERN)}. Instrum. Exp. Tech- \textbf{48}, 303--307 (2005). https://doi.org/10.1007/s10786-005-0055-5.

\bibitem{ME15} P.~MengucciP, G.~Andr\'{e}, E.~Auffray, G.~Barucca, R.~Chipaux, C.~Cecchi, A.~Cousson, F.~Dav\'{\i},  N.~Di Vara, D.~Rinaldi and E.~Santecchia \emph{Structural, mechanical and light yield characterisation of heat treated {LYSO:Ce} single crystals for medical imaging applications}, Nuclear Instrument and Methods in Physics Research-A \textbf{785}, 2015, 110--116.

\bibitem{LAB} D.J.~Singh \emph{Structure and optical properties of high light output halide scintillators}, Phys. Rev.B\textbf{82}, (2010), 155145.

\bibitem{AKL02} A.A.~Annenkova, M.V.~Korzhik, P.~Lecoq, \emph{Lead  tungstate scintillation material},  Nuclear Instrument and Methods in Physics Research-A \textbf{490} (2002), 30--50.

\bibitem{BH54} M.~Born and K.~Huang, Dynamical Theory of Crystal Lattices, \emph{International Series of Monographs on Physics}, Clarendon Press, 1954. 

\bibitem{DO93} M.T.~Dove, \emph{Introduction to Lattice Dynamics}. Cambridge University Press, 1993.

\bibitem{CA00} G.~Capriz, \emph{Continua with {M}icrostructure}, Springer Tracts in Natural Philosophy, Vol. \textbf{35}, Springer-Verlag, New York, 1989.

\bibitem{ER99} A.C.~Eringen, \emph{Microcontinuum {F}ield {T}heories. I: Foundations and solids}. Springer, New York, 1999.

\bibitem{MI64} R~.D.~Mindlin, \emph{Micro-structure in {L}inear {E}lasticity}, Arch. Rat. Mech. Anal. \textbf{16},  1964, 51--77.

\bibitem{BBEN11} A.~Berezovsky, J.~Engelbrecht, M.~Berezovsky, \emph{Waves in microstructured solids: A unified viewpoint of modeling} Acta Mechanica \textbf{220}(1):349--363, March 2011.

\bibitem{NE13} P.~Neff, I-D.~Ghiba, A.~Madeo, L.~Placidi, G.~Rosi.\emph{A unifying perspective: the relaxed linear micromorphic continuum}. Continuum Mech. Thermodyn. \textbf{26}(5), 639--681, 2014, doi: 10.1007/s00161-013-0322-9

\bibitem{CPW82}
G.~Capriz, P.~Podio-Guidugli, W.~Williams, \emph{On balance equations for materials with affine structure}, Meccanica \textbf{17}, 1982, 80--84.

\bibitem{NKDM11} R.W.~Novotny, D.~Bremer, V.~Dormenev, P.~Drexler, T.~Eissner, T.~Kuske, M.~Moritz and the PANDA collaboration, \emph{High-quality PWO crystals for the PANDA-EMC}, XIV International Conference on Calorimetry in High Energy Physics (CALOR 2010), Journal of Physics: Conference Series \textbf{293}, 012003, 2011. doi:10.1088/1742-6596/293/1/012003

\bibitem{BANE17} G.~Barbagallo, A.~Madeo, M.V.~d'Agostino, R.~Abreu, I.-D.~Ghiba and P.~Neff, \emph{Transparent anisotropy for the relaxed micromorphic model: macroscopic consistency conditions and long wave length asymptotics}. Int. J. of Solids and Structures, \textbf{120}, 7--30, 2017, doi.org/10.1016/j.ijsolstr.2017.01.030.

\bibitem{MS19} H.~Moosavian, H.~M.~Shodja, \emph{Mindlin-Eringen anisotropic micromorphic elasticity and lattice dynamics representation}, Philosophical Magazine, DOI: 10.1080/14786435.2019.1671998, (2019).

\bibitem{MS20} H.~M.~Shodja, H.~Moosavian,  \emph{Weakly nonlocal micromorphic elasticity for diamond structures vis-a-vis lattice dynamics}, to appear on Mechanics of Materials, 2020. doi.org/10.1016/j.mechmat.2020.103365

\bibitem{ER68} A.C.~Eringen, \emph{Mechanics of {M}icromorphic {C}ontinua}, in: Kr\"{o}ner E. (ed.), Mechanics of Generalized Continua. IUTAM Symposia. Springer-Verlag, Berlin, Heidelberg, 1968, 18--35.

\bibitem{BANE16} P.~Neff, A.~Madeo, G.~Barbagallo, M.V.~d'Agostino, R.~Abreu and I.-D.~Ghiba, \emph{Real wave propagation in the isotropic relaxed micromorphic model}. Proc. Royal Soc. A\textbf{473}, 20160790,  2017. http://dx.doi.org/10.1098/rspa.2016.0790.

\bibitem{GU72} M~E.~Gurtin, 1972, The linear theory of elasticity. In: C. Truesdell (ed.) {\it Handbook of Physics}, Vol. VIa/2. Berlin, Springer Verlag.

\bibitem{MANE18} A.~Madeo, P.~Neff, I.-D.~Ghiba, L.~Placidi and G.~Rosi, \emph{Wave propagation in relaxed micromorphic continua: modelling metamaterials with frequency band-gaps}, Continuum Mechanics and Thermodynamics, \textbf{27}(4), 551--570, 2015. doi.org/10.1007/s00161-013-0329-2

\bibitem{LA64}
P.~Lancaster, \emph{On {E}igenvalues of {M}atrices {D}ependent on a {P}arameter}. Numerische Mathematik, \textbf{6}, (1964). 377--387. 

\bibitem{MANE16} A.~Madeo, P.~Neff, M.V.~d'Agostino, and G.~Barbagallo,  \emph{Complete band gaps including non-local effects occur only in the relaxed micromorphic model}. C. R. Mecanique \textbf{344}, 784-796, 2016. doi.org/10.1016/j.crme.2016.07.002.

\bibitem{MANE17} A.~Madeo, P.~Neff, \emph{Dispersion of {W}aves in {M}icromorphic {M}edia and {M}etamaterials} in Handbook of Nonlocal Continuum Mechanics for Materials, G.Z. Voyiadjis (ed.),  Springer Int. Publ. AG, 2017.  doi.org/10.1007/978-3-319-22977-5 12-1

\bibitem{SA41} Z. Sadaki, 1941, \emph{Elastic waves in crystals}. Proc. Phys. Math. Soc. Japan (3) {\bf 23}, 539--547.

\bibitem{AC73} J.~D.~Achenbach,  Wave propagation in elastic solids. {\it N. H. Series in Applied Mathematics and Mechanics\/}, Vol. 16, North-Holland Publishing Company, Amsterdam, New York, Oxford, 1973.

\bibitem{ABNE18} M.~V.~d'Agostino, G.~Barbagallo, I.~D.~Ghiba, B.~Eidel, P.~Neff, A.~Madeo, \emph{Effective Description of Anisotropic Wave Dispersion in Mechanical Band-Gap Metamaterials via the Relaxed Micromorphic Model}, J. Elasticity, (2019).

\bibitem{CRC}  D.~Zwillinger (Ed.), \emph{Standard {M}athematical {T}ables and {F}ormulae, 31$^{\rm{st}}$ {E}dition}, CRC Press Company, Boca Raton, 2003.

\bibitem{AU03}
A.~Authier (Ed.), \emph{International {T}ables for {C}rystallography. {V}olume {D}: {P}hysical {P}roperties of {C}rystals}, Kluwer Academic Publ. Dordrecht, 2003.

\bibitem{FIFU53} R.~Fieschi, F.~G.~Fumi, \emph{High-order {T}ensors in {S}ymmetrical {S}ystems}, Il Nuovo Cimento, vol X, N.7, serie nona,  865--882, (1953).

\bibitem{BOU64} J.~Bouman, \emph{On the {C}onstruction of {M}atter {T}ensors in {C}rystals}, Acta Crystallographica, \textbf{17}, 15--20, (1953).

\bibitem{OLA13a} M.~Olive, N.~Auffray, \emph{Symmetry classes for odd-order tensors}. Journal of Applied Mathematics and Mechanics / Zeitschrift f\"ur Angewandte Mathematik und Mechanik, \textbf{94}(5), 421--447, (2014).

\bibitem{OLA13b} M.~Olive, N.~Auffray, \emph{Symmetry classes for even-order tensors}, Mathematics and Mechanics of Complex Systems, vol. \textbf{1}(2), 177--210, (2013).

\bibitem{ALH18} N.~Auffray, Q.-C.~He, H.~Le~Quang,.\emph{Complete symmetry classification and compact matrix representations for {3D} strain gradient elasticity}. Int. J. of Solids and Structures, \textbf{159}, 197--210, (2019).

\bibitem{QHE10} H.~Le~Quang, Q.-C.~He, \emph{The number and types of all possible rotational symmetries for flexoelectric tensors}, Proc. Royal Society A, \textbf{467}, 2369--2386, (2011).
 
\end{thebibliography}
\end{document}